\documentclass{pinchcr}   
\usepackage{graphicx}
\usepackage{amsmath,amssymb,amsfonts}

\title{Glassy dynamics of electrons near the metal-insulator transition}

\author{Dragana Popovi\'c}
\affiliation{National High Magnetic Field Laboratory, Florida State University\\ Tallahassee, Florida 32310, USA}

\begin{document}

\maketitle

\preface

This review first describes the evidence that strongly suggests the existence of the metal-insulator transition (MIT) in a two-dimensional electron system in Si regardless of the amount of disorder.  Extensive studies of the charge dynamics demonstrate that this transition is closely related to the glassy freezing of electrons as temperature $T\rightarrow 0$.  Similarities to the behavior of three-dimensional materials raise the intriguing possibility that such correlated dynamics might be a universal feature of the MIT regardless of the dimensionality.

\tableofcontents

\maintext

\section{Introduction}
\label{intro}

\begin{figure}[b]
	\centering
		\includegraphics[width=0.55\textwidth]{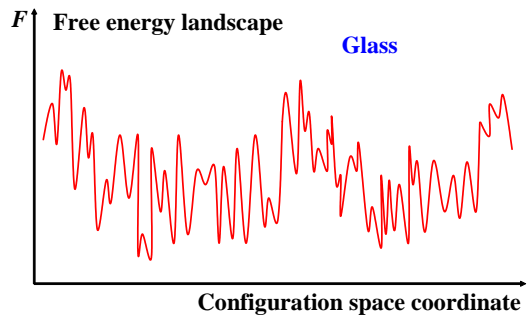}
		\caption{The free-energy ($F$) landscape of a glassy system.  The horizontal axis represents the one-dimensional projection of the configurational coordinates of the degrees of freedom.}
	\label{fig:FEL}
\end{figure}

The discovery of many novel materials over the last couple of decades has revived interest in the metal-insulator transition (MIT), one of the longstanding, fundamental problems of condensed matter physics.  Many such materials, including manganites, diluted magnetic semiconductors and high-temperature superconductors, are created by doping an insulating host and thus find themselves close to a conductor-insulator transition.  There is substantial evidence that, in many cases, strong electronic correlations (Mott localization) and disorder (Anderson localization) both play an important role in this regime (see \shortciteNP{Mir-Dob-review}
for review).  In general, the competition between the Coulomb repulsion, which favors a uniform distribution of electrons (Wigner crystallization), and disorder, which favors a random one, leads to the frustration in the system.  This means that electrons are unable to satisfy these requirements simultaneously, resulting in the absence of a unique ground state and the emergence of a large number of metastable states or minima (``valleys'') in the free energy landscape.  Metastable states, which correspond to different charge configurations, are separated by high barriers, leading to slow dynamics, divergence of the equilibration time, and breaking of ergodicity.  Such a system is called a Coulomb glass.  However, there are many other types of glassy systems.  A common denominator in all of them is their complex or ``rugged'' energy landscape (Fig.~\ref{fig:FEL}).  The most extensively studied and best known glasses in condensed matter physics are probably conventional spin glasses \shortcite{Binder}, such as Cu:Mn.  On the other hand, Coulomb glasses were first anticipated theoretically almost three decades ago \shortcite{eglass1,eglass2,eglass3,eglass4,eglass5} in situations where electrons are strongly localized due to disorder, but experimental studies have been scarce.  Some of the recent work on electron glasses in such strongly localized regime, away from the MIT, has been reviewed by \shortciteN{ArielAmir-review}.  In the opposite limit of well-delocalized electrons, one expects a single, well-defined ground state and the absence of glassiness.  Not surprisingly, it is the behavior in the intermediate region, near the MIT, that has been most difficult to understand.

In the presence of disorder, the local electron density undergoes strong spatial fluctuations.  Therefore, it is plausible to expect that, in the vicinity of the MIT, the local density in some areas may be higher than the average and correspond to the metallic state, whereas in the remaining regions, the local density may be lower than the average and the electrons are localized.  In fact, it has been proposed that, in weakly doped Mott insulators near the MIT, the system will settle for a nanoscale phase separation\footnote{Global phase separation is not possible as it would violate charge neutrality.} between a conductor and an insulator \shortcite{phasesep1,phasesep2,phasesep3}.  New powerful experimental techniques, including imaging by scanning tunneling microscopy \shortcite{STM-Seamus},
have indeed firmly established the existence of charge inhomogeneities in strongly correlated electron systems, such as cuprates \shortcite{Millis,Elbio}.  This obviously leads to the possibility for a myriad of competing charge configurations and the emergence of the associated glassy dynamics.  Recent studies \shortcite{IR_SPIE07,IR_PRL,Glenton,IR-inplane} of a lightly doped, insulating La$_{2-x}$Sr$_x$CuO$_4$, $x=0.03$, have found several clear signatures of glassy charge dynamics as temperature $T\rightarrow 0$, consistent with an underlying glassy ground state that results from Coulomb interactions.  Further work is needed, however, to see how this glassy state evolves as the system approaches the transition from an insulator to a conductor.

It is also desirable to extend studies of charge dynamics to other types of materials to explore a possible generality of glassy freezing in strongly correlated systems near the MIT \shortcite{Darko-glass,Pankov-scr,Mir-Dob-review}. Many materials do exhibit complex behavior due to the existence of several competing ground states \shortcite{Elbio}.  In fact, the frustration caused by the competition of interactions on different length scales may give rise to glassy dynamics even in the absence of disorder \shortcite{Schmalian}, while even a small amount of disorder may favor glassiness over various static charge-ordered states \shortcite{Pankov-scr}.  Even though the emergence of glassiness thus appears to be ubiquitous at low temperatures, Coulomb glasses and out-of-equilibrium systems in general remain poorly understood.  Experimentally, studies of charge dynamics near the MIT in many materials are often complicated by the accompanying changes in magnetic or structural symmetry, as well as by the glassy freezing of spins.  Doped semiconductors, such as Si:P and Si:B, are free from such complications.  They have been used extensively to study the critical behavior near the MIT \shortcite{Mir-Dob-review,Myriam_review}
but, in spite of some early hints of glassiness in the insulating regime \shortcite{Rosenbaum,Monroe1,Monroe2}, charge dynamics was not studied further until recently \shortcite{Kar,Armitage}, as discussed below.

Two-dimensional electron systems (2DES) in semiconductors~\shortcite{AFS}, such as Si, provide another relatively simple system for exploring the interplay of electronic correlations and disorder (for a review of disordered electronic systems, see \shortciteNP{LR}).  They have an additional advantage that all the relevant parameters, carrier concentration, disorder and interactions, can be varied relatively easily.  This review will describe some of the work that has demonstrated that the 2DES in Si is an excellent model system not only for studying the MIT in two dimensions (2D), but also for investigating the nearly universal nonequilibrium behavior exhibited by a large class of both three-dimensional (3D) and 2D systems (\textit{e.g.} spin glasses, supercooled liquids, granular films).  In fact, the work on the 2DES in Si provides additional strong evidence that many such universal features are robust manifestations of glassiness, regardless of the dimensionality of the system.

The dimensionality
plays an important role in the MIT.  In 2D, for example, the very existence of the metal and the MIT had been questioned for many years.  Recently, considerable experimental evidence  has become available in favor of such a transition.  Some of that work has been described in several review papers (\textit{e.g.} \shortciteNP{Sergey-review})
with a focus on very clean (low-disordered) samples.  This review will first extend that discussion to the cases of higher disorder in order to (a) demonstrate that the 2D MIT is observed in all 2DES in Si regardless of the amount of disorder, (b) point out important similarities to the MIT in 3D systems, and (c) set the stage for the discussion of glassy dynamics near the MIT.  As summarized below, the 2DES in Si exhibits all the main manifestations of glassiness: slow, correlated dynamics; nonexponential relaxations; diverging equilibration time, as $T\rightarrow 0$; aging and memory.  The results provide strong support to theoretical proposals describing the 2D MIT as the melting of a Coulomb glass~\shortcite{Darko-glass,MIT-glassothers1,MIT-glassothers2,MIT-glassothers3,Vlad-MITglass1,Vlad-MITglass2}.  The review concludes by further comparison to other complex materials, both 2D and 3D, and by identifying some open directions for future research.

\section{Metal-insulator transition in two dimensions}
\label{mit}

In 2D systems in semiconductor heterostructures, the electron density $n_s$ can be varied easily over two orders of magnitude simply by applying voltage $V_g$ to the so-called gate electrode.  At low $n_s$, the 2DES is strongly correlated: $r_s\gg 1$, where $r_s=E_{C}/E_{F}\propto n_{s}^{-1/2}$ ($E_C$ is the average Coulomb energy per electron and $E_F$ is the Fermi energy; see also \shortciteNP{Sergey-review}
for more details).  At the same time, the electrons ``feel'' a random potential (disorder) caused by charged impurities that are located away from the 2DES.  Some of the striking experimental evidence for the MIT that occurs in a variety of ``clean'', \textit{i.e.} low-disordered 2D systems has been presented also in the chapter by S. V. Kravchenko (this volume).

This section reviews work that demonstrates the existence of the 2D MIT regardless of the amount or type of disorder.  The disorder, however, does affect the values of $n_c$, the critical density for the MIT, and the precise form of the temperature dependence of conductivity $\sigma(T)$.  For the sake of clarity, it is important to recall that a qualitative distinction between a metal and an insulator exists only at $T=0$: $\sigma(T=0)\neq 0$ in the metal and $\sigma(T=0)=0$ in the insulator.  The experiments were performed on a 2DES in Si metal-oxide-semiconductor field-effect transistors (MOSFETs), where the low density regime can be reached more easily compared to other semiconductors \shortcite{Sergey-review}.
In Si MOSFETs, the (Drude) mobility $\mu=\sigma/(n_{s}e)$ peaks as a function of $n_s$ because, at very high $n_s$ that are not of interest here, the scattering due to the roughness of the Si-SiO$_2$ interface becomes increasingly important \shortcite{AFS}.  The peak mobility at 4.2~K is commonly used as a rough measure of the amount of disorder.

\subsection{Effects of disorder}
\label{disorder}

\subsubsection{High-mobility (low-disordered) samples}
\label{highmobility}

\begin{figure}
	\centering
		\includegraphics[width=0.55\textwidth]{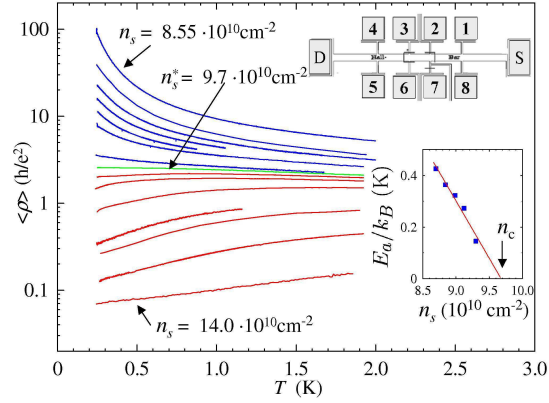}
		\caption{High-mobility sample ($\mu\approx 2.5$~m$^2$/Vs): resistivity $\rho=1/\sigma$ \textit{vs.} $T$ for $n_{s}(10^{10}$cm$^{-2})=$8.55, 8.70, 8.84, 8.99, 9.13, 9.27, 9.56, 9.71, 9.85, 9.99, 10.4, 11.2, 11.6, 12.9, 14.0 (from the top) \protect\shortcite{JJ_PRL02}.  Insets: a schematic of the sample (top view), and activation energies \textit{vs.} $n_s$; $n_c\approx n_{s}^{\ast}$ (here $r_s\approx 17$).}
	\label{fig:highmu}
\end{figure}

In high-mobility Si MOSFETs where, roughly speaking, the 4.2~K peak $\mu>2$~m$^2$/(Vs), the low-temperature resistivity $\rho$ exhibits a strong, metallic drop ($d\rho/dT>0$) with decreasing $T$ for $n_s>n_{s}^{\ast}$ (Fig.~\ref{fig:highmu}) \shortcite{Sergey-review}.
$d\rho/dT$ changes sign at a density $n_{s}^{\ast}$ and becomes insulatinglike ($d\rho/dT<0$) for $n_s<n_{s}^{\ast}$.  Even though $d\rho/dT<0$ does not necessarily imply that $\rho$ diverges at zero temperature (\textit{i.e.} that $\sigma(T=0)=0$), the density $n_{s}^{\ast}$ has been often assumed to represent the critical density for the MIT.  However, other, more appropriate methods have been also used to identify $n_c$.  For example, $n_c$ was determined based on both a vanishing activation energy and a vanishing nonlinearity of current-voltage characteristics when extrapolated from the insulating phase \shortcite{nc1,nc2}.  It was established that $n_c\approx n_{s}^{\ast}$, although a small but systematic difference of a few percent has been reported such that $n_c < n_{s}^{\ast}$ \shortcite{nc3,nc4}.  Fig.~\ref{fig:highmu} inset illustrates the use of activation energies, determined from the fits of the data at the lowest $n_s$ and $T$ to the simply activated form\footnote{Close enough to $n_c$, the data can be fitted almost equally well to the variable-range hopping law.  This gives the same values of $n_c$ within the error of the fit.} $\rho\propto \exp(E_A/k_{B}T)$.

In the vicinity of $n_c$ (or $n_{s}^{\ast}$), the resistivity curves can be collapsed onto the same scaling function of a single parameter $T_0$, \textit{i.e.} $\rho\,(T,n_s)=\rho_{c}f(T/T_0)$ \shortcite{Krav-scal1}.  $T_0$ is the same function of $\delta_n\equiv(n_s-n_c)/n_c$ on both the metallic and the insulating sides of the transition, $T_0\propto |\delta_n|^{z\nu}$ ($z$ -- dynamical exponent, $\nu$ -- correlation length exponent), $z\nu=1.6$ \shortcite{Krav-scal2}.  Here $\rho_c(n_c)$ is the so-called ``separatrix'', the temperature independent critical resistivity curve.  On the metallic side, the scaling does not work at higher $T$, in the regime where $\rho(T)$ goes through a maximum and then decreases as $T$ is raised further \shortcite{Sergey-review}.
On general grounds, the scaling of $\rho(n_s,T)$ at low enough $T$ represents a strong indication for the existence of the metal-insulator transition at $T=0$ \shortcite{Sachdev-book}.

At somewhat higher $n_s$ in the metallic phase, it becomes apparent that, after the initial rapid drop, $\rho(T)$ becomes a very weak function as $T\rightarrow 0$ \shortcite{nc3,nc5}.  In fact, it is so weak that, on a logarithmic scale, it appears to saturate.  As a result of this ``saturation'' of $\rho(T)$, single-parameter scaling fails at the lowest temperatures \shortcite{nc3}.  However, a careful analysis of the data shows \shortcite{njk1,njk2,njk3} that all of the $\sigma (n_s,T)$ (or $\rho(n_s,T)$) curves can be scaled according to the more general form \shortcite{Belitz} $\sigma(n_s,T)=\sigma_c(T)f(T/T_0)$, where the critical conductivity $\sigma_c=\sigma(n_c,T)\propto T^x$, \textit{i.e.} it vanishes as $T\rightarrow 0$.  In other words, the critical conductivity is not the ``separatrix'', but instead it belongs to the insulating family of curves: $n_c<n_{s}^{\ast}$, consistent with other studies \shortcite{nc3,nc4}.  The exponent $x$, however, is very small: $x= 0.1-0.2$ \shortcite{njk1,njk2,njk3}, so that a direct observation of $\sigma_c(T)$ is very difficult at experimentally accessible temperatures.  A small value of $x$ is also the reason for the apparent success of the single-parameter scaling in the limited range of $T$ and $n_s$.  It is important to note that scaling with $x\neq 0$, also observed in 3D materials near the MIT \shortcite{Myriam_review}, does not contradict \shortcite{Belitz} any fundamental principle for 2D systems.  Indeed, violations of ``Wegner scaling'' \shortcite{Belitz}, where $x=(D-2)/z$ ($D$ -- dimensionality), were predicted for certain microscopic models \shortcite{noWegner1,noWegner2,noWegner3} with strong spin-dependent components of the Coulomb interactions.

While the general scaling form describes the data satisfactorily as $T\rightarrow 0$, it would be desirable to explore scaling and the precise form of $\sigma_c(T)$ over a wider range of $T$.  It turns out that the situation becomes simpler in samples with more disorder, where the onset of ``saturation'' of $\sigma$ (or $\rho$) in the metallic phase is pushed to higher $T$.

\subsubsection{Intermediate-mobility samples and effects of local magnetic moments}
\label{localmoments}

Although the metallic temperature dependence of conductivity, $d\sigma/dT<0$ (\textit{i.e.} $d\rho/dT>0$), is less pronounced in samples with moderate mobility (4.2~K peak $\mu\sim 1$~m$^2$/Vs), it was demonstrated early on \shortcite{DP_PRL} that scaling $\sigma(n_s,T)=\sigma_{c}f(T/T_0)$ is obeyed with exactly the same value of the exponent $z\nu=1.6$ as in high-mobility samples.  While the value of $n_c$, defined as $n_{s}^{\ast}$, was higher than in high-mobility devices\footnote{A study of Si MOSFETs with different peak mobilities indicated a systematic increase of $n_c$ with disorder;  $n_c$ was defined as the ``separatrix'' $n_{s}^{\ast}$ \shortcite{nc3}.}, the corresponding $r_s\approx 13$ was still comparably large.

\begin{figure}
\centering
\includegraphics[width=0.55\textwidth]{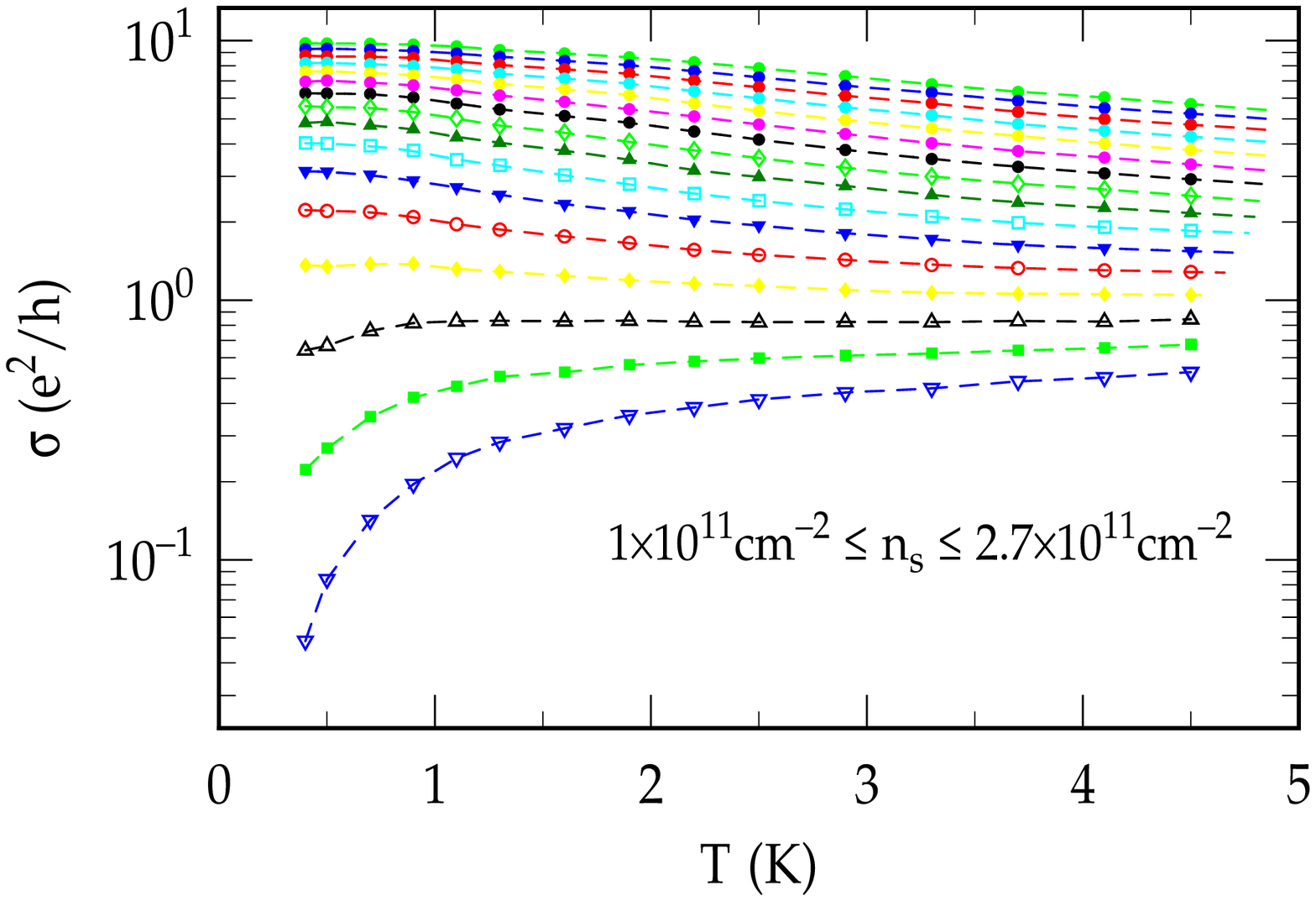}\\
\includegraphics[width=0.55\textwidth]{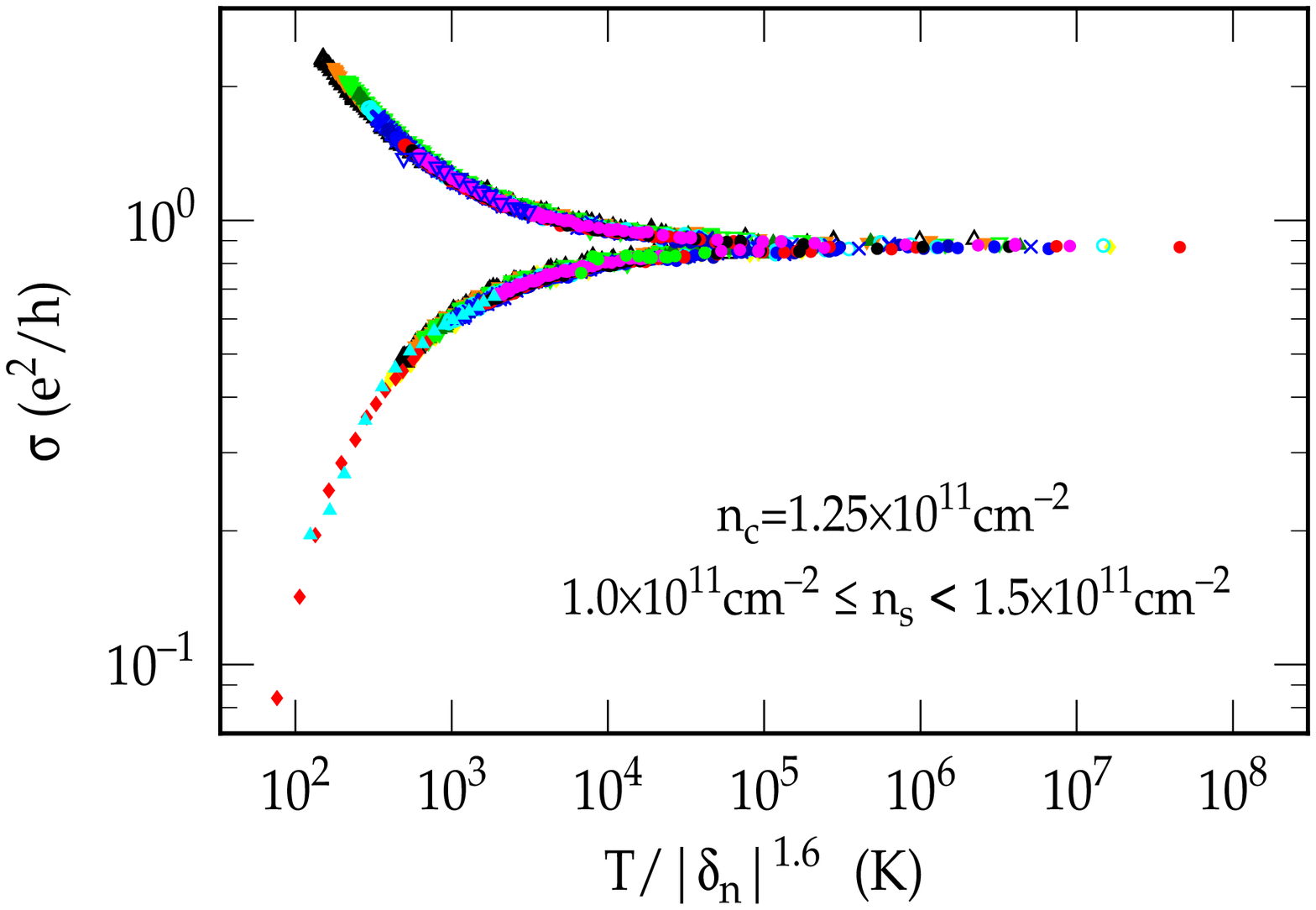}
\caption{Intermediate-mobility sample ($\mu\approx 1$~m$^2$/Vs; sample 12): conductivity $\sigma$ \textit{vs.} $T$ for different $n_s$ (top), and scaling of $\sigma$ with temperature for the same data (and other $n_s$ not shown) in the $n_s$ range given on the plot (bottom).  $\delta_n\equiv(n_s-n_c)/n_c$ is the reduced density; $n_c\approx n_{s}^{\ast}$ (the corresponding $r_s\approx 15$).  For $n_s\geq n_{s}^{\ast}$, the scaling works only for $T\geq 1.1$~K.  The substrate (back-gate) bias on the sample was $V_{sub}=-40$~V \protect\shortcite{Feng-moments}.}
\label{fig:medmu}
\end{figure}

By extending the measurements to lower $T$ \shortcite{Feng-moments}, however, the ``saturation'' of $\sigma(T)$ was observed to take place in the metallic phase at $T$ as high as $\sim 1$~K (Fig.~\ref{fig:medmu} top).  Moreover, the separatrix acquired an insulatinglike ($d\sigma/dT>0$) temperature dependence.  As a result, the simple, single-parameter scaling around the separatrix works only at $T>1.1$~K on the metallic side (Fig.~\ref{fig:medmu} bottom).  Thus the data are qualitatively similar to those on high-mobility samples, indicating that $n_c<n_{s}^{\ast}$.  However, since $\sigma(T)$ in the metallic phase ($n_s>n_c$) does not have a simple enough form over a sufficiently wide range of $T$, it is still difficult to make reliable extrapolations to $T=0$.
Fortunately, in Si MOSFETs it is also possible to change the type of the disorder in the same sample.  As shown below, this results in a surprisingly simple, precise form of $\sigma(T)$ over a very wide range of $T$, allowing reliable zero-temperature extrapolations and striking scaling behavior.

For a 2DES in Si, the disorder is due to the oxide charge scattering (scattering by ionized impurities randomly distributed in SiO$_2$ within a few \AA\, of the interface) and to the roughness of the Si-SiO$_2$ interface~\shortcite{AFS}.  For a fixed $n_s$, it is possible to change the disorder by applying bias $V_{sub}$ to the Si substrate (back gate).  In particular, the reverse (negative) $V_{sub}$ moves the electrons closer to the
interface, which increases the disorder.  It also increases the splitting between the subbands since the width of the triangular potential well at the interface is reduced by applying negative $V_{sub}$.  Usually, only the lowest
subband is occupied at low $T$, giving rise to the 2D behavior.  In sufficiently disordered samples, however, the band tails associated with the upper subbands can be so long that some of their strongly localized states may be populated even at low $n_s$, and act as additional scattering centers for 2D electrons.  In particular, since at least some of them must be singly populated due to a large on-site Coulomb repulsion (tens of meV), they may act as local magnetic moments.  Clearly, the negative $V_{sub}$ reduces this type of scattering by depopulating the upper subband.

\begin{figure*}
\normalsize{(a)}\hspace*{-0.5cm}\includegraphics[width=0.47\textwidth,height=5.5cm]{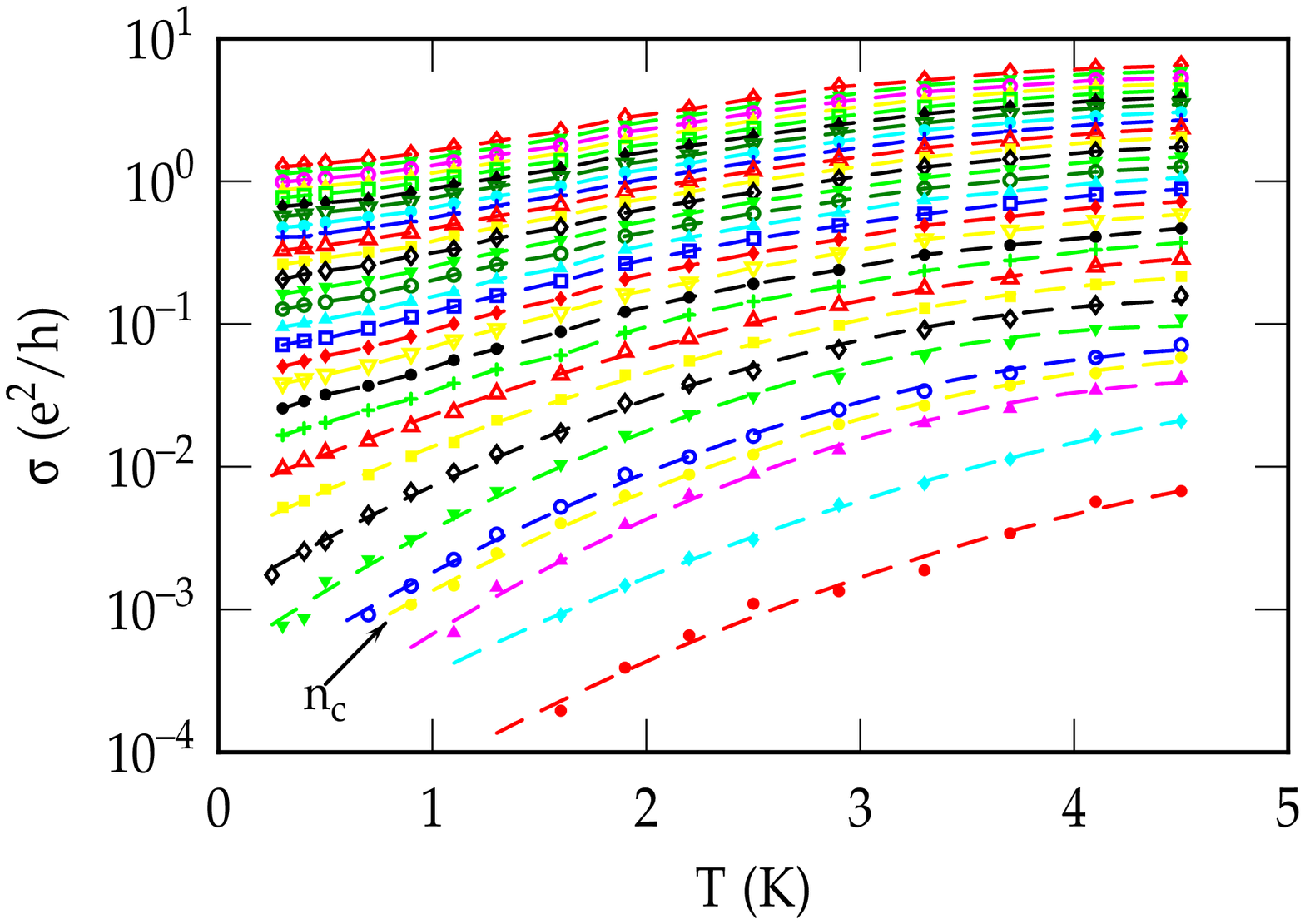}%
\hspace*{0.8cm}\normalsize{(b)}\hspace*{-0.8cm}\includegraphics[width=0.47\textwidth,height=5.5cm]{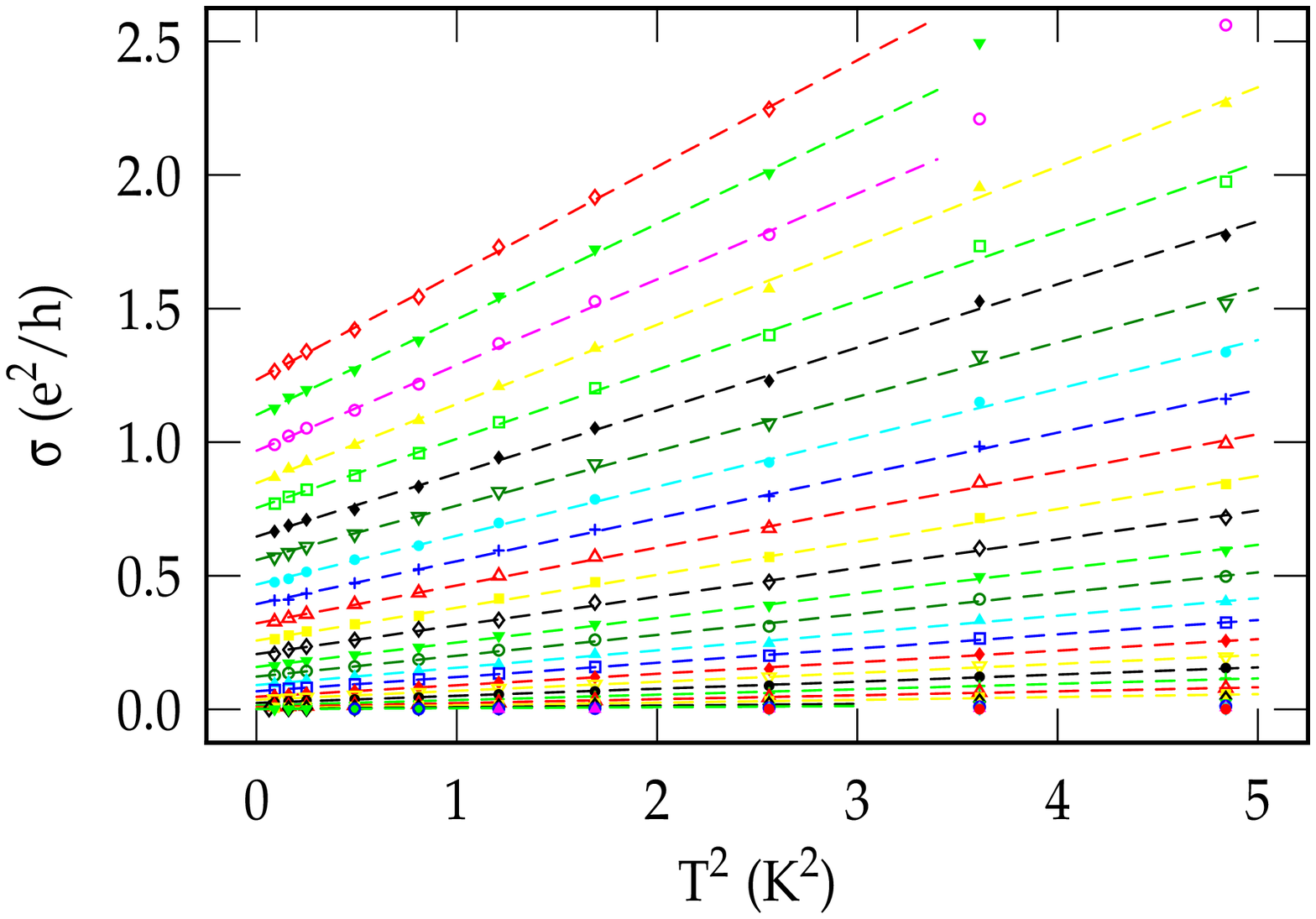}
\begin{minipage}[c]{0.47\textwidth}
\normalsize{(c)}\hspace*{-0.5cm}\includegraphics[width=6.0cm,height=4.5cm]{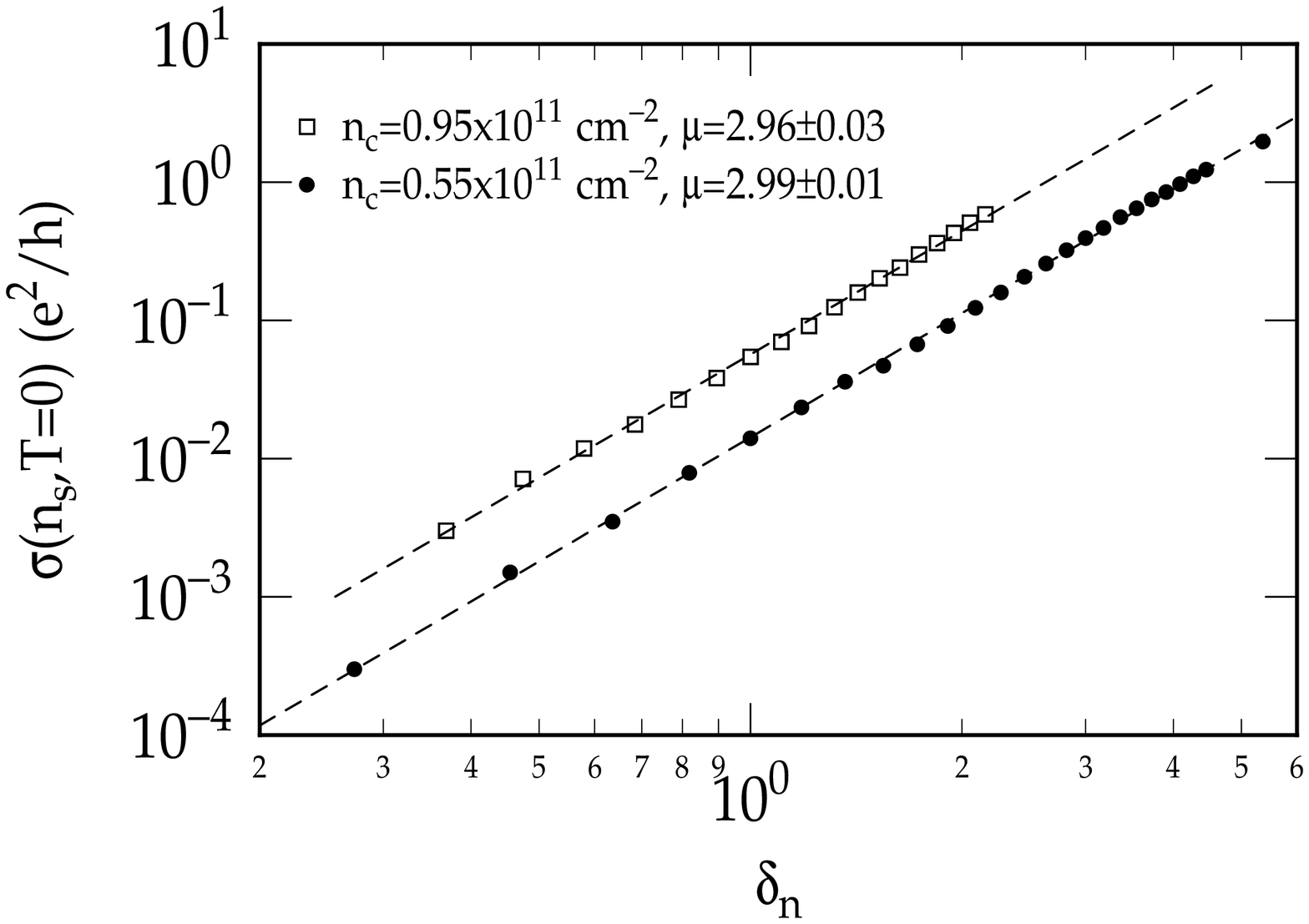}\\
\normalsize{(d)}\hspace*{-0.5cm}\includegraphics[width=6.0cm,height=5.0cm]{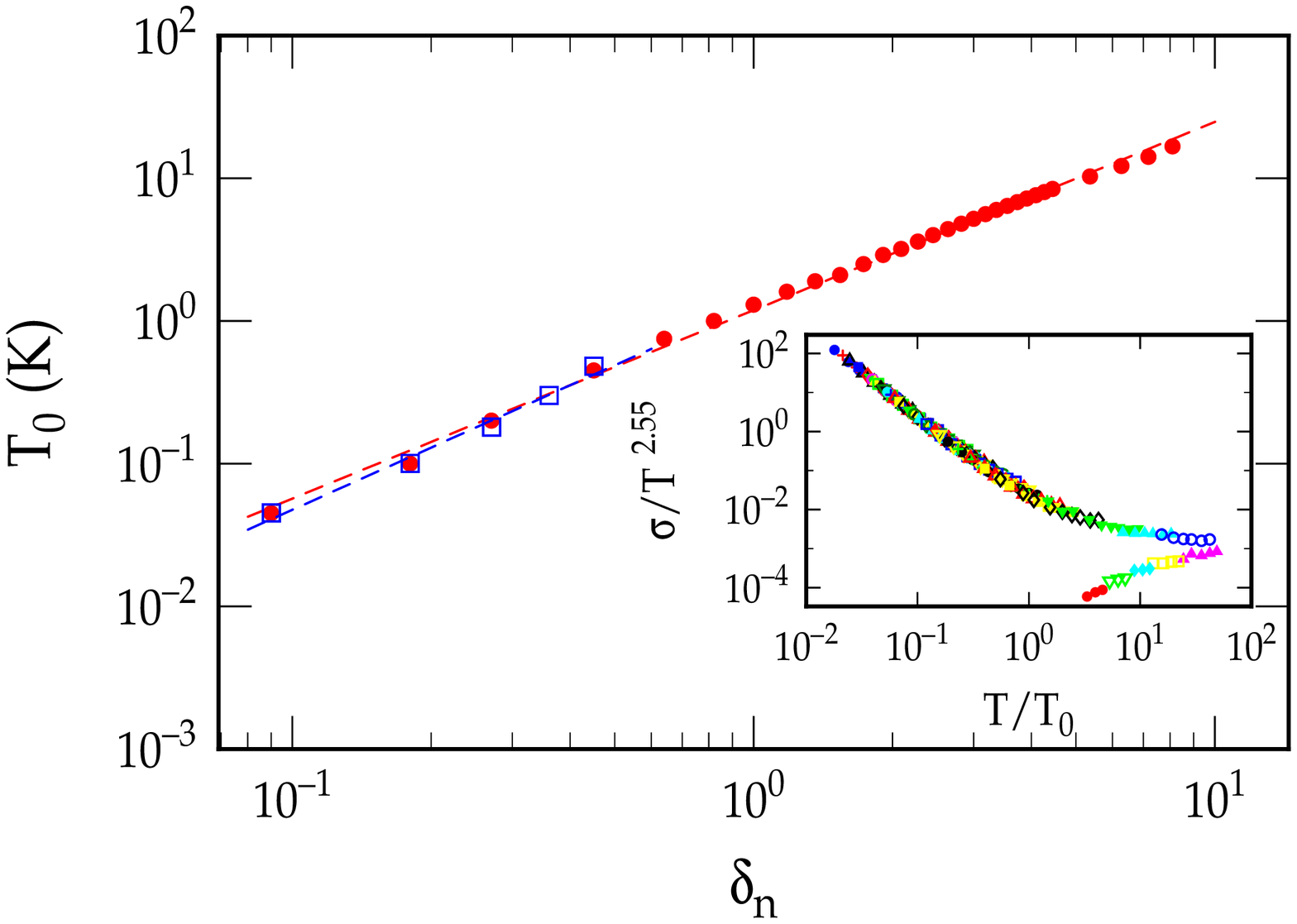}
\end{minipage}
\begin{minipage}[c]{0.47\textwidth}
\hspace*{0.8cm}\normalsize{(e)}\hspace*{-1.3cm}\includegraphics[width=6.7cm,height=9.5cm]{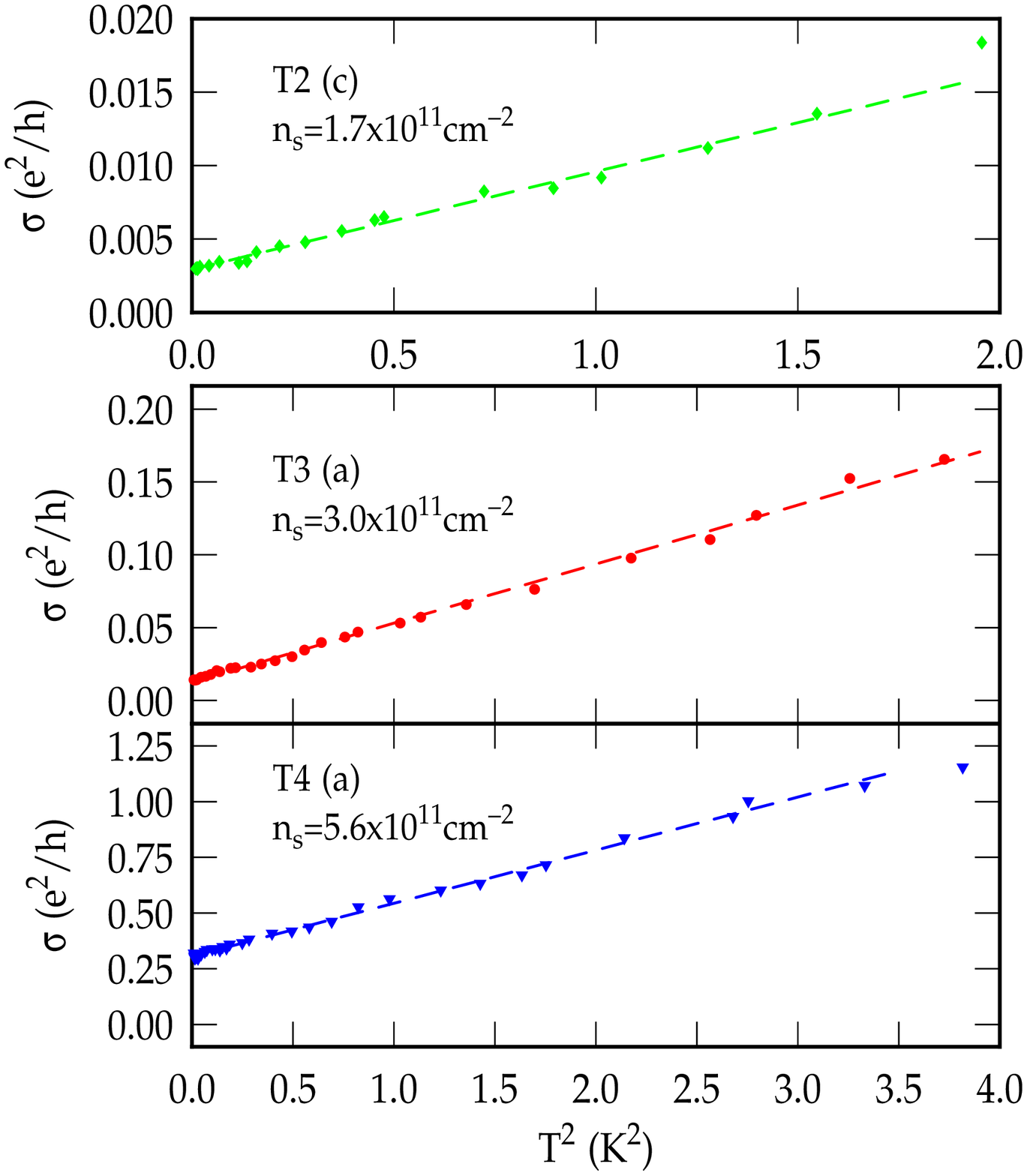}
\end{minipage}
\caption{(a)-(d) Adapted from \protect\shortciteN{Feng-novel}; the same intermediate-mobility sample (sample 12) as in Fig.~\ref{fig:medmu}, but here $V_{sub}=+1$~V.
(a) $\sigma(T)$ for $0.3\leq n_s(10^{11}$cm$^{-2})\leq3.0$ (bottom to top) in steps of $0.1\times 10^{11}$cm$^{-2}$.  (b) The same $\sigma(T)$ data plotted \textit{vs.} $T^2$; the lowest $n_s=0.7\times 10^{11}$cm$^{-2}$ and $0.3\leq T \leq2.2$~K. Other measurements show that this $\sigma(T)$ holds at least down to 0.020~K \protect\shortcite{Feng-novel,Kevin_PRL}.  (c) $\sigma(n_s,T=0)$ \textit{vs.} $\delta_n$ for samples 12 (dots) and 9 (squares).  The dashed lines are fits with the slopes equal to the critical exponent $\mu$. At the MIT, the corresponding $r_s\approx 22$ and 17 for the two samples, respectively.  (d) Scaling parameter $T_0$ as a function of $|\delta_n|$ for sample 12; open symbols: $n_s<n_c$, closed symbols: $n_s>n_c$.  The dashed lines are fits with slopes $1.4\pm 0.1$ and $1.32\pm 0.01$, respectively.  Inset: scaling of raw data $\sigma/\sigma_c\sim\sigma/T^x$ in units of $e^2/h$K$^{2.55}$ for all $n_s$ shown in (a) and $T<2$~K.  (e) Adapted from \protect\shortciteN{Stiles1-T2} and \protect\shortciteN{Stiles2-T2}: $\sigma(T)$ \textit{vs.} $T^2$ for three different samples and densities.  The 4.2~K peak mobility was between 0.15 and 0.33~m$^2$/Vs.}
\label{fig:kondo}
\end{figure*}

Therefore, by varying $V_{sub}$, it is possible to study the effect of local magnetic moments on the transport properties of the conduction electrons in a systematic and controlled way.  It has been established \shortcite{Feng-moments} that scattering by local magnetic moments suppresses the metallic behavior with $d\sigma/dT<0$.  Indeed, the data suggest that in the $T\rightarrow 0$ limit, the $d\sigma/dT<0$ (\textit{i.~e.} $d\rho/dT>0$) behavior is suppressed by an arbitrarily small amount of scattering of the conduction electrons by disorder-induced local moments \shortcite{Feng-moments}.  Figures~\ref{fig:medmu}(top) and \ref{fig:kondo}(a) illustrate the striking effect of local moments on $\sigma(T)$ in the same sample: while the ``usual'', metallic behavior with $d\sigma/dT<0$ is observed in the absence of local moments [Fig.~\ref{fig:medmu} top], $\sigma(T)$ curves  become insulatinglike ($d\sigma/dT>0$) for all $n_s$ after many local moments are introduced with $V_{sub}$ [Fig.~\ref{fig:kondo}(a)].
However, this does not necessarily indicate the destruction of the metallic phase.  In disordered 3D metals, for example, it is well known that the derivative $d\sigma/dT$ can be either negative or positive near the MIT \shortcite{LR}.  On the other hand, the metallic behavior where $\sigma$ decreases but does not go to zero (as expected for an insulator) when $T\rightarrow 0$ is new and unexpected in 2D.

The analysis of the insulatinglike $\sigma(T)$ curves in Fig.~\ref{fig:kondo}(a) shows \shortcite{Feng-novel} that they follow a very simple and precise form over a broad (two decades) range of $T$ and $n_s$: $\sigma(n_s,T)=\sigma(n_s,T=0)+A(n_s)T^2$ [Fig.~\ref{fig:kondo}(b)].  The high quality of the fits allows a reliable extrapolation of $\sigma (n_s,T=0)$, whose finite ({\it i.~e.} non-zero) values mean that, in spite of the decrease of $\sigma (n_s,T)$ with decreasing $T$, the 2D system is in the metallic state.  In particular, the zero-temperature conductivity $\sigma(n_s,T=0)$ is a power law function of $\delta_n$ [Fig.~\ref{fig:kondo}(c)]: $\sigma (n_s,T=0)\propto\delta_{n}^{\mu}$ ($\mu\approx 3$), as expected in the vicinity of a quantum critical point~\shortcite{Goldenfeld}, such as the MIT.  The power law holds over a very wide range of $\delta_n$ (up to 5) similar to what has been observed~\shortcite{Rosenbaum2} in Si:P near the MIT.  In addition, even though the MIT occurs at different $n_c$ in different samples, the critical exponents $\mu$ are the same [Fig.~\ref{fig:kondo}(c)], as expected from general arguments~\shortcite{Goldenfeld}.  It has been also demonstrated \shortcite{Feng-novel} that, near the MIT, the data obey dynamical scaling $\sigma (n_s,T)=\sigma_c(T)f(T/\delta_{n}^{z\nu})$ [Fig.~\ref{fig:kondo}(d)] with a temperature dependent critical conductivity $\sigma_c=\sigma(n_s=n_c,T)\propto T^x$ ($z\nu=1.3\pm 0.1$, $x\approx 2.6$, $\mu=x(z\nu)=3.4\pm0.4$), both in agreement with theoretical expectations near a quantum phase transition~\shortcite{Belitz} and consistent with the extrapolations of $\sigma (T)$ to $T=0$.

The $T^2$ form of $\sigma(T)$ is well established for metals containing local magnetic moments, and it is believed to result from the Kondo
effect~\shortcite{Hewson}.  In fact, there is no other known mechanism that results in an \textit{increase} of $\sigma$ as $T^2$.  Here this feature provides the most direct evidence of the presence of local magnetic moments.  It should be noted that, in general, one expects the $T^2$ behavior for a quantum impurity embedded in a Fermi liquid in any dimension.  While the nature of this novel metallic state in 2D may require further study, its simple $\sigma(T)$ allows for an unambiguous extrapolation to $T=0$.  The zero-temperature conductivity $\sigma (n_s,T=0)$ decreases continuously, and follows a distinct
power-law behavior as the MIT is approached. In particular, metallic $\sigma$ as small as $10^{-3} e^2/h$ has been observed \shortcite{Feng-novel}, in a striking contrast to anything that has been reported in other 2D systems when $d\sigma/dT < 0$.  A similar observation in 3D systems~\shortcite{Rosenbaum2} has demonstrated the absence of minimum metallic conductivity, and has had a profound impact on shaping the theoretical ideas about the MIT.

2DES in Si MOSFETs have been studied extensively for more more than four decades, so it is interesting that the $T^2$ behavior has been identified only relatively recently \shortcite{Feng-novel}.  A thorough examination of the early literature reveals, however, that the samples discussed here are representative of a broad class of Si MOSFETs historically (and somewhat unfairly) known as ``nonideal'' samples \shortcite{AFS}.  ``Nonideal'' samples could be made more ``ideal'' by applying $V_{sub}$ and \textit{vice versa} \shortcite{AFS}, consistent with recent studies \shortcite{Feng-moments,Feng-novel} discussed above. Figure~\ref{fig:kondo}(e) shows an example of $\sigma(T)$ measured on samples with a modest peak mobility between 0.15 and 0.33~m$^2$/Vs \shortcite{Stiles1-T2,Stiles2-T2}.  The ``saturation'' (on a log scale) of $\sigma(T)$ observed for $T<1$~K was puzzling at the time, but the replotted data show clear $T^2$ behavior.

\subsubsection{Low-mobility (highly disordered) samples}
\label{lowmobility}

Samples with very low 4.2~K peak mobility ($\mu < 0.1$~m$^2$/Vs) have attracted less attention because they do not exhibit a pronounced, if any, $d\sigma/dT<0$ metallic behavior.  However, they not only exhibit the 2D MIT, but also other similarities to the behavior of``clean'' 2DES, such as the onset of glassy charge dynamics (Section \ref{glass}).  In some ways, it is even advantageous to investigate samples with a lot of disorder.  For example, the relevant electron densities, such as $n_c$, are pushed to higher values (higher $E_F$), so that it is possible to reach much lower effective temperatures $T/T_F$ ($T_F$ -- Fermi temperature) than in high-mobility samples.  Therefore, comparative studies of samples with varying amounts of disorder should provide valuable insights into the physics of systems near the MIT.

\begin{figure*}
\vspace*{-0.75in}
\begin{minipage}[c]{0.5\textwidth}
\includegraphics[width=8.2cm,clip]{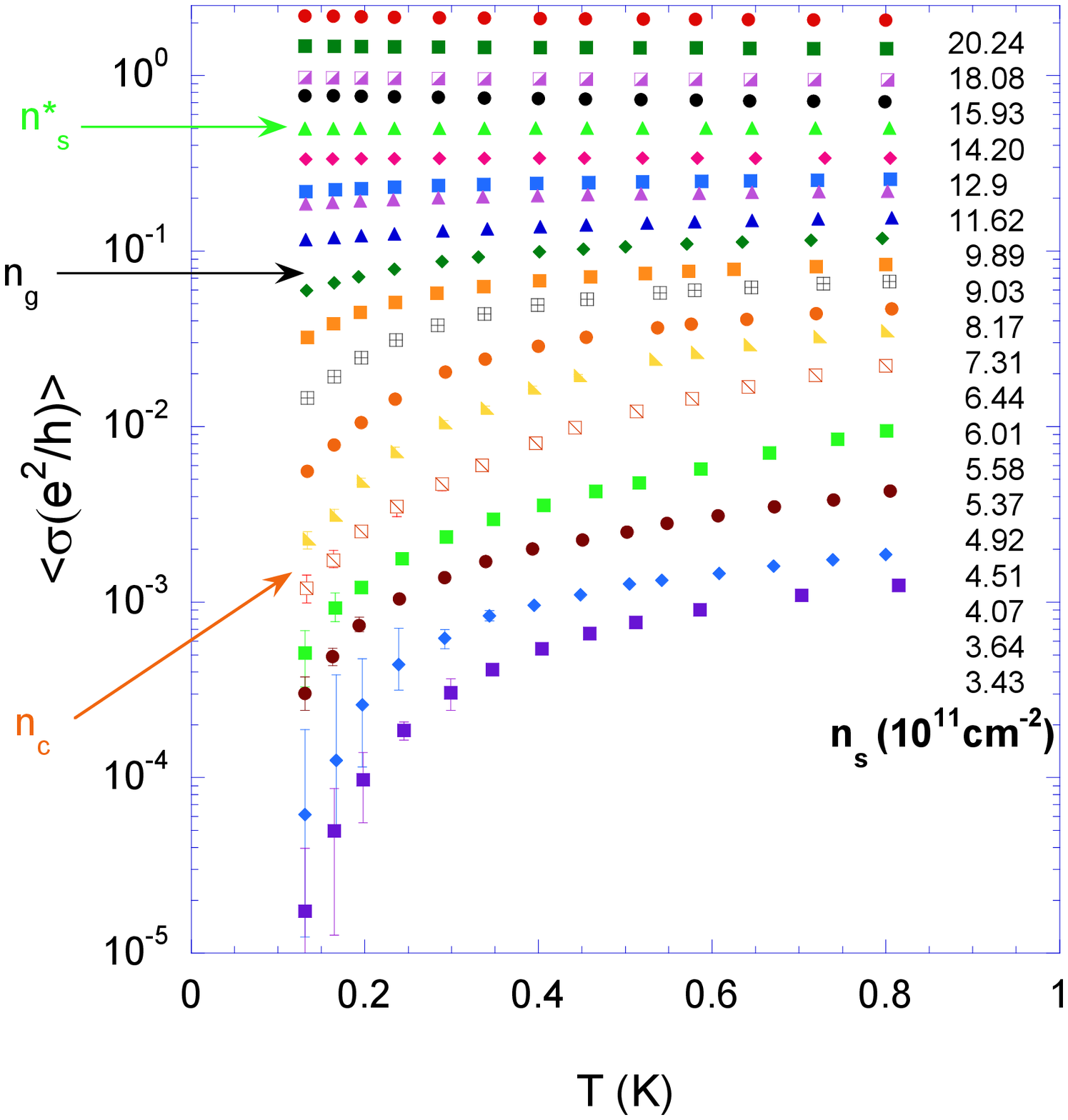}
\end{minipage}
\begin{minipage}[c]{0.45\textwidth}
\hspace*{-0.25cm}\includegraphics[height=9.5cm,clip]{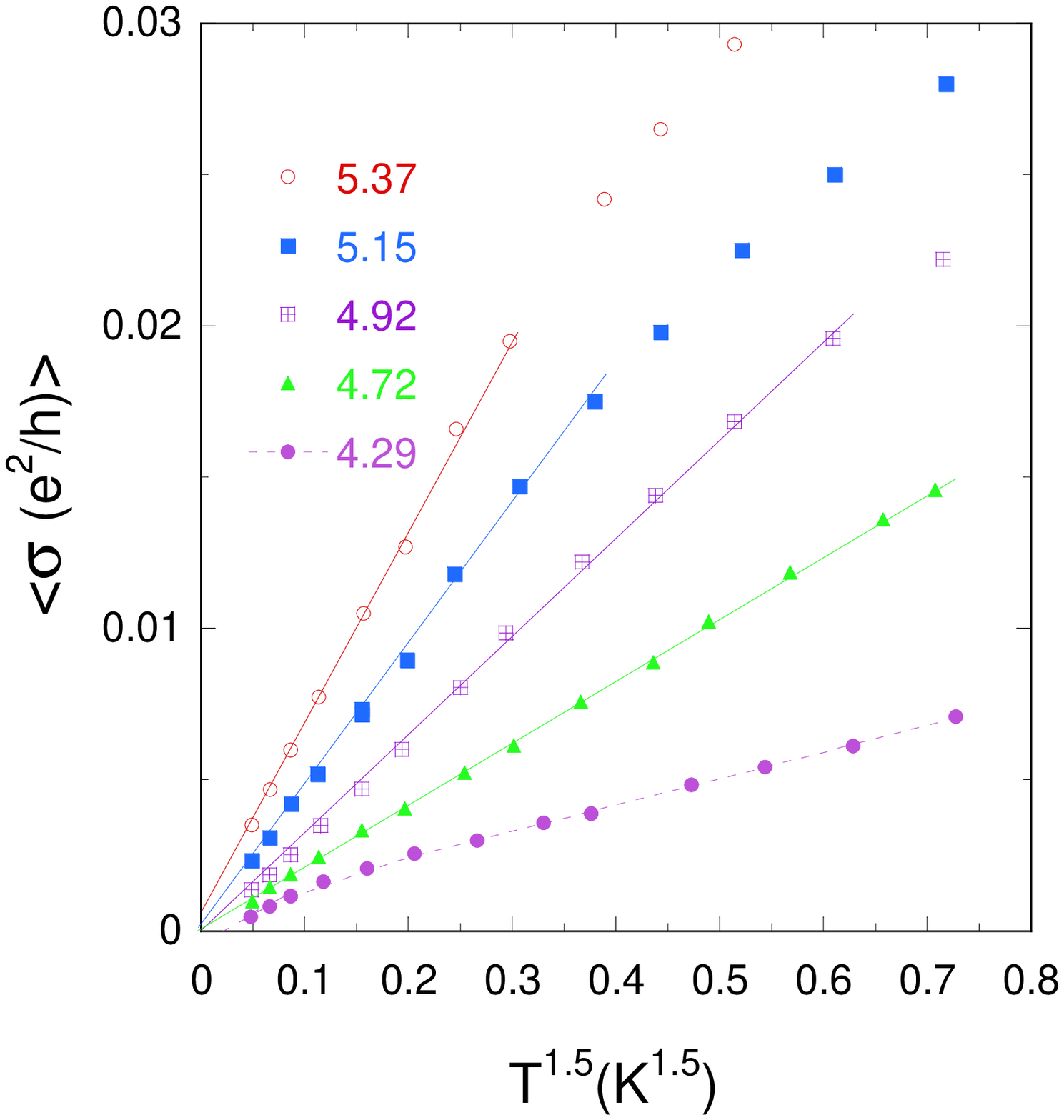}
\end{minipage}
\vspace*{-0.65in}
\caption{Low-mobility sample ($\mu\approx 0.06$~m$^2$/Vs) \protect\shortcite{SBPRL}.  Left: time-averaged $\langle\sigma\rangle$ {\it vs.} $T$ for different $n_s$.  The error bars show the size of the fluctuations with time.  $n_{s}^{\ast}$, $n_g$, and $n_c$ are marked by arrows ($n_g$ -- glass transition density).  Right: $\langle\sigma\rangle$ {\it vs.} $T^{1.5}$ for a few $n_s (10^{11}$cm$^{-2})$ near $n_c$.  The solid lines are fits; the dashed line is a guide to the eye, clearly
showing insulating behavior [$\langle\sigma(T\rightarrow 0)\rangle=0$].}
\label{fig:saverage}
\end{figure*}

Detailed studies of transport and electron dynamics near the MIT have been performed on a set of Si MOSFETs with the 4.2~K peak mobility of only 0.06~m$^2$/Vs with the applied $V_{sub}=-2$~V \shortcite{SBPRL}.  This value of $V_{sub}$ maximizes the peak mobility by removing the contribution of scattering by local magnetic moments (see Sec.~\ref{localmoments} above), at least in the experimental $T$-range.  Because of the glassy fluctuations of $\sigma$ with time $t$ at low $n_s$ and $T$ (see Sec.~\ref{glass}), the carrier density had to be changed at relatively high $T$ (here $\sim 0.8$~K) in small steps in order to obtain reproducible values of the time-averaged conductivity $\langle\sigma\rangle$.  The behavior of $\langle\sigma(n_s,T)\rangle$ (Fig.~\ref{fig:saverage} left) turns out to be quite similar to that of high-mobility Si MOSFETs.  At the highest $n_s$, for example, the devices exhibit metallic behavior with $d\langle\sigma\rangle/dT<0$.  The change of $\langle\sigma\rangle$ in a given $T$ range, however, is small (only 6\% for the highest $n_s=20.2\times10^{11}$cm$^{-2}$) as observed in other Si MOSFETs with a large amount of disorder~\shortcite{nc3}.  Even though the density $n_{s}^{\ast}=12.9\times10^{11}$cm$^{-2}$ at the separatrix \shortcite{SBPRL,SB_PhysE} is much higher than in ``clean'' samples, the value of $\langle\sigma(n_{s}^{\ast})\rangle=0.5~e^{2}/h$ is similar, which, according to Drude formula, corresponds to $k_{F}l\lesssim 1$ ($k_{F}$ -- Fermi wave vector, $l$ -- mean free path).
Likewise, at the lowest $n_s$, the data are best described by the simply activated form $\langle\sigma\rangle\propto\exp(-E_{A}/k_{B}T)$.

Perhaps the most striking difference between high- and low-mobility samples first becomes apparent when the vanishing of activation energy $E_A$ is used to determine $n_c$.  While in ``clean'' samples this gives $n_c\lesssim n_{s}^{\ast}$ (Sec. \ref{highmobility}), here $E_A$ vanishes at $n_c\approx5\times10^{11}$cm$^{-2}$, which is more than a factor of two smaller than $n_{s}^{\ast}$ (Fig.~\ref{fig:saverage} left).  This suggests that there is a wide range of $n_s$ on the metallic side of the MIT where $\sigma(T)$ is insulatinglike.  Indeed, close to $n_c$, the best phenomenological fit to the data is the metallic power-law behavior $\langle\sigma(n_s,T)\rangle=a(n_s)+b(n_s)T^{x}$ with $x\approx1.5$ (Fig.~\ref{fig:saverage} right) \shortcite{SBPRL}.  The fitting parameter $a(n_s)$ is relatively small
and, in fact, vanishes for $n_s (10^{11}$cm$^{-2})=4.72$ and 4.92.  Therefore, $n_c=(5.0\pm 0.3)\times 10^{11}$cm$^{-2}$ ($r_s\sim 7$) based on the data on both metallic and insulating sides of the MIT.  Of course, a simple power law $\langle\sigma(n_c,T)\rangle\propto T^x$ is consistent not only with general expectations near the MIT and the behavior observed in 3D systems~\shortcite{Belitz}, but also with the careful analysis of high-mobility 2DES \shortcite{njk1,njk2,njk3} (Sec. \ref{highmobility}) and those with local magnetic moments \shortcite{Feng-novel} (Sec. \ref{localmoments}).  Here the exponent $x$ takes a distinctly different value, presumably reflecting the different universality classes of those situations.

The surprising non-Fermi liquid $T^{3/2}$ behavior is consistent with theory \shortcite{Vlad-MITglass2} for the transition region between a Fermi liquid and an (insulating) electron glass.  Indeed (see Sec. \ref{glass}), the transition into a charge (Coulomb) glass in low-mobility samples takes place as $T\rightarrow 0$ at a density $n_g$, such that $n_c<n_g<n_{s}^{\ast}$ (Fig.~\ref{fig:saverage} left).  The $T^{3/2}$ correction is characteristic of transport in the intermediate, $n_c < n_s < n_g$ region where the dynamics is glassy, but where $\sigma$ is still metallic [$\sigma(T\rightarrow 0)\neq 0$] albeit so small that $k_{F}l < 1$.  Such ``bad metals'' include a variety of strongly correlated materials with unusual properties~\shortcite{bad}.  Interestingly, the $T^{3/2}$ behavior can be revealed also in high-mobility 2DES by applying a parallel magnetic field.

\subsection{2D metal-insulator transition in a parallel magnetic field}
\label{mitinB}

Since magnetic field $B$ applied parallel to the 2DES plane couples only to electrons' spins, it is often used to probe the importance of spin, as opposed to charge, degrees of freedom.  Some of the intriguing results that have been obtained in parallel $B$ in the vicinity of the zero-field 2D MIT have been described in the review by \shortciteN{Sergey-review}.
One of the main issues has been the fate of the metallic phase in a parallel $B$.  From the insulating side, the critical density $n_c(B)$ can be determined by extrapolating to zero the activation energy and nonlinearity of current-voltage characteristics (Sec. \ref{highmobility}).  For $n_s>n_c(B)$, however, the metallic $d\sigma/dT<0$ behavior observed in high-mobility samples is suppressed by $B$, making it even more difficult to determine the critical density from the metallic side.  Nevertheless, a careful analysis reveals exactly the same metallic $T^{3/2}$ correction in high-mobility samples in parallel magnetic fields as in highly disordered samples in zero magnetic field, confirming the existence of the MIT in those two cases.

\begin{figure*}
\normalsize(a)\hspace*{-0.7cm}\includegraphics[width=7.0cm,clip]{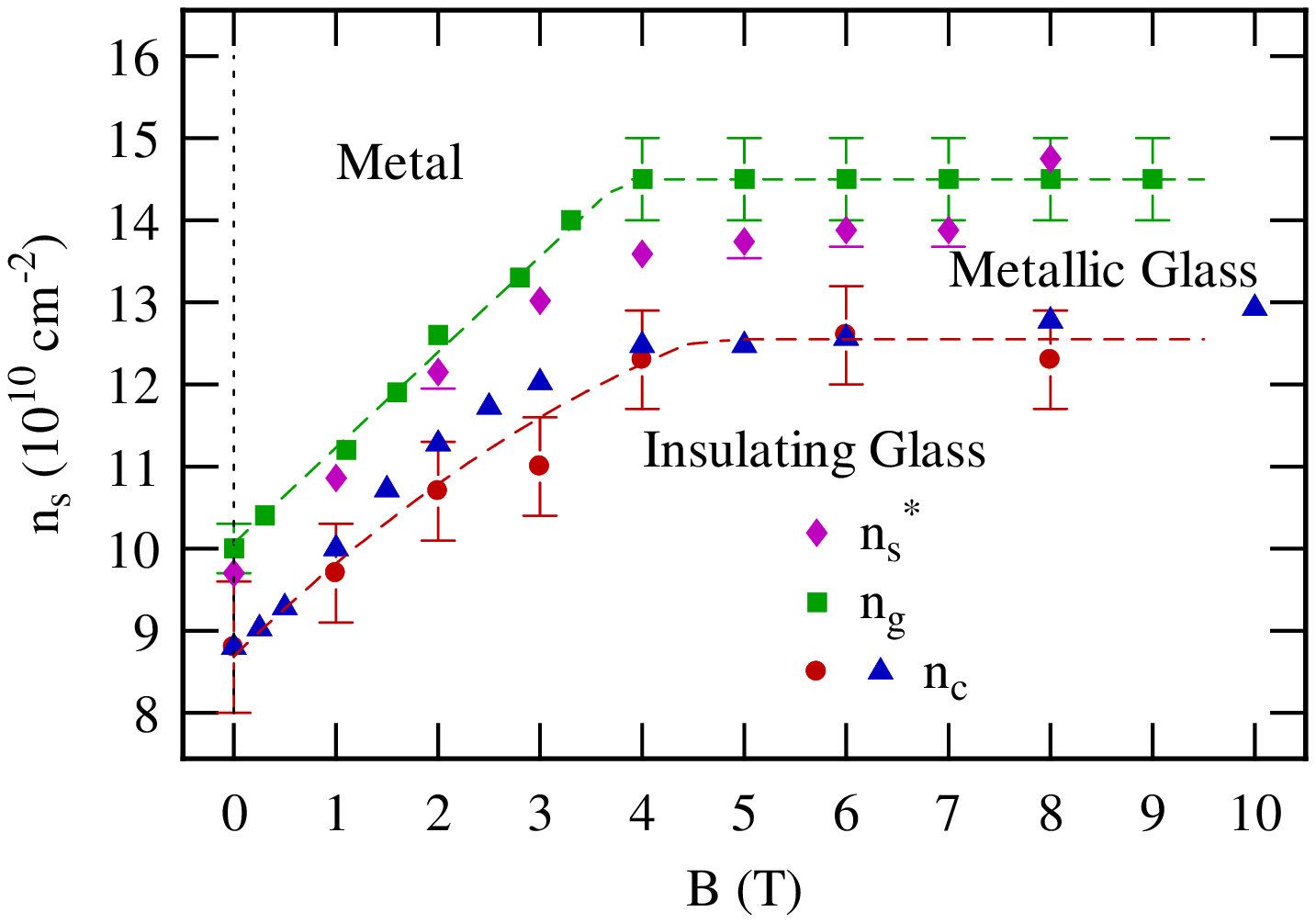}\hspace*{-1.0cm}
\normalsize(b)\hspace*{-0.3cm}\includegraphics[height=4.5cm,clip]{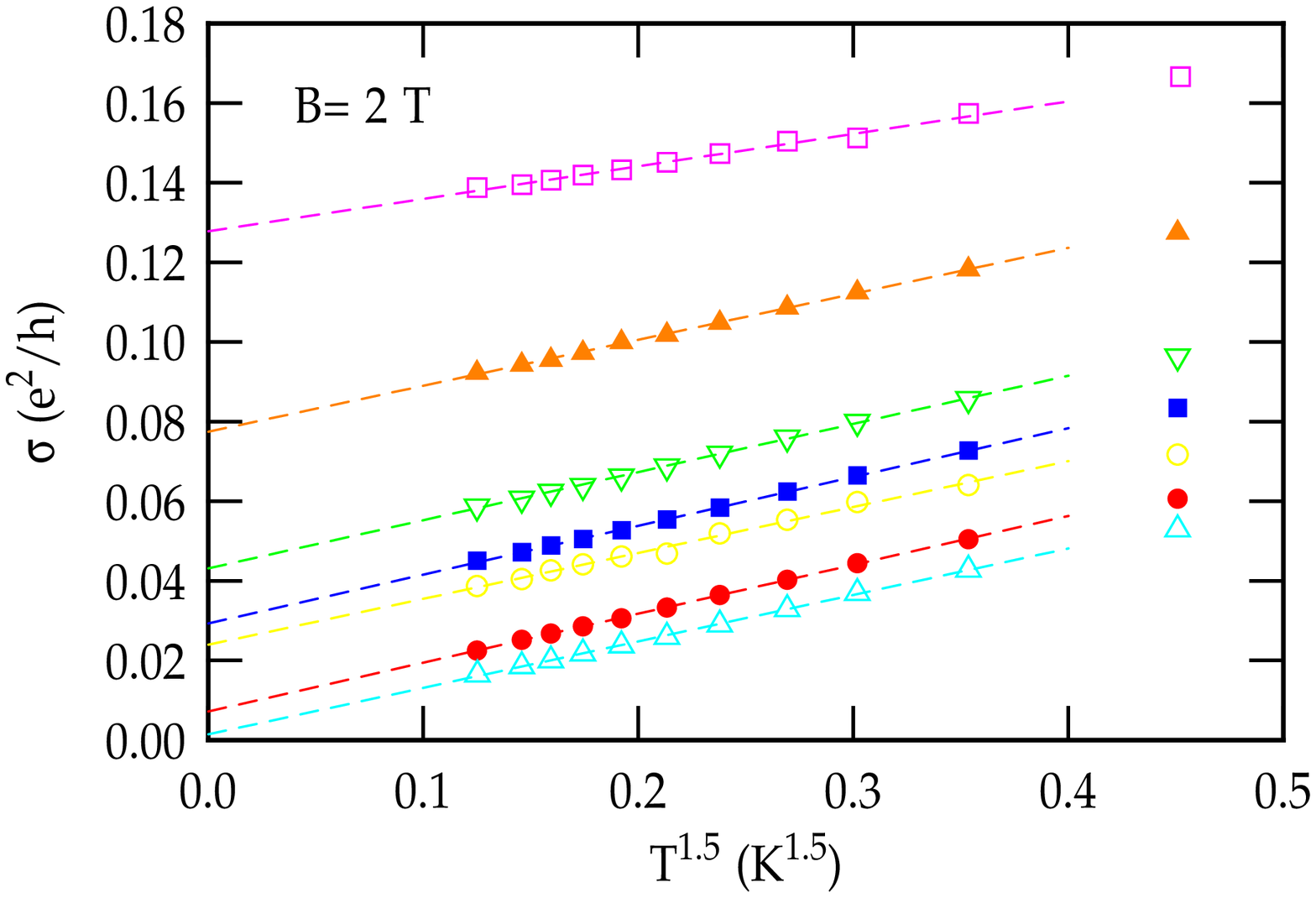}
\normalsize(c)\hspace*{-0.5cm}\includegraphics[width=6.5cm,clip]{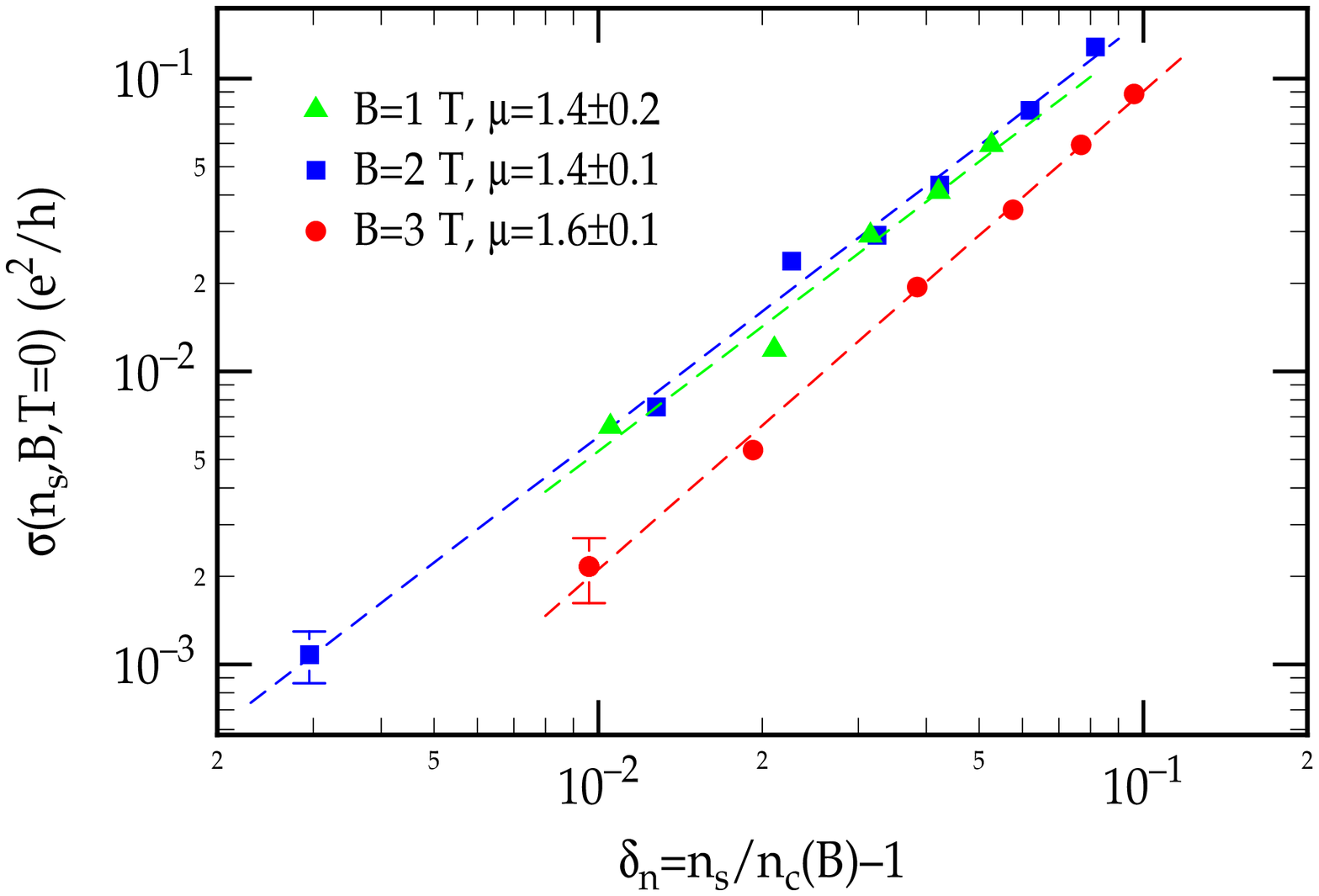}\hspace*{-0.5cm}
\normalsize(d)\hspace*{0cm}\includegraphics[width=6.5cm,clip]{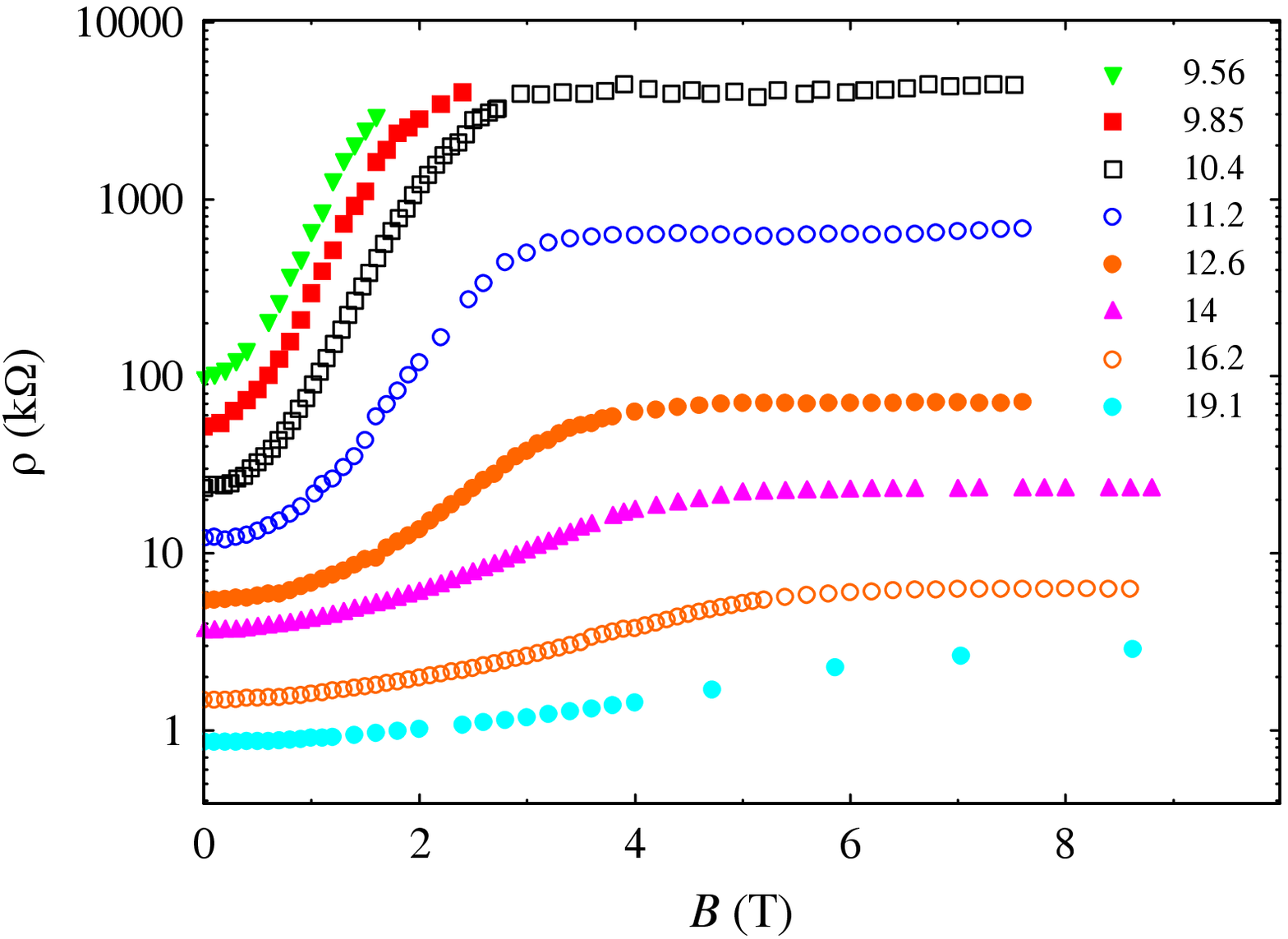}
\caption{High-mobility Si MOSFET.  (a) $T=0$ phase diagram in a parallel magnetic field \protect\shortcite{JJ_noiseB}.  The dashed lines guide the eye.  The $n_c$ values are from \protect\shortciteN{JJ_noiseB} (dots) and \protect\shortciteN{nc2} (triangles).  The glass transition takes place at $n_g(B)>n_c(B)$, giving rise to an intermediate, metallic glass phase.  The density at the separatrix $n_{s}^{\ast}\approx n_g$ within the error for all $B$.  (b) $\sigma$ \textit{vs.} $T^{1.5}$ for several $n_s$ in the metallic glass phase for $B=2$~T ($n_s(10^{10}$cm$^{-2})=11.9, 11.6, 11.3, 11.2, 11.0, 10.9, 10.7$ from top; $n_c(B=2$~T$)=10.67\times10^{10}$cm$^{-2}$) for the same sample.  Dashed lines are fits.  (c) Zero-temperature conductivity $\sigma(T=0)\propto\delta_{n}^{\mu}$ obtained from the fits shown in (b). (d) The positive magnetoresistance for the same sample \protect\shortcite{JJ_noiseB} and different densities $n_s(10^{10}$cm$^{-2}$), as shown.  A strong increase in MR reflects a magnetic-field driven MIT at $n_c(B)$.}
\label{fig:phased-inB}
\end{figure*}

Figure \ref{fig:phased-inB}(a) shows the $(n_s,B,T=0)$ phase diagram obtained for a high-mobility sample \shortcite{JJ_noiseB}.  The critical densities $n_c(B)$ were first determined using the activation energy method.  Good agreement was found with the $n_c(B)$ dependence obtained on very similar samples using both activation energies and nonlinear current-voltage characteristics\footnote{Because of the small sample to sample differences in the amount of disorder, the data from \shortciteNP{nc2} have been shifted up by $0.85\times 10^{10}$cm$^{-2}$ to make the $n_c(B=0)$ values coincide.} \shortcite{nc2}.  At low fields, $n_c$ increases with $B$, and then it saturates for $B\gtrsim 4$~T, consistent with other studies that show that the 2DES is here fully spin polarized \shortcite{polarization1,polarization2,polarization3}.  Even though $\sigma(T)$ is very weak at higher $n_s$, it is interesting to attempt to determine the separatrix, where $d\sigma/dT=0$.  The corresponding density $n_{s}^{\ast}(B)>n_c(B)$ and, most intriguingly, within the measurement error, it coincides with $n_g(B)$, the glass transition density \shortcite{JJ_SPIE04}.  $n_g(B)$ was determined independently, based on the measurements of the fluctuations in $\sigma(t)$ \shortcite{JJ_noiseB} (Sec. \ref{fluct}).

The important question to address is the form of $\sigma(T)$ in the narrow $n_c(B)<n_s<n_g(B)$ region.  Here
the data are best described by the metallic ($\langle\sigma(T=0)\rangle\neq 0$) power-law form $\langle\sigma(n_s,B,T)\rangle=\langle\sigma(n_s,B,T=0)\rangle +b(n_s,B)T^{1.5}$ [Fig.~\ref{fig:phased-inB}(b)], similar to what was observed in the metallic glassy phase of highly disordered samples at $B=0$ (Sec. \ref{lowmobility}). The extrapolated $T=0$ conductivities go to zero precisely at $n_c(B)$, in a power-law fashion $\langle\sigma(n_s,B,T=0)\rangle\propto\delta_{n}^{\mu}$ with $\mu\approx 1.5$ [Fig.~\ref{fig:phased-inB}(c); $\delta_{n}=n_s/n_c(B)-1$] \shortcite{JJ_noiseB} that is in agreement with theoretical expectations near a quantum phase transition \shortcite{Belitz}.  Interestingly, there is some evidence \shortcite{Fletcher} of similar behavior at $B=0$ with $\mu\sim 1-1.5$, obtained by extrapolating to $T=0$ the ``saturation'' of $\sigma(T)$ in the $d\sigma/dT<0$ regime of different high-mobility Si MOSFETs.  The striking power-law behavior shown in Fig.~\ref{fig:phased-inB}(c) and the remarkable agreement between $n_c(B)$ obtained from $\sigma(T)$ on both insulating and metallic sides of the MIT are strong evidence for the survival of the MIT and the metallic phase in parallel $B$.

In low-disordered (``clean'') samples at $B=0$, the critical density $n_c\lesssim n_{s}^{\ast}\approx n_g$ [Sec. \ref{highmobility}, Fig.~\ref{fig:phased-inB}(a)] and the intermediate, metallic glass phase is practically absent. A parallel $B$, however, increases its width, allowing the emergence of the characteristic $T^{3/2}$ correction to $\sigma$.  The increase of both $n_g$ and $n_c$, and the broadening of the metallic glass phase with $B$ can be understood to result from the suppression of screening by a parallel $B$ \shortcite{Bscreening1,Bscreening2}, which increases the effective disorder. This, in turn, favors glassiness, consistent with theoretical expectations \shortcite{Darko-glass}, and makes the behavior of ``clean'' samples more similar to that of highly disordered ones (Sec. \ref{lowmobility}).  The existence of the glass transition in high parallel $B$, where the 2DES is spin polarized, provides evidence that charge, not spin, degrees of freedom are responsible for glassy ordering.  This result clearly imposes a strong constraint on the types of theories that can be formulated to describe this phenomenon.  Likewise, the broadening of the metallic glass phase with $B$ also indicates that its existence is not due to spin.

For $n_s$ near $n_c(B=0)$, 2DES in various semiconductors exhibit a strong, positive magnetoresistance (MR)
\shortcite{Sergey-review},
which has been a subject of great interest.  The MR saturates at the field that corresponds to the full spin polarization \shortcite{polarization1,polarization2,polarization3,polarization4}.  As shown in Fig.~\ref{fig:phased-inB}(d) for the same high-mobility Si MOSFET discussed above, the MR jump is strong only for $n_s$ not too far from $n_c(B=0)$ since it takes place as the system undergoes a magnetic-field driven MIT at $n_c(B)$ [Fig.~\ref{fig:phased-inB}(a)].  In the metallic phase where $n_s\gg n_c(B>4$~T), the MR is weak.

In low-mobility Si MOSFETs, where the metallic glass phase is clearly observable already at $B=0$, the $(n_s,B,T=0)$ phase diagram has not been studied.  However, it is plausible to expect that the parallel $B$ will broaden the metallic glass phase even further.

\begin{figure*}
\centering
\includegraphics[width=8.5cm]{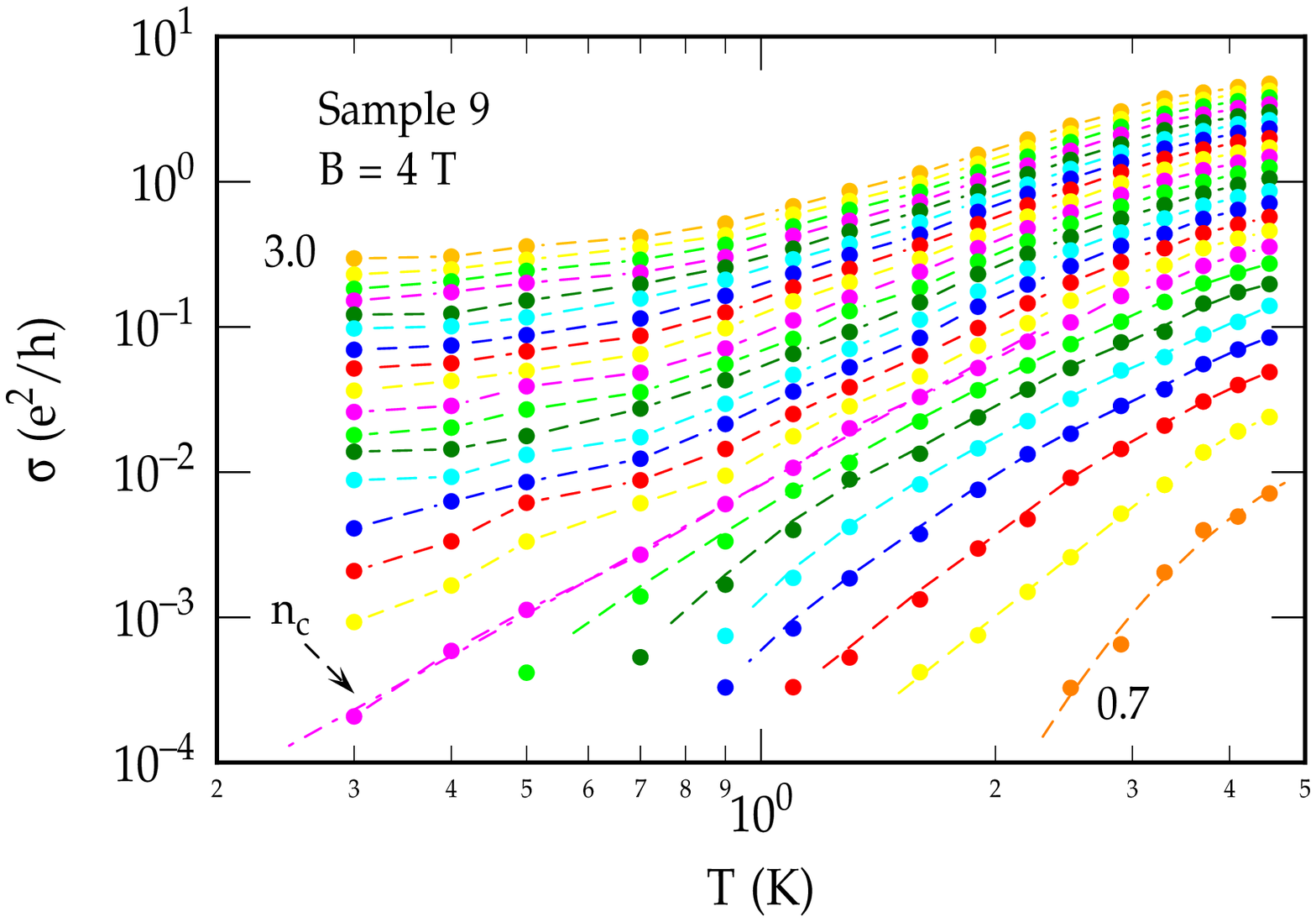}\\
\includegraphics[width=8.5cm]{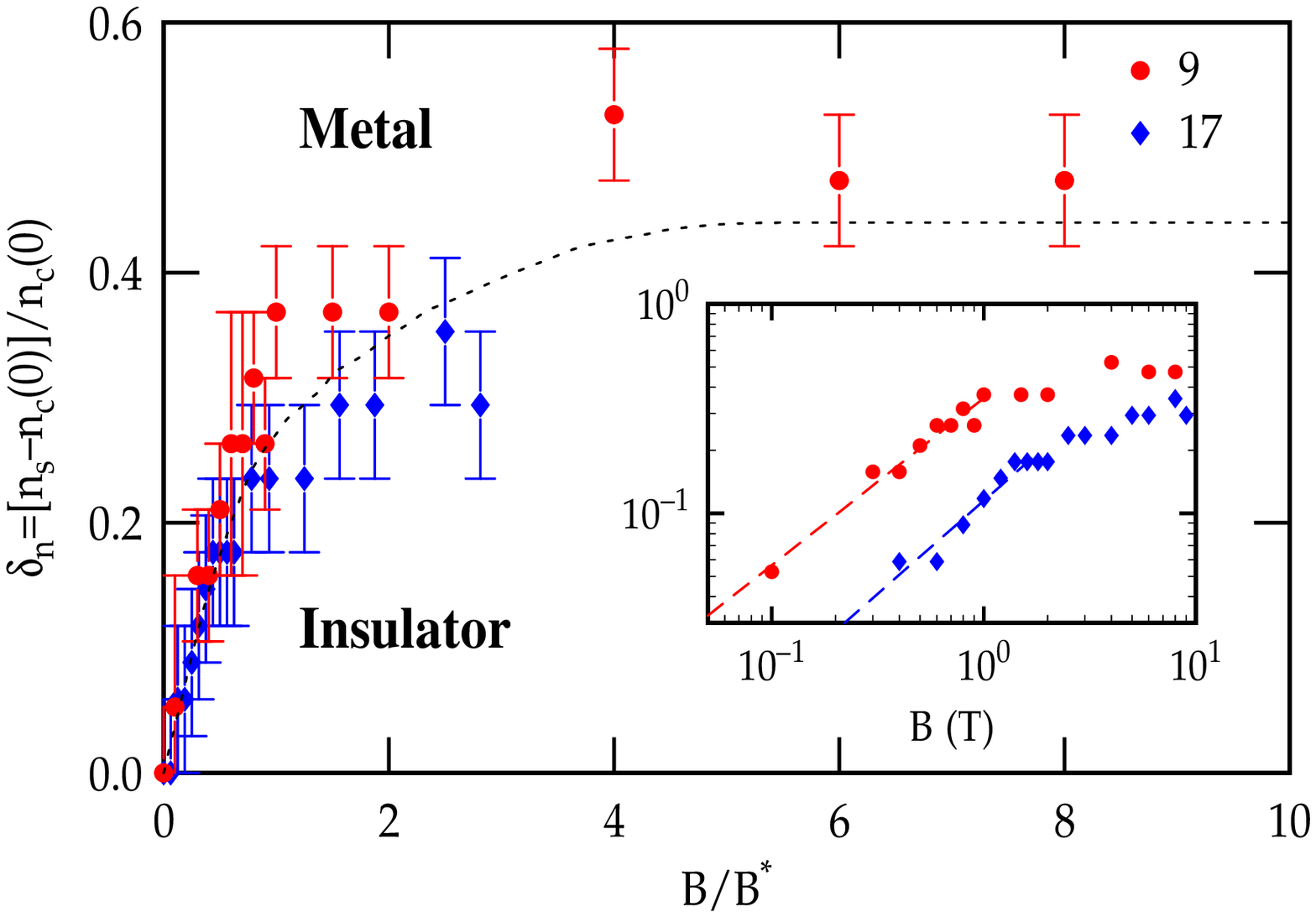}
\caption{Intermediate-mobility sample ($\mu\approx 1$~m$^2$/Vs) with local magnetic moments ($V_{sub}=+1$~V) \protect\shortcite{Kevin_PRL}.  Top panel: $\sigma (T)$ for sample 9 at $B=4$~T.  $n_s$ varies from $3.0\times 10^{11}$cm$^{-2}$ (top) to $0.7\times 10^{11}$cm$^{-2}$ (bottom) in steps of $0.1\times 10^{11}$cm$^{-2}$.  $n_c(B=4$~T$)=1.4\times 10^{-11}$cm$^{-2}$ and is marked by the arrow.  The dashed lines guide the eye.  $\sigma_c$ clearly follows a simple power-law dependence on $T:\sigma_c \propto T^x$.  Bottom panel: $T=0$ phase diagram for two samples.  The dashed line guides the eye.  The boundary between metallic and insulating phases is described by a power-law relation (see inset) $[n_c(B)-n_c(0)]/n_c(0)\propto
(B/B^{\ast})^{\beta}$ at low fields, with the same crossover exponent $\beta\approx 0.9$ for both samples ($B^{\ast}=1$~T for sample 9).  Inset: the same data {\it vs.} $B$ on a log-log scale.  The dashed lines are fits with
the slopes equal to $\beta$.  At $B=0$, $n_c(10^{11}$cm$^{-2})=0.95\pm 0.05$ ($r_s\approx 17$) and $0.85\pm 0.05$ ($r_s\approx 18$) for samples 9 and 17, respectively.}
\label{fig:eng-phased}
\end{figure*}

Finally, in intermediate-mobility 2DES with local magnetic moments, it is relatively easy to map out the $(n_s,B,T=0)$ phase diagram because of the simple form of $\sigma(T)$ that holds over a wide range of $T$.  In parallel $B$, $\sigma(n_s,T)$ data (Fig. \ref{fig:eng-phased} top) are
qualitatively similar to the $B=0$ case (Sec. \ref{localmoments}) \shortcite{Kevin_PRL}.  At the lowest $n_s<n_c$, for example, $\sigma$ decreases exponentially with decreasing $T$, indicating an insulating state at $T=0$.  For $n_s>n_c$, $\sigma (T)$ is weaker and its \textit{curvature} is the opposite from the one expected for an insulating state.  It clearly extrapolates to a finite value as $T\rightarrow 0$, indicating a metallic phase (Sec. \ref{localmoments}).  $n_c$ is identified  as the density where $\sigma_c =\sigma(n_s=n_c,T) \propto T^x$ (Fig. \ref{fig:eng-phased} top), consistent with the $B=0$ case and in agreement with general arguments \shortcite{Belitz}.  The exponent $x=2.7\pm 0.4$ remains constant as a function of $B$.  The critical density $n_c$ determined in this way at each given $B$ allows the mapping of the $(n_s,B,T=0)$ phase diagram (Fig. \ref{fig:eng-phased} bottom).  It should be noted that, at low fields ($B\lesssim 2$~T), $\sigma(n_s,B,T)$ curves exhibit beautiful scaling \shortcite{Kevin_PRL} in agreement with $T\rightarrow 0$ extrapolations and general considerations \shortcite{Belitz}, providing additional strong evidence for a quantum phase transition in this system in parallel $B$.  In analogy with high-mobility samples, the MR exhibits a strong increase in the region of the magnetic-field driven MIT near $n_c(B)$ \shortcite{Kevin_Proc}.  For $B>2$~T, where 2DES is spin polarized, $n_c(B)$ saturates, indicating that the MIT occurs between a spin-polarized metal and a spin-polarized insulator.  The existence of the metallic phase at fields up to 18~T has been confirmed by the measurements of $\sigma(T)$, which retains a simple power-law form, albeit with a different exponent \shortcite{Kevin_PRL}.  The charge dynamics has not been studied in these samples yet.

It is interesting that the $n_c(B)$ dependence in samples with local moments (Fig. \ref{fig:eng-phased} bottom) is quite similar to that obtained on high-mobility samples [Fig.~\ref{fig:phased-inB}(a)].  In both cases, at low fields $n_c(B)$ increases with $B$ in a power-law fashion: $[n_c(B) - n_c(0)]/n_c(0)\propto B^{\beta} $.  The crossover exponent $\beta\approx0.9-1$ in Fig. \ref{fig:eng-phased} (bottom panel) and $\beta=1.1\pm 0.1$ in Fig.~\ref{fig:phased-inB}(a).  In other words, $n_c$ increases approximately linearly with $B$ at low fields.  It is remarkable that this dependence is essentially the same in both cases, even though the $d\sigma/dT$ behaviors in the metallic phase at $B=0$ are strikingly different.  The key features of the $n_c(B)$ phase diagram have been reproduced theoretically based on a scenario of quantum melting of a Wigner crystal as the mechanism of the MIT in sufficiently clean samples \shortcite{Vlad-Nature}.  In general, a power-law shift of $n_c$ with $B$ is expected to occur in the case of a true MIT~\shortcite{Belitz}, and has been observed in several 3D systems \shortcite{3Dbeta-1,3Dbeta-2,3Dbeta-3,3Dbeta-4}.

\section{Glassy freezing of electrons in two dimensions}
\label{glass}

Understanding the dynamics of glasses and other systems out of equilibrium is one of the most challenging and rapidly evolving topics in condensed matter research (see \shortciteNP{LesHouches-2002}).  Since vastly different types of systems exhibit similar behavior, it is tempting to search for common organizing principles and unified theoretical approaches.  However, despite some progress, there are still no well established theoretical frameworks for treating nonequilibrium behavior, which ``remains largely uncharted territory'' \shortcite{CMMP-report}.  For example, although glassy behavior may dominate the low-temperature properties of many complex materials near quantum phase transitions \shortcite{Mir-Dob-review} (Sec. \ref{intro}), such as the MIT, quantum glasses are even less understood than their classical counterparts.  Experimental studies of charge or Coulomb glasses \shortcite{eglass1,eglass2,eglass3,eglass4,eglass5}, which are of particular relevance to the MIT, have been relatively scarce \shortcite{Monroe1,films1-Zvi,films1a-Zvi,films2-Zvi,films3-Zvi,films4-Zvi,films5-Zvi,films6-Zvi,films7-Zvi,films3-a,films3-b,films3-c,films3-d,films3-e,films3-f,films3-g,Kar,Armitage}, and mostly limited to insulating systems far from the MIT.  Recent
observations \shortcite{SBPRL,SB_PhysE,JJ_PRL02,JJ_PhysE02,SB_SPIE03,JJ_noiseB,JJ_SPIE04,relax-PRL,tw-PRL,DP-aging} of
glassiness in a 2DES in Si MOSFETs near the MIT open up opportunities for exploring glassy phenomena in this important regime over a wide range of all the relevant parameters.

There are two basic ways to explore the dynamics of glassy systems.  The first one is to measure the response of the system to some kind of a perturbation.  In a spin glass, for example, this would typically involve a study of the relaxation of the magnetization following some combination of rapid cooling and a change in the applied magnetic field (see \shortciteN{Vincent-school} for a pedagogical review).  The second one is to measure the fluctuations of an observable with time (\textit{i.e.} noise), which provides information on correlations\footnote{In equilibrium systems, the connection between spontaneous fluctuations of a variable and the response of such a variable to a small perturbation in its conjugated field is given by the fluctuation-dissipation relation.  See \shortciteN{nieuw-book} for the review and discussion of thermodynamics of out-of-equilibrium systems.}.  In spin glasses, transport noise measurements were required in order to provide definitive information on the details of glassy ordering and dynamics \shortcite{Weiss93}.  Both approaches have been used to probe the dynamics of the 2DES in Si, focussing on the conductivity $\sigma$ as the variable most relevant to the MIT.  Measurements were performed on both high- and low-mobility Si MOSFETs, which also differ substantially in their geometry, size, and many other fabrication details, spanning essentially the entire range of Si technology.  Thus the emergence of glassy dynamics proves to be a universal phenomenon in Si inversion layers, at least in the absence of disorder-induced local magnetic moments.  The effect of local moments on charge dynamics still remains to be investigated.

\subsection{Fluctuations of conductivity}
\label{fluct}

The experimental protocol for measuring the $\sigma(t)$ fluctuations is simple, in principle: the measurement of $\sigma$ at a given $V_g$ or electron density $n_s$ is set up at high temperatures, the sample is then cooled to the measurement $T$, and $\sigma$ is measured as a function of time.  At the end of the measurement, the sample is warmed up to a temperature that is so high that a subsequent cool-down to the same measurement $T$ would result in a reproducible value of the time-averaged $\langle\sigma(t)\rangle$ within the experimental uncertainty.  The carrier density $n_s$ is changed at a high temperature, and the protocol is repeated for different $n_s$ and $T$.  In practice, the measurement set-up itself is often somewhat complicated \shortcite{SBPRL,JJ_PRL02,SB_SPIE03} because it is important to rule out various extraneous effects, such as the fluctuations of $T$, $V_g$, or contacts, as possible sources of the measured noise.  However, there are now a number of standard methods to accomplish this (see, \textit{e.g.}, \shortciteN{Sco87}, \shortciteN{Verbruggen89}).

\begin{figure*}
\vspace*{-0.35in}
\begin{minipage}[c]{0.4\textwidth}
\hspace*{-0.4in}\includegraphics[width=7.5cm,clip]{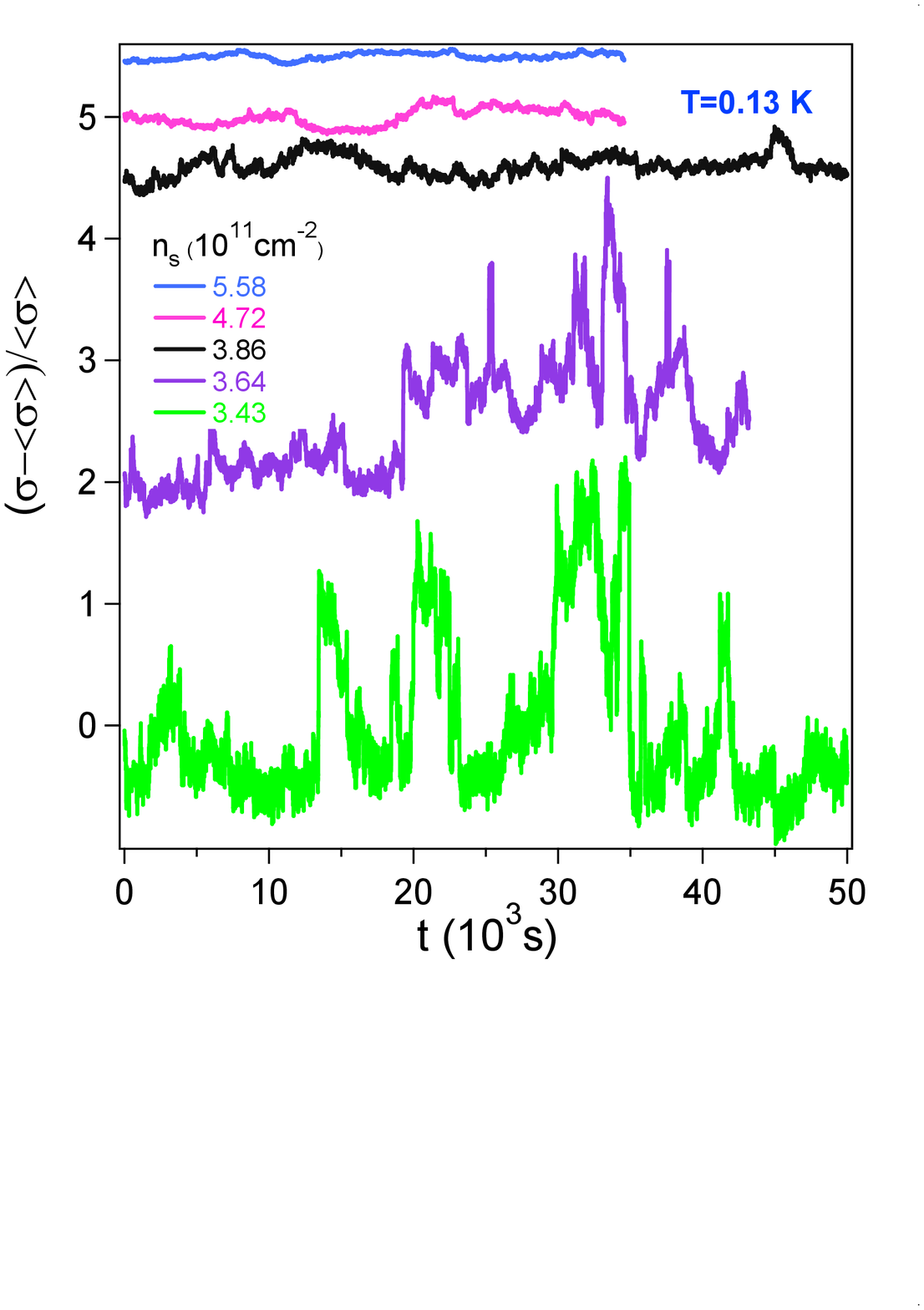}
\end{minipage}
\begin{minipage}[c]{0.4\textwidth}
\vspace*{-2.0cm}\hspace*{0cm}\includegraphics[height=4.7cm,clip]{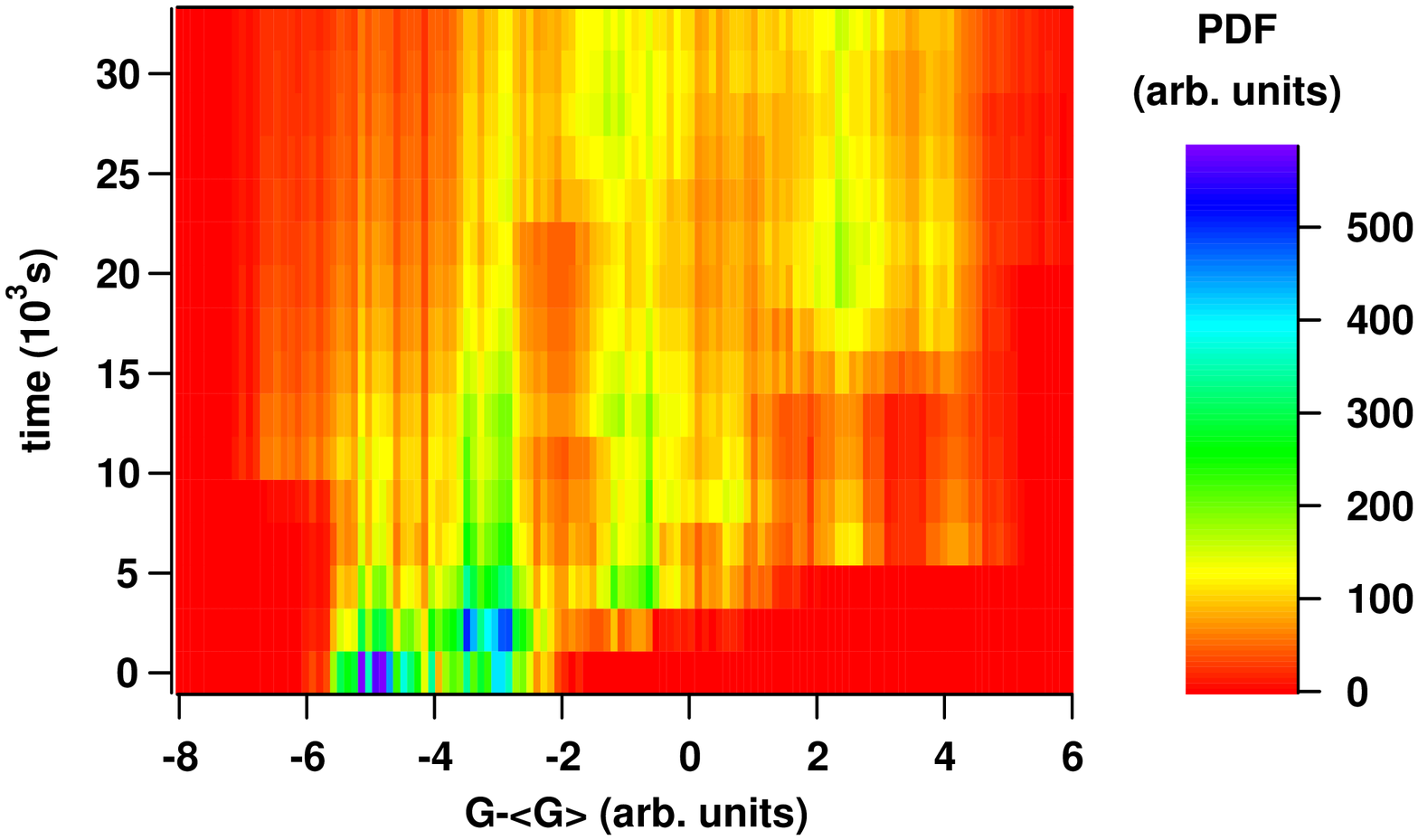}
\end{minipage}
\vspace*{-1.15in}
\caption{Low-mobility sample (\protect\shortciteNP{SBPRL} and Fig.~\ref{fig:saverage}).  Left: Relative fluctuations of $\sigma$ {\it vs.} time for different
$n_s$ at $T=0.13$~K.  Different traces have been shifted for clarity; the lowest $n_s$ is at the bottom and the highest $n_s$ at the top.  Right: The color map of the probability density function (PDF) of the conductance ($G$) fluctuations as a function of sampling time $t$ for the $n_c<n_s(10^{11}$cm$^{-2})=5.58<n_g$ data shown on the left.}
\label{fig:datasb}
\end{figure*}

In both low- and high-mobility samples, $\sigma$ exhibits strong fluctuations with time at low $n_s$ and $T$.  Figure \ref{fig:datasb} left
shows the fluctuations of $(\sigma-\langle\sigma\rangle)/\langle\sigma\rangle$ in a low-mobility sample\footnote{$\langle\ldots\rangle$ represents averaging over time intervals of, typically, several hours.} for several $n_s$ at $T=0.13$~K.  It is striking that, for
the lowest $n_s$, the fluctuation amplitude is of the order of 100\%. In addition to rapid, high-frequency fluctuations, slow changes over periods of several hours are also evident.
The probability density function (PDF) of the fluctuations illustrates how the sample explores the free energy landscape (FEL).  For $n_s < n_g$, the PDF is always not only non-Gaussian but also has a very complex structure that changes with time (Fig. \ref{fig:datasb} right).  The PDF broadens with the increasing  sampling time $t$, as the system has more time to explore the FEL.  However, it explores it so slowly that it remains nonergodic on experimental time scales.  For $n_s>n_g$, on the other hand, the PDFs become perfectly Gaussian on much shorter time scales, suggesting that the system reaches equilibrium.  The amplitude of the fluctuations decreases dramatically from $\sim$100\% to less than 1\% with increasing either $n_s$ (Fig.~\ref{fig:datasb} left) or $T$.

\begin{figure*}
\vspace*{-0.5in}
\begin{minipage}[c]{0.4\textwidth}
\hspace*{-0.4in}\includegraphics[width=8.5cm,clip]{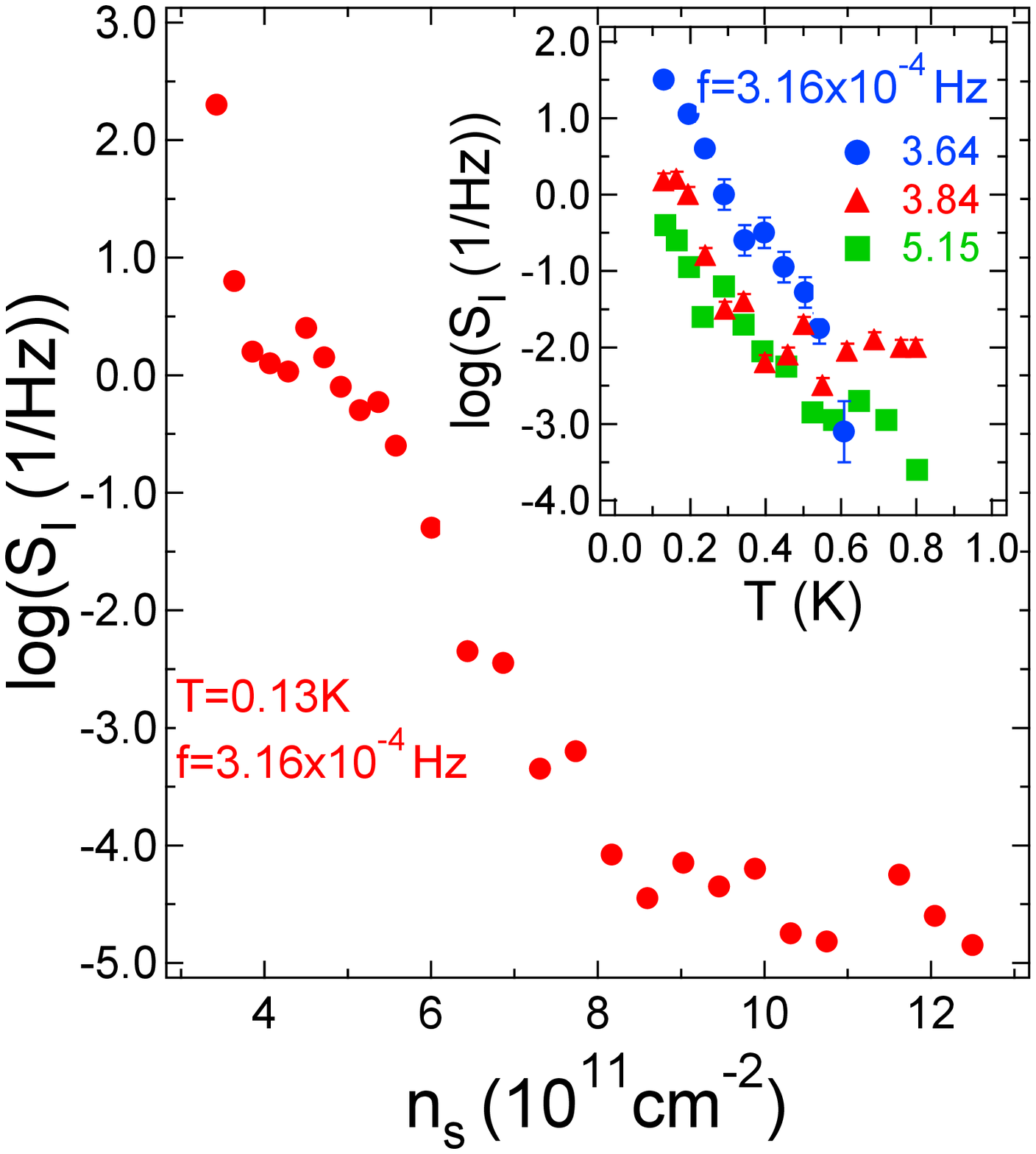}
\end{minipage}
\begin{minipage}[c]{0.4\textwidth}
\vspace*{0.0cm}\hspace*{1.0cm}\includegraphics[width=6.5cm,clip]{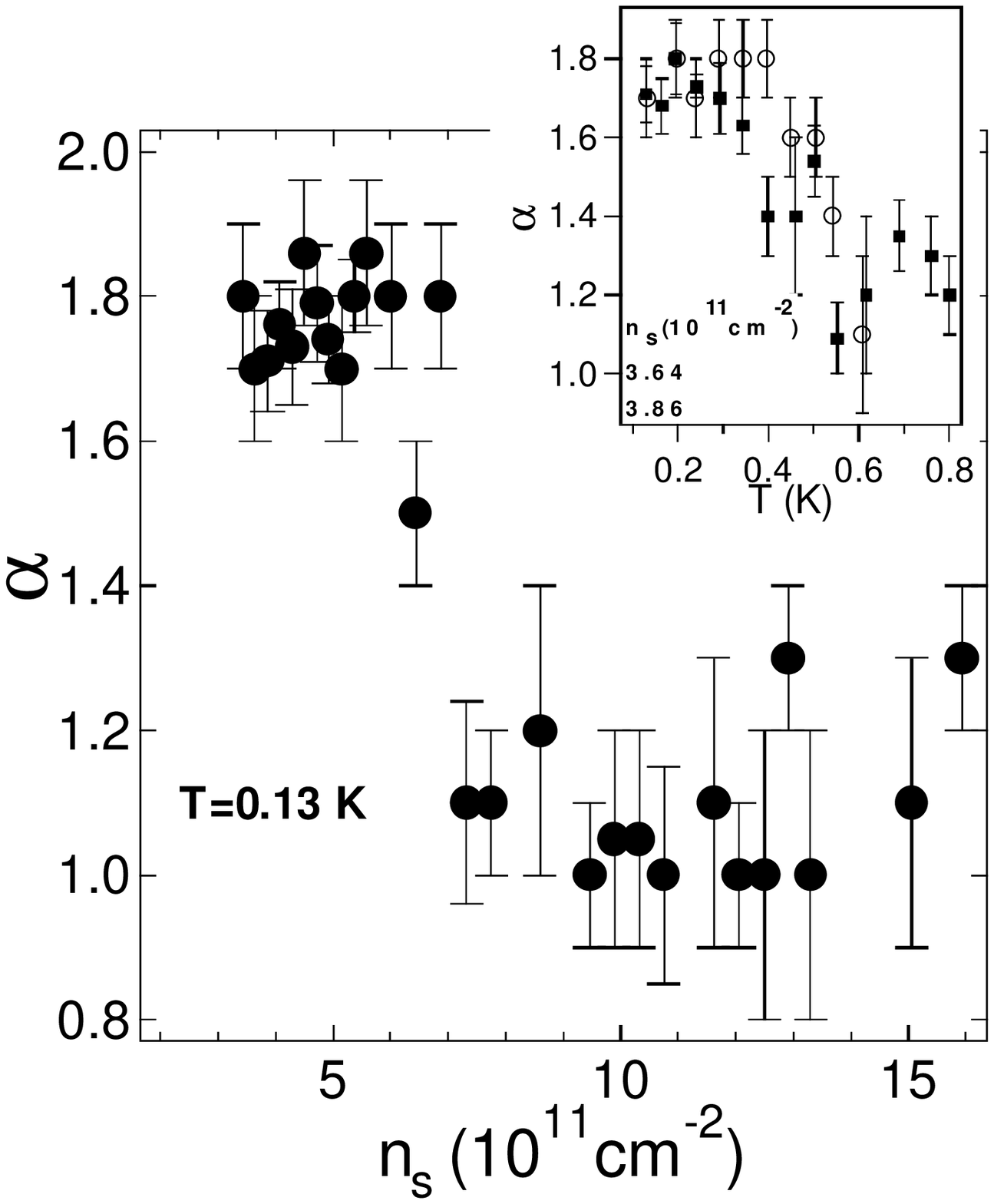}
\end{minipage}
\vspace*{-0.5in}
\caption{Low-mobility sample (\protect\shortciteN{SBPRL} and Fig.~\ref{fig:saverage}).  Left: The normalized noise power $S(f=3.16\times 10^{-4}$Hz) {\it
vs.} $n_s$ at $T=0.13$~K \protect\shortcite{SB_PhysE}.  Below $n_g\approx 7.5\times 10^{11}$cm$^{-2}$, the noise increases exponentially with decreasing $n_s$.  Inset: $S$ \textit{vs.} $T$ for three different $n_s(10^{11}$cm$^{-2})$ given on the plot.  Right: At $n_s\approx n_g$, the exponent $\alpha$ exhibits a sharp jump from $\approx 1$ at high $n_s$ (``pure'' $1/f$ noise) to $\approx1.8$ at low $n_s$ \protect\shortcite{SBPRL}.  Inset: $\alpha$ \textit{vs.} $T$ for two different $n_s(10^{11}$cm$^{-2})$ (3.64 -- open symbols, 3.86 -- solid symbols) in the glassy phase \protect\shortcite{SB_SPIE03}.}
\label{fig:spectra}
\end{figure*}

An even more dramatic density dependence of the noise at low $T$ is revealed by the study of the power spectra $S(f)$ ($f$--frequency) of the relative changes in the conductivity $(\sigma(t)-\langle\sigma\rangle)/\langle\sigma\rangle$.  Most of the spectra were obtained in the $f=(10^{-4}-10^{-1})$~Hz bandwidth, where they follow the well-known empirical law $S\propto 1/f^{\alpha}$ \shortcite{Hooge,Weiss88}.  At the highest $n_s$, $S(f)$ does not depend on $n_s$ (Fig.~\ref{fig:spectra} left) but, as $n_s$ is reduced below $n_g$, $S$ increases enormously, by up to six orders of magnitude at low $f$. Moreover, for a given $n_s<n_g$, $S(f)$ increases exponentially with decreasing $T$ (Fig.~\ref{fig:spectra} left, inset).  The observed $dS/dT<0$ makes it possible to rule out various simple models of noise (see, \textit{e.g.}, \shortciteN{SBPRL}, \shortciteN{SB_SPIE03} for discussion).  The most striking feature of the data, however, is the sharp jump of the exponent $\alpha$ at $n_s\approx n_g$ (Fig. \ref{fig:spectra} right).  While $\alpha\approx1$ (``pure'' $1/f$ noise) for $n_s>n_g$, $\alpha\approx 1.8$ below $n_g$, reflecting a sudden shift of the spectral weight towards lower frequencies.  Similar large values of $\alpha$ have been observed in some spin glasses above the MIT~\shortcite{jjprl98,Neutt}, and in submicron wires~\shortcite{Wrobel}.  Both the increase in the magnitude of the noise at low $f$  and the jump in $\alpha$ reflect a sudden and dramatic slowing down of the electron dynamics at $n_g$, indicating glassy freezing.
The onset of glassy dynamics on the metallic side of the MIT, \textit{i.e.} at $n_g>n_c\approx 5\times 10^{11}$cm$^{-2}$, implies the existence of the metallic glass phase for $n_c<n_s<n_g$, which is consistent with the predictions of the model of interacting electrons near a disorder-driven MIT \shortcite{Darko-glass}.  Since, in the glassy phase, $\alpha$ decreases with increasing $T$ (Fig.~\ref{fig:spectra} right, inset),  the large values of $\alpha$ and the jump in $\alpha(n_s)$ are observable only at relatively low $T$.

\begin{figure*}
\begin{minipage}[c]{0.4\textwidth}
\hspace*{0cm}\normalsize{(a)}\hspace*{0cm}\includegraphics[width=7.0cm,clip]{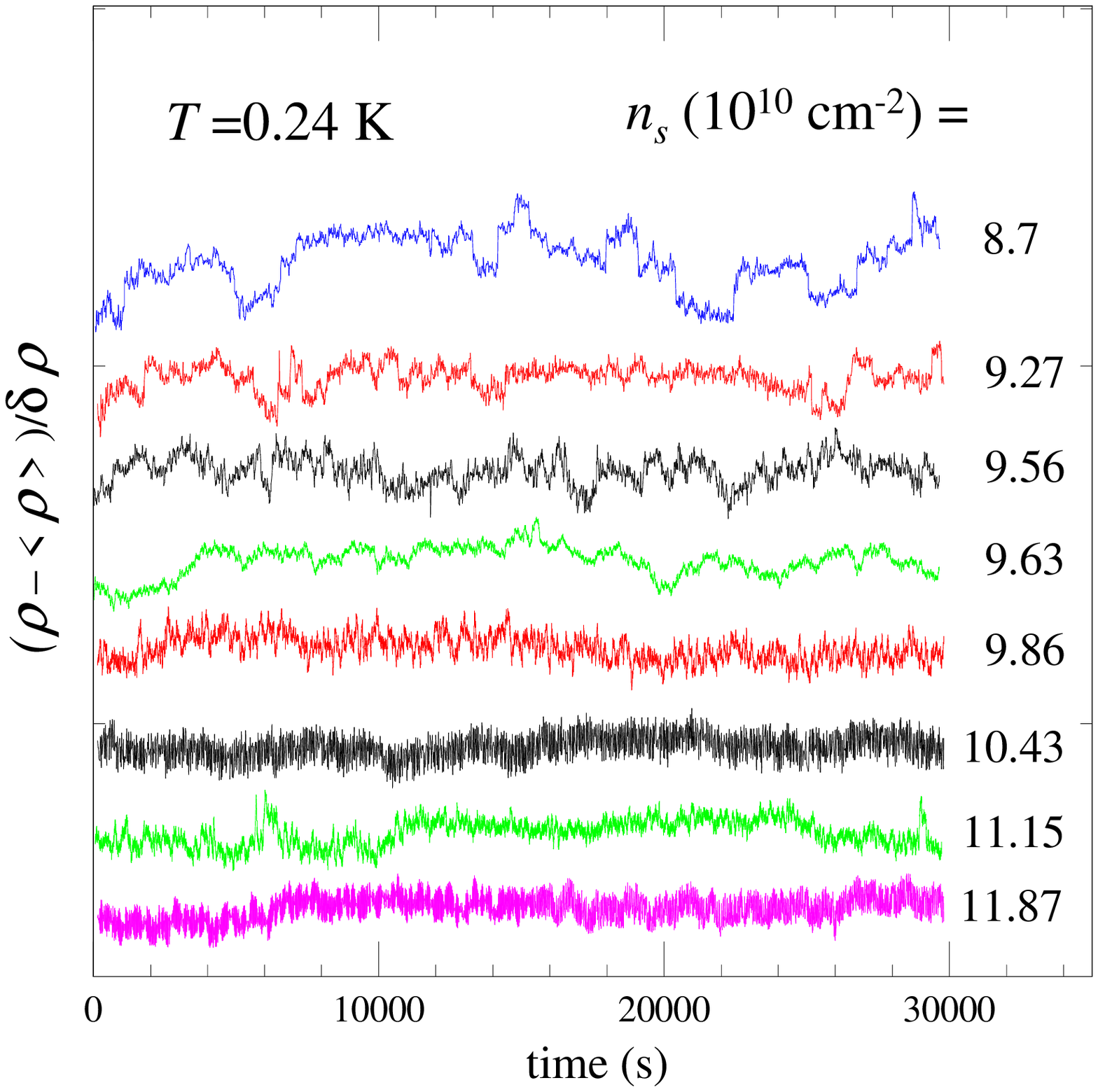}
\end{minipage}
\begin{minipage}[c]{0.4\textwidth}
\vspace*{0.0cm}\hspace*{2.2cm}\includegraphics[width=5.5cm,clip]{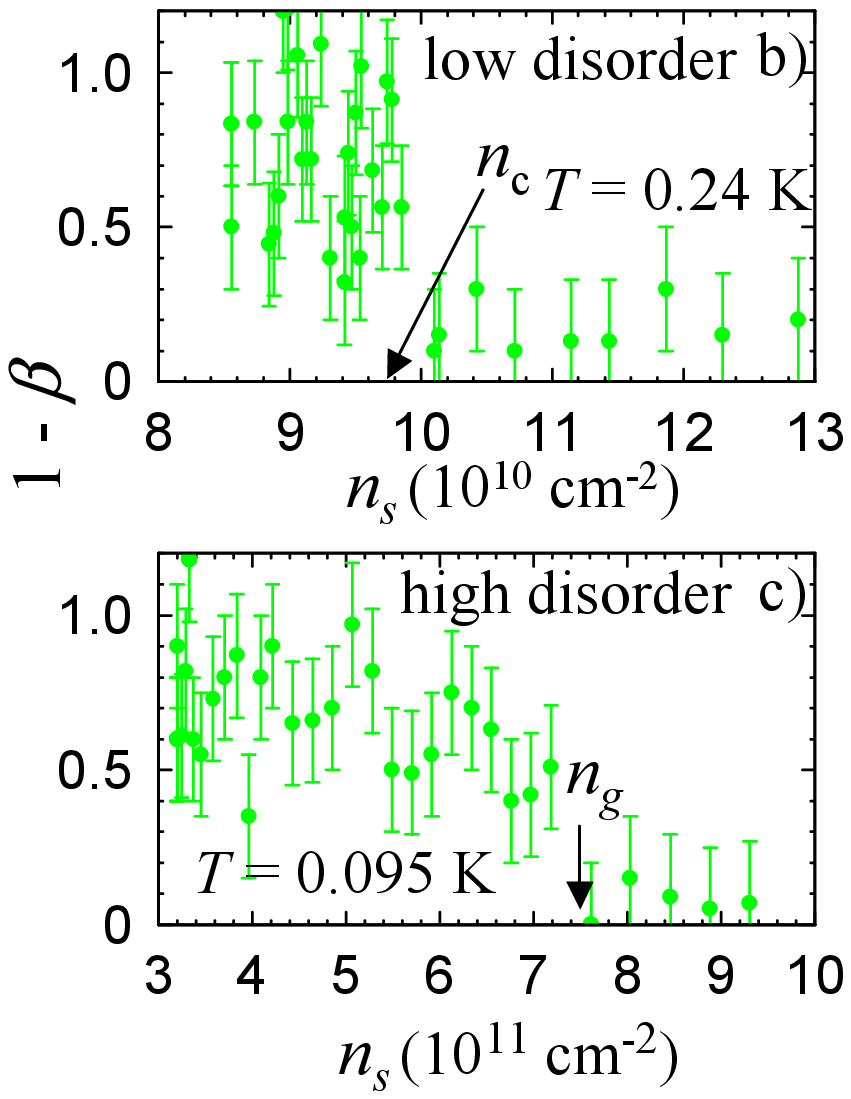}
\end{minipage}
\caption{(a) Resistance noise in a high-mobility (low-disordered) sample for several $n_s$ shown on the plot (adapted from \protect\shortciteNP{JJ_PRL02}).  $(\rho-\langle\rho\rangle)/\delta\rho$ is plotted ($\delta\rho^{2}=$variance, $\rho$ -- resistivity) in order to make the change in the character of the noise with $n_s$ more apparent. Different traces have been shifted vertically for clarity.  Exponent $1-\beta$, a measure of correlations, {\textit vs.} $n_s$ for (b) high-mobility [$n_c\approx 9.7\times 10^{10}$cm$^{-2}$ \protect\shortcite{JJ_PRL02}] and (c) low-mobility samples [$n_c\approx 5.0\times 10^{11}$cm$^{-2}$ \protect\shortcite{SBPRL}].}
\label{fig:ss}
\end{figure*}

Qualitatively the same behavior has been observed in the resistance (or conductance) noise of high-mobility samples, except that the glassy freezing takes place at $n_g\approx n_c$ \shortcite{JJ_PRL02}.  Importantly, in both types of samples, the character of the noise also changes with density: while at low $n_s$ both the ``shape'' and the variance of the noise exhibit random, nonmonotonic changes with time, at high enough $n_s$ the noise always ``looks'' the same.  This is illustrated in Fig.~\ref{fig:ss}(a) for a high-mobility sample.  The figure shows $(\rho-\langle\rho\rangle)/\delta\rho$, where $\delta\rho=\langle(\rho-\langle\rho\rangle)^{2}\rangle^{1/2}$, in order to make it easier to compare the signals.  A quantitative measure of the spectral wandering with time, such as that observed at low $n_s $, is the
so-called second spectrum $S_2(f_2,f)$, which is a fourth-order noise statistic.  $S_2$ is the power spectrum of the fluctuations of $S(f)$ with time \shortcite{Weiss93,WeissMMM}, {\textit i.~e.} the Fourier transform of the autocorrelation function of the time series of $S(f)$. In general, if the fluctuators (\textit{e.g.} two-level systems) are not correlated, $S_2(f_2,f)$ is white (independent of $f_2$) \shortcite{WeissMMM,Weiss88,Weiss93} and equal to the square of the first spectrum.  Such noise is called Gaussian. On the other hand, $S_2$ has a nonwhite character, $S_2\propto 1/f_{2}^{1-\beta}$, for interacting fluctuators \shortcite{WeissMMM,Weiss88,Weiss93}. Therefore, the deviations from Gaussianity, or the exponent $(1-\beta)$, provide a direct probe of correlations between fluctuators.  Indeed, $S_2$ has been an important tool in studies of other glasses.

A detailed dependence of the exponent $(1-\beta)$ on $n_s$ has been determined for both high- and low-mobility samples
(Figs.~\ref{fig:ss}(b) and (c), respectively) \shortcite{JJ_PRL02}. In both cases, $S_2$ is white [$(1-\beta)=0$] for $n_s>n_g$, indicating that the observed $1/f$ noise results from uncorrelated fluctuators.  It is quite remarkable that $S_2$ changes its character in a dramatic way exactly at $n_g$ in both types of samples.  For $n_s<n_g$, $S_2$ is strongly non-Gaussian, which demonstrates that the fluctuators are strongly correlated and provides an unambiguous evidence for the onset of glassy dynamics in a 2DES at $n_g$.

\begin{figure*}
\centering
\includegraphics[height=6.0cm,clip]{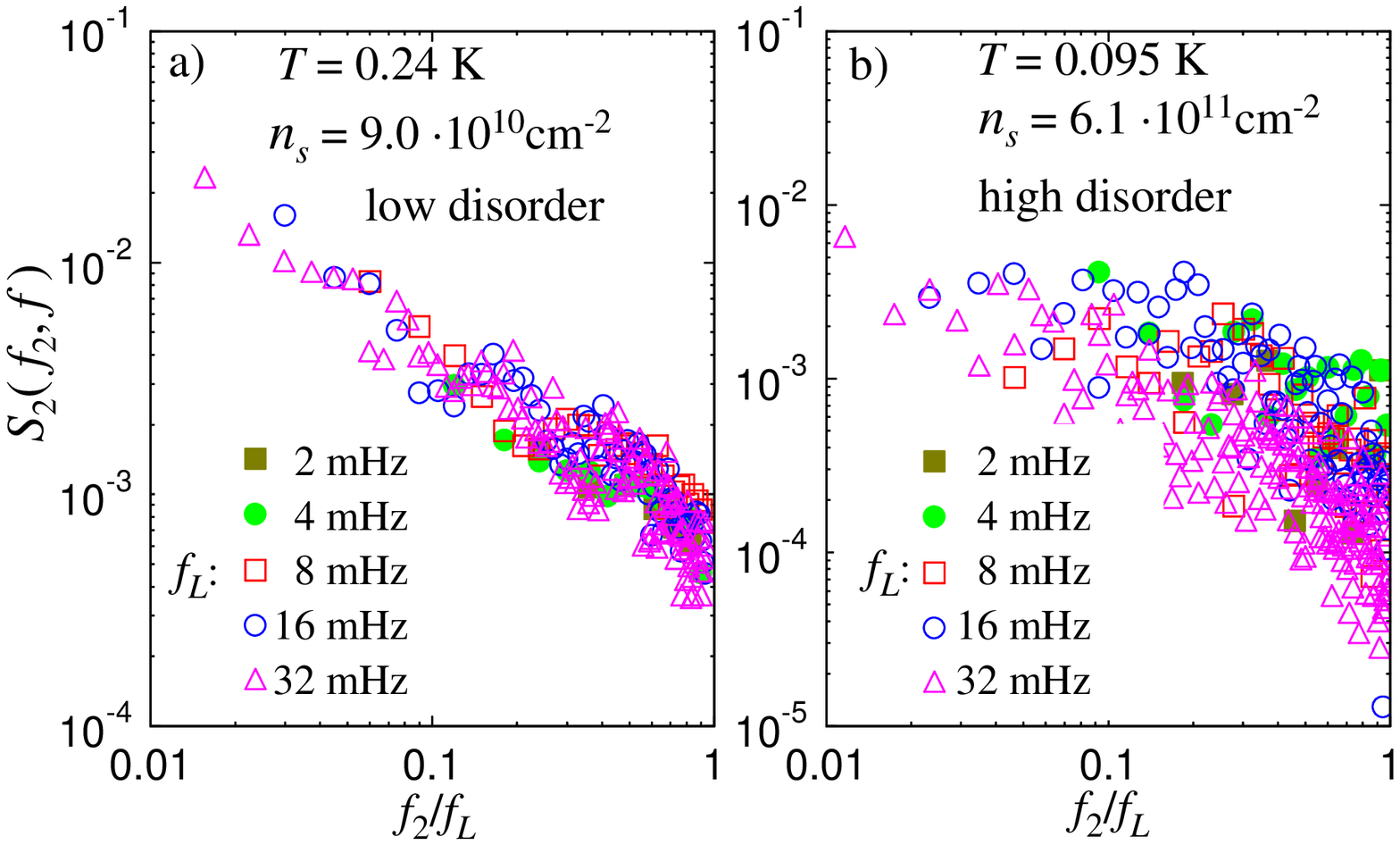}\\
\vspace*{0.2cm}\includegraphics[height=5.5cm,clip]{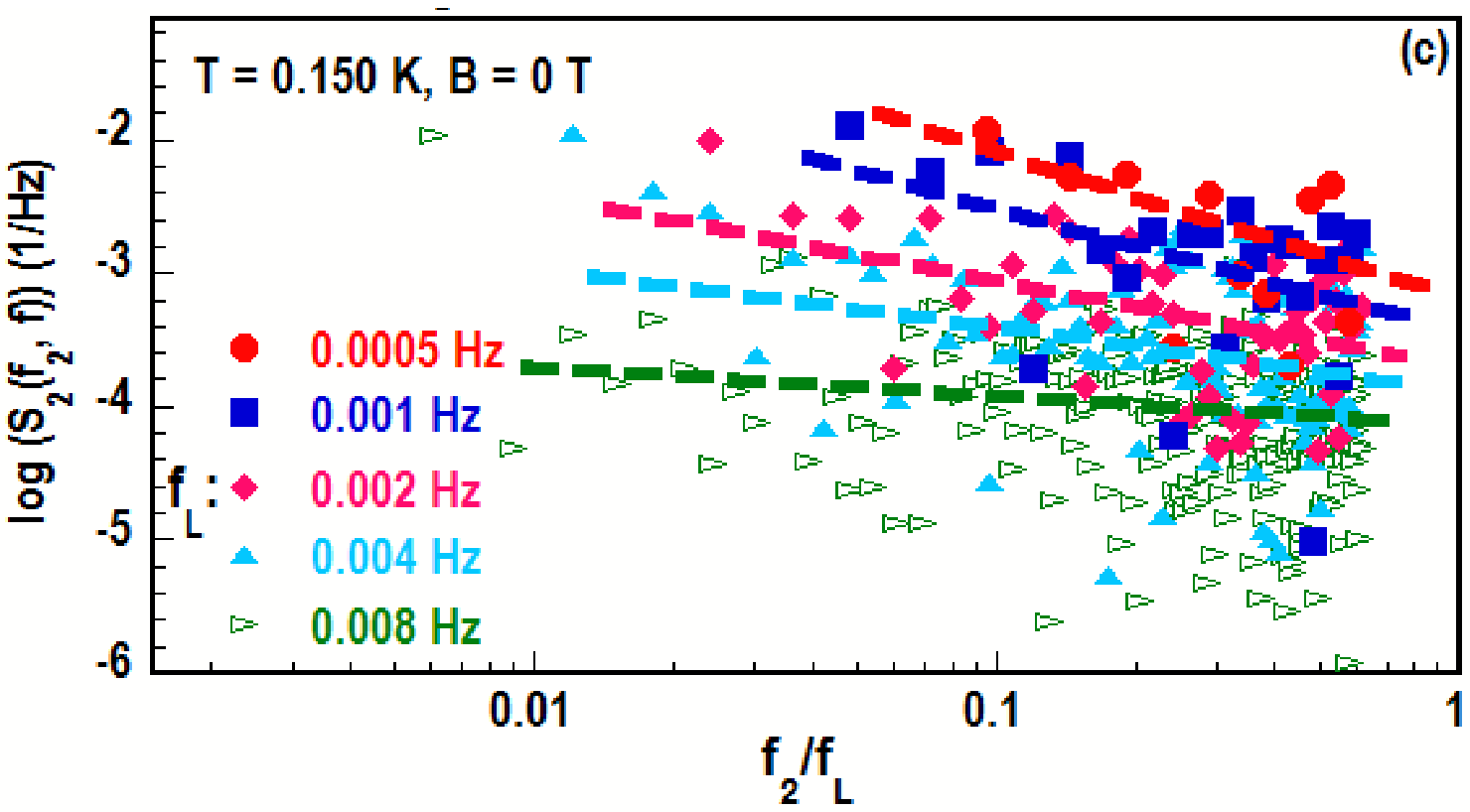}
\caption{Scaling of the second spectra $S_2$ measured at frequency octaves $f=(f_L,2f_L)$ for a given $n_s$ in (a) high-mobility (for $n_s<n_c$) and (b) low-mobility (for $n_c<n_s<n_g$) 2DES \protect\shortcite{JJ_PRL02}.  There is no characteristic time (or energy) scale observed, in agreement with hierarchical models.  This is in contrast to (c) the insulating La$_{1.97}$Sr$_{0.03}$CuO$_4$, where $S_2$ clearly shows the presence of some characteristic energy scale in that system \protect\shortcite{IR_PRL,IR-inplane}, suggesting that charge carriers form a cluster glass state.  Dashed lines are linear fits to guide the eye.}
\label{fig:scaling}
\end{figure*}

In the studies of spin glasses, the scaling of $S_2$ with respect to $f$ and $f_2$ has been used \shortcite{WeissMMM,Weiss93} to unravel the glassy dynamics and, in particular, to distinguish generalized models of interacting droplets or clusters from hierarchical pictures. In the former case, the low-$f$ noise comes from a smaller number of large elements because they are slower, while the higher-$f$ noise comes from a larger number of smaller elements, which are faster.  In this picture, which assumes compact droplets and short-range interactions between
them, big elements are more likely to interact than small ones and, hence, non-Gaussian effects and $S_2$ will be stronger for lower $f$.  $S_2(f_2,f)$, however, need to be compared for fixed $f_2/f$, {\textit i.~e.} on time scales determined by the time scales of the fluctuations being measured, since spectra taken over a fixed time interval average the high-frequency data more than the low-frequency data.  Therefore, in the interacting ``droplet'' model, $S_2(f_2,f)$ should be a decreasing function of $f$ at constant $f_2/f$.  In the hierarchical picture, on the other hand, $S_2(f_2,f)$ should be scale invariant: it should depend only on $f_2/f$, not on the scale $f$ \shortcite{WeissMMM,Weiss93}. Figures~\ref{fig:scaling}(a) and (b) show that no systematic dependence of $S_2$ on $f$ is seen in either high- or low-mobility 2DES \shortcite{JJ_PRL02}.  The observed scale invariance of $S_2(f_2,f)$ signals that the system wanders collectively between many metastable states related by a kinetic hierarchy.  Metastable states correspond to the local minima or ``valleys'' in the FEL (Fig. \ref{fig:FEL}), separated by barriers with a wide, hierarchical distribution of heights and, thus, relaxation times. Intervalley transitions, which are reconfigurations of a large number of electrons, thus lead to the observed strong, correlated, $1/f$-type noise, remarkably similar to what was observed in spin glasses with a long-range correlation of spin configuration \shortcite{WeissMMM,Weiss93}.
On the other hand, in systems where both long-range and short-range interactions are present, such as lightly doped La$_{2-x}$Sr$_x$CuO$_4$, $S_2(f_2,f)$ reveals clearly \shortcite{IR_PRL,IR-inplane} [Fig.~\ref{fig:scaling}(c)] the presence of some characteristic energy scale, indicating that hierarchical models are not applicable and suggesting instead the formation of a cluster glass state.  This is exactly what is expected in Coulomb systems with competing interactions (Sec. \ref{intro}).

In the resistance noise measurements, the glass transition in a 2DES is thus manifested by (a) a sudden and dramatic increase of $S(f)$ and a jump of $\alpha$ from $\approx 1$ to $\approx 1.8$, indicating the slowing down of the dynamics, and (b) a change of the exponent $(1-\beta)$ from a white (zero) to a nonwhite (nonzero) value, indicating an abrupt change to a correlated statistics, consistent with the hierarchical pictures of glassy dynamics \shortcite{Binder}.  For high-mobility 2DES in Si MOSFETs, low-$T$ noise measurements were performed also in parallel $B$ \shortcite{JJ_noiseB}.  By adopting the same criteria for the glass transition in $B$, it was possible to determine $n_g(B)$ shown in Fig.~\ref{fig:phased-inB}(a) and to establish that charge, not spin, degrees of freedom are responsible for glassy ordering (see Sec.~\ref{mitinB}).  Measurements in both $B=0$ and $B\neq 0$ show that glassy behavior generally emerges before the electrons localize (\textit{i.e.} $n_g>n_c$), consistent with theory \shortcite{Darko-glass}.  The glassy signatures in the noise become gradually stronger as $T$ decreases (\textit{e.g.} Fig.~\ref{fig:spectra} insets), suggesting that the glass transition takes place as $T\rightarrow 0$.  Strong evidence for $T_g=0$ and further support for $n_g$ as the glass transition density is provided by the measurements of the response of the 2DES to a strong perturbation.

\subsection{Glassy response}
\label{response}

In MOSFET structures, the easiest and the most obvious way to excite the system is a sudden change of $V_g$.  This method has been applied to several electron glasses \shortcite{films1-Zvi,films1a-Zvi,films2-Zvi,films3-Zvi,films4-Zvi,films5-Zvi,films6-Zvi,films7-Zvi,films3-a,films3-b,films3-e,films3-g}, including a 2DES in Si \shortcite{relax-PRL,tw-PRL,DP-aging,aging-PhysB}.  So far, only low-mobility samples discussed above (Secs. \ref{lowmobility} and \ref{fluct}) have been probed in this way.  In particular, three different experiments have been carried out, as described below.

\subsubsection{Relaxations of conductivity after a rapid change of carrier density}
\label{relax}

\begin{figure*}
\centering
\includegraphics[height=4.6cm,clip]{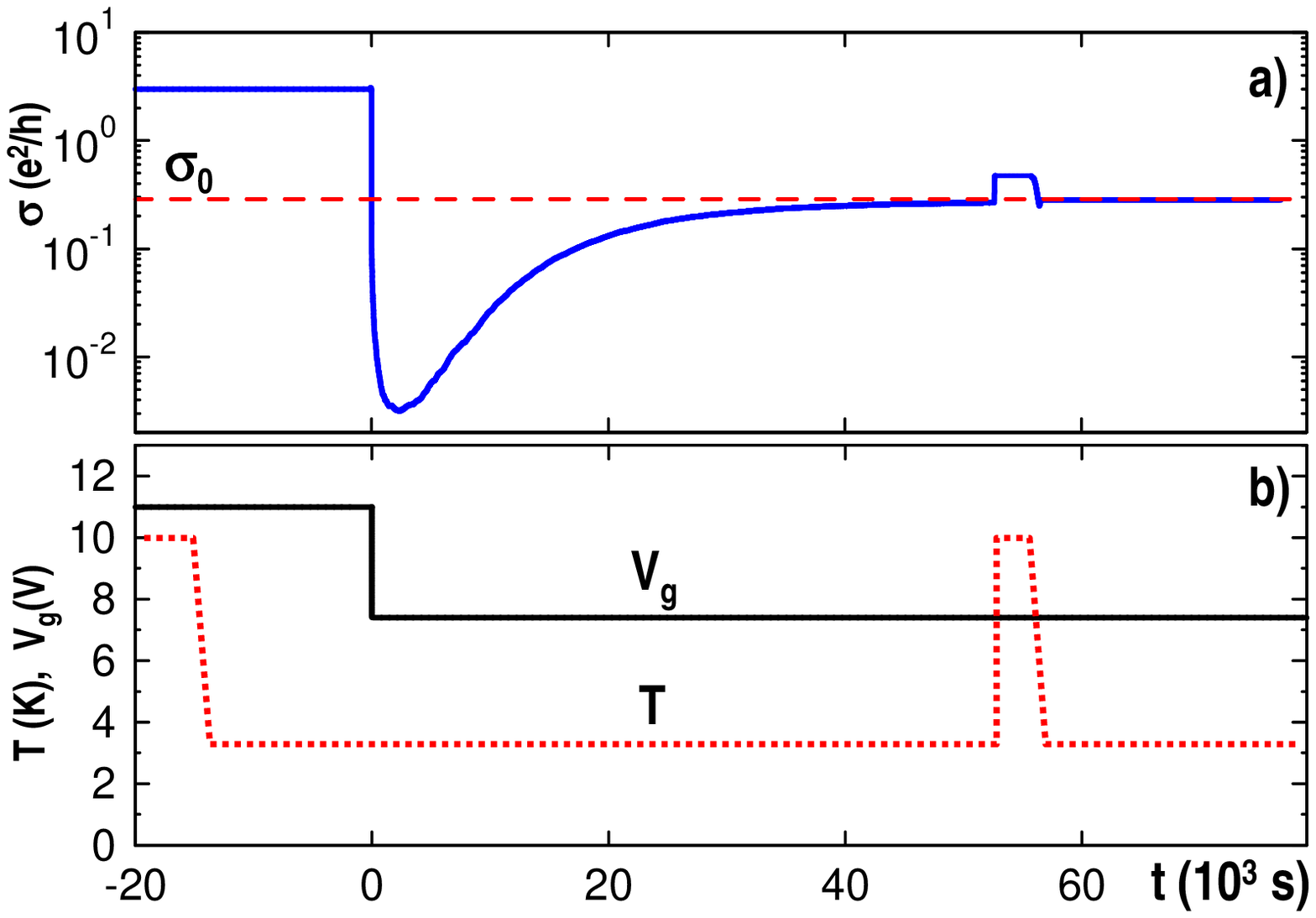}\\
\vspace*{0.0cm}\hspace*{0.0cm}\includegraphics[height=4.6cm,clip]{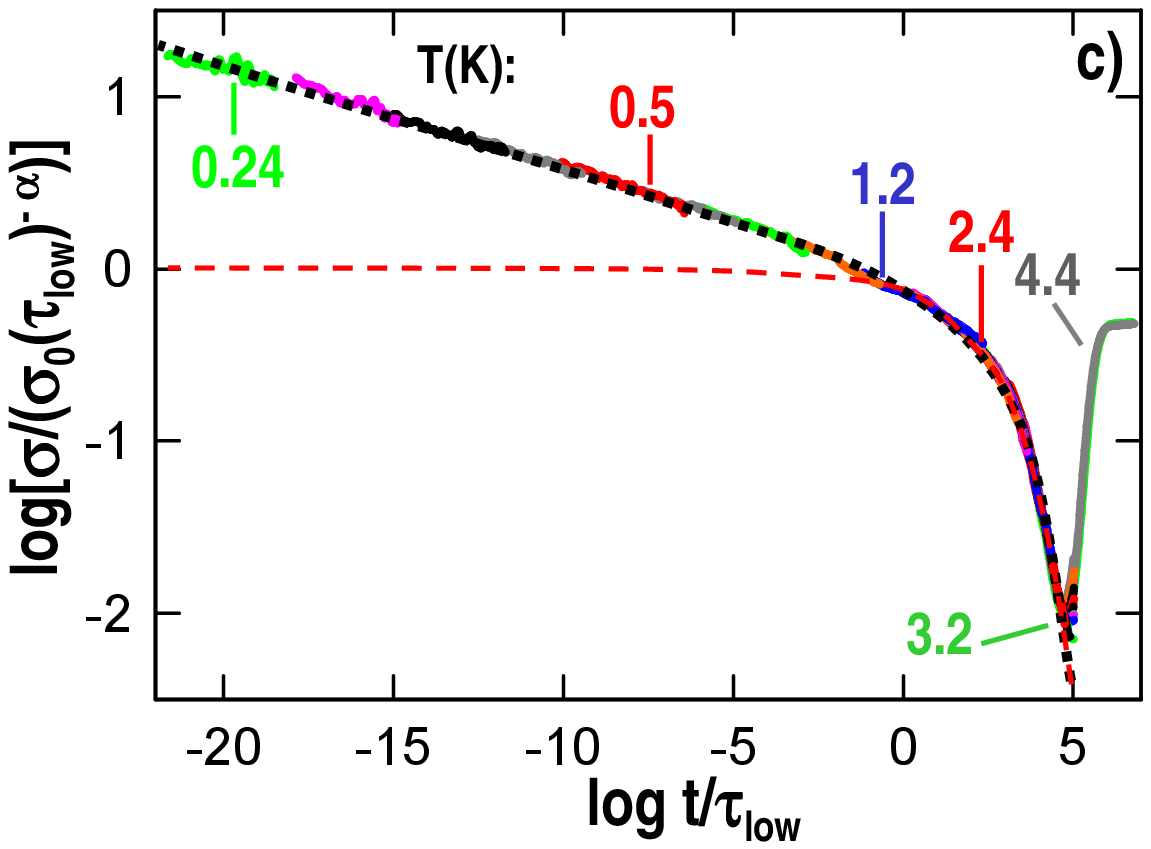}
\caption{Low-mobility sample with $n_g\approx 7.5\times 10^{11}$cm$^{-2}$ and $n_c\approx 4.5\times 10^{11}$cm$^{-2}$.
(a) $\sigma(t)$ for $V_{g}^{i}=11$~V [$n_s(10^{11}$cm$^{-2})=20.26$], $V_{g}^{f}=7.4$~V [$n_s(10^{11}$cm$^{-2})=4.74$], and $T=3.3$~K. (b) Experimental protocol: $V_g(t)$ and $T(t)$.  (c)  All the data at short times, (\textit{i.e.} before the minimum) are consistent with the Ogielski relaxation, $\sigma(t,T)/\sigma_{0}(T)\propto(\tau_{low})^{-\alpha}(t/\tau_{low})^{-\alpha}\exp[-(t/\tau_{low})^{\beta}]$ (dotted line).  The lowest $T$ data clearly deviate from the stretched exponential dependence (dashed line).  The data have been collapsed with respect to the 2.4~K curve.  Adapted from \protect\shortciteNP{relax-PRL}.}
\label{fig:relax}
\end{figure*}

A systematic study of the relaxations (\textit{i.e.} time dependence) of the conductivity $\sigma(t)$ has been performed at different $n_s$ and $T$ after a large, rapid change of $n_s$ (controlled by $V_g$) \shortcite{relax-PRL}.  The sample was first cooled from high temperature (10~K) to the measurement temperature $T$ with an initial gate voltage $V_{g}^{i}$.  Then, at $t=0$, the gate voltage was switched rapidly (within 1~s) to a final value $V_{g}^{f}$, and $\sigma(t,V_{g}^{f},T)$ was measured.  In general, the results did not depend on any of the following: initial temperature, as long as it was $\geq 10$~K; $V_{g}^{i}$; the cooling time, which was varied between 30 minutes and 10 hours;  the time the sample was kept at 10~K (from 5 minutes to 8 hours); the time the sample spent at the measurement $T$ before $V_g$ was changed (from 5 minutes to 8 hours).  Figures~\ref{fig:relax}(a) and (b) show a typical experimental run with $\sigma(t>0)$ exhibiting a rapid ($<10$~s) initial drop followed by a slower
relaxation\footnote{The RC charging times of the device and the circuit were at most 10~ms.}. After going through a minimum, $\sigma(t)$ increases and approaches a value $\sigma_0(V_{g}^{f},T)$. A subsequent warm-up to $10$~K and a cool down to the same measurement $T=3.3$~K, while keeping the gate voltage fixed at $V_{g}^{f}$, shows that $\sigma_0(V_{g}^{f},T)$ represents the equilibrium conductivity corresponding to the given $V_{g}^{f}$ and $T$.  It is interesting that, even though initially it drops to a value close to $\sigma_0$, $\sigma$ first goes away from equilibrium before it starts approaching $\sigma_0$ again.  At the end of the run, the sample was warmed up to 10~K, gate voltage changed back to the same $V_{g}^{i}$, and the experiment was repeated at a different $T$ for the same $V_{g}^{f}$.  The whole process was then repeated for different values of $V_{g}^{f}$, in order to map out the density dependence of various relaxation parameters.  It was established \shortcite{relax-PRL} that, for $n_s<n_g$, the relaxations have the following properties.

After a sufficiently long $t$, $\sigma$ relaxes exponentially to its equilibrium value $\sigma_0$.  The corresponding equilibration time obeys the simply activated form $\tau_{eq}\propto\exp(E_{act}/T)$, $E_{act}\approx 57$~K.  While the microscopic origin of the activation energy $E_{act}$ is not known yet, the activation to an upper subband in Si MOSFETs or to Si-SiO$_2$ interface traps has been ruled out \shortcite{tw-PRL}. However, regardless of the equilibration mechanism, the important result is that $\tau_{eq}\rightarrow \infty$ as $T\rightarrow 0$, so that, strictly speaking, the system cannot reach equilibrium only at $T=0$.  In other words, the glass transition takes place at $T_{g}=0$.  Remarkably, a Monte Carlo study of the 2D Coulomb glass model has also found \shortcite{Grempel} an exponential divergence of $\tau_{eq}$, signaling a glass transition at $T_g=0$.  There are, however, some differences in the detail between the model and the experiment, indicating a need for further refinement of theory.

For short enough $t$, $\sigma(t)$ obeys a nonexponential, Ogielski form \shortcite{Ogielski} $\sigma(t,T)/\sigma_{0}(T)\propto t^{-\alpha}\exp[-(t/\tau_{low})^{\beta}]$ ($0<\alpha(n_s)<0.4$, $0.2<\beta(n_s)<0.45$) [Fig.~\ref{fig:relax}(c)], which is a product of a power law and a stretched exponential function.  Both types of relaxations are considered to be typical signatures of glassy behavior and reflect the existence of a broad distribution of relaxation times.  The scaling parameter $\tau_{low}\propto\exp(E_a/kT)$ ($E_a\approx 20$~K) so that, as $T\rightarrow 0$, $\tau_{low}\rightarrow\infty$ and the relaxations attain a pure power-law form $\propto t^{-\alpha}$ .  The Ogielski form, the divergence of $\tau_{low}$, and the resulting power law relaxation at $T_g$ are consistent with the general scaling arguments \shortcite{scaling} near a continuous phase transition occurring at $T_g=0$.  These results are very similar to scaling observed in spin glasses\footnote{$T_g$ is finite in spin glasses.} above $T_g$ \shortcite{Pappas}.  In a 2DES, $\alpha\rightarrow 0$ as $n_s\rightarrow n_g$, providing further evidence for $n_g$ as the glass transition density.

$\tau_{low}$ exhibits a very pronounced and precise dependence on the density: $\tau_{low}\propto\exp(\gamma n_{s}^{1/2})$ ($\gamma$--a proportionality constant).  Since $1/r_s=E_F/U\propto n_{s}^{1/2}$ in 2D (Sec.~\ref{mit}), this particular form of $\tau_{low}(n_s)$ provides strong evidence that the observed out-of-equilibrium behavior is dominated by the Coulomb interactions between 2D electrons.

Perhaps the most peculiar finding is that the 2DES equilibrates only after it first goes \textit{farther away} from equilibrium.  While the precise
mechanism for nonmonotonic relaxation remains controversial, studies of other materials [spin glasses \shortcite{sg-nonmon1,sg-nonmon2}, manganites \shortcite{mang-nonmon}, insulating granular metals \shortcite{Aviad-nonmon}, biological systems \shortcite{bio-nonmon}] and some theoretical models \shortcite{th-nonmon1,th-nonmon2} suggest that it may be a general feature of systems that are far from equilibrium.

It should be noted that it is quite remarkable that this one experiment has provided so many important results.  To put things into a perspective, for example, it has not been possible to determine $T_g$ in other electron glasses so far, and scaling in spin glasses \shortcite{Pappas} has been observed only relatively recently, in spite of many more years of study.

\subsubsection{Relaxations of conductivity after a waiting time protocol: aging and memory loss}
\label{tw}

\begin{figure}
	\centering
		\includegraphics[height=5.5cm,clip]{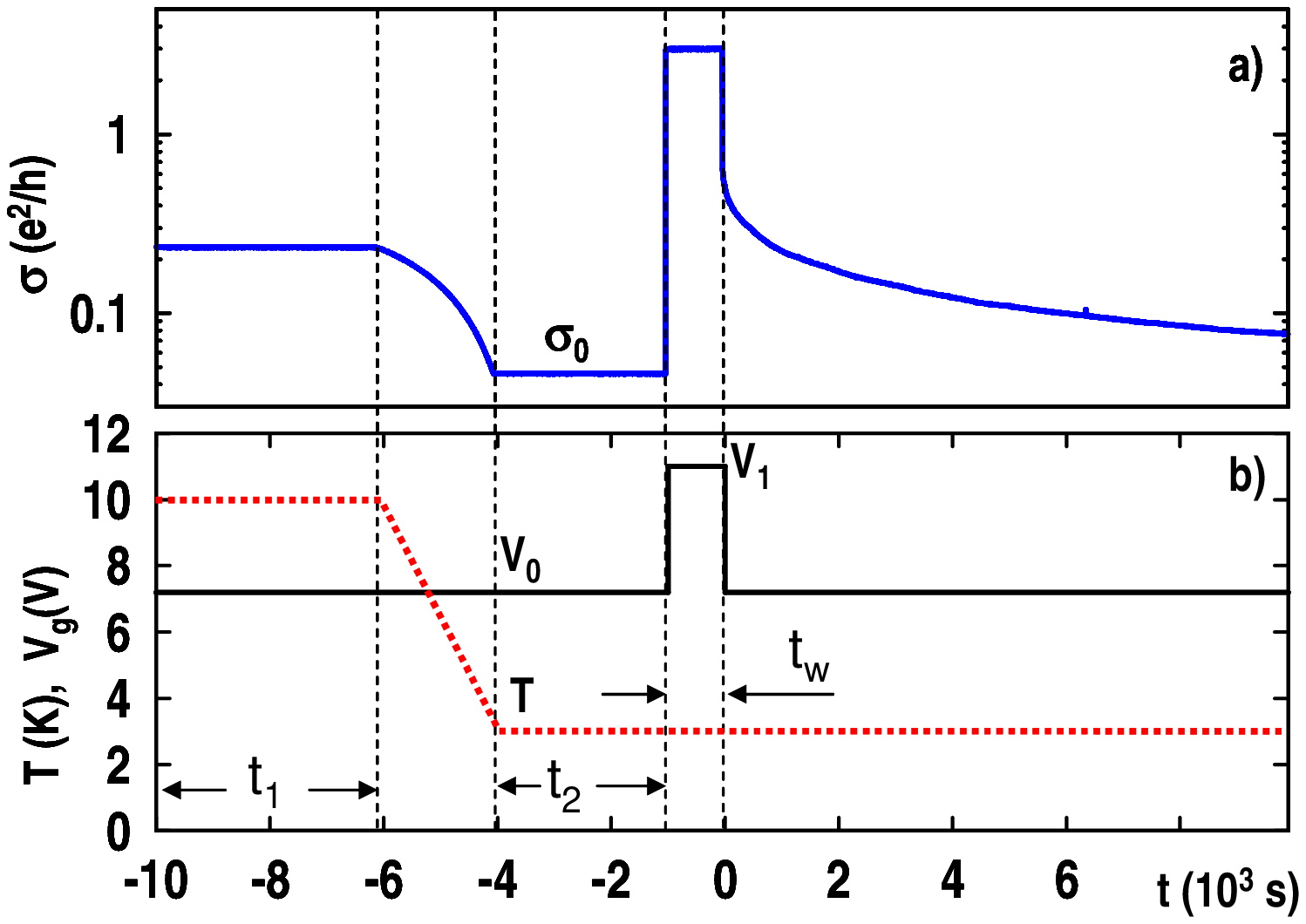}
\vspace*{0.0cm}\includegraphics[height=5.5cm,clip]{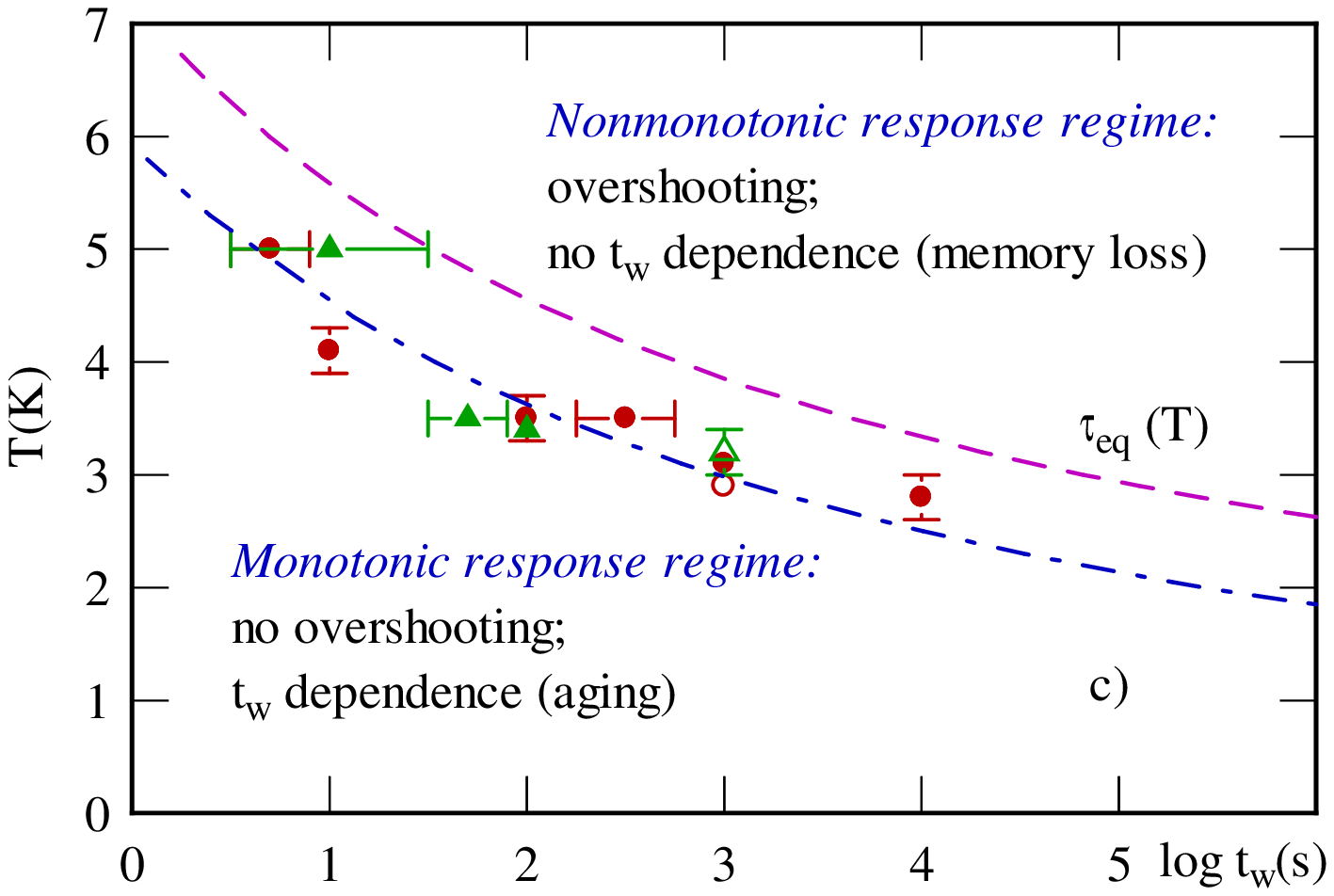}
		\caption{(a) $\sigma(t)$ for $V_{0}=7.2$~V [$n_s(10^{11}$cm$^{-2})=3.88<n_c$], $V_{1}=11$~V [$n_s(10^{11}$cm$^{-2})=20.26$], $t_{w}=1000$~s,
and $T=3.0$~K. (b) The corresponding experimental protocol: $V_g(t)$ and $T(t)$. The results do not depend on the cooling time (varied from 30 minutes to 10 hours), nor on $t_1$ and $t_2$ (varied from 5 minutes to 8 hours each).  (c) The symbols show the values of $(T,t_w)$ where the overshooting vanishes, \textit{i.e.} the conditions that separate the two regimes.  Open and solid symbols correspond to different samples; the blue dot-dashed line guides the eye.  The purple dashed line represents $\tau_{eq}(T)$.  Adapted from \protect\shortciteN{tw-PRL}.}
	\label{fig:T-tw}
\end{figure}

A key characteristic of relaxing glassy systems is the loss of time translation invariance, reflected in aging effects \shortcite{aging1,aging2,glasses}.  The system is said to exhibit aging if its response to an external excitation depends on the system history in addition to the time $t$.  In a systematic study of the history dependence in a 2DES \shortcite{tw-PRL}, $\sigma(t)$ was measured after the temporary change of $n_s$ during the waiting time $t_w$ [Figs.~\ref{fig:T-tw}(a) and (b)].  The history was varied by changing $t_w$ and $T$ for several initial (final) $n_s$.

Two types of response have been observed: 1) monotonic, for relatively ``small'' excitations, where $\sigma(t)$ depend on $t_w$, \textit{i.e.} 2DES shows aging; 2) nonmonotonic, for sufficiently ``large'' excitations, where $\sigma(t)$ ``overshoots'' $\sigma_0$ (\textit{i.e.} it first goes farther away from equilibrium) and relaxations no longer depend on $t_w$ (memory loss).  The monotonic relaxations [Fig.~\ref{fig:T-tw}(a)] are consistent with a power-law form at the shortest times (or lowest $T$) and, at the longest $t$, with a simple exponential approach to equilibrium.

The aging is observed when $t_w\ll\tau_{eq}(T)$, \textit{i.e.} if the system is unable to equilibrate under the new conditions during $t_w$ [Fig.~\ref{fig:T-tw}(c)].  In that case, $\sigma(t)$ depends also on $t_w$: the system has a \textit{memory} of the time $t_w$.  This is very similar to spin glasses, where $T$ or $B$ play a role analogous to that of $n_s$.  In the opposite case, when $\tau_{eq}(T)\ll t_w$ and the 2DES equilibrates at a new state, the relaxations do not depend on $t_w$ since the system, naturally, has no memory of the waiting time.  Therefore, the overshooting is observed when the system is excited out of thermal equilibrium.  This is analogous to the experiment described above (Sec.~\ref{relax}) and thus sheds some light on that intriguing phenomenon.

The gate voltage changes $\Delta V_g$ employed in the relaxation experiments in a 2DES have been relatively large.  For example, $n_s$ was changed up to a factor of 7, and thus the 2DES could go from the conducting to the insulating regime.  Such large $\Delta V_g$ are expected to trigger major rearrangements of the electron configuration since the corresponding shifts of the Fermi energy\footnote{$E_F$[K]$=7.31 n_s[10^{11}$cm$^{-2}]$ for electrons in Si~\shortcite{AFS}.} $\Delta E_F\gg k_{B}T$ \shortcite{Muller-French}.  It is possible to speculate that such large perturbations might be somehow responsible for the peculiar overshooting effect.  Considerable charge rearrangements, coupled with possibly substantial changes in the screening of the 2DES across the MIT \shortcite{Vlad-MITglass1,Muller-scr,Pankov-scr}, present a fundamentally different situation from the cooling of the 2DES at a fixed $n_s$ when $k_{B}\Delta T\ll E_F$, where no relaxations have been observed.  On the other hand, smaller perturbations of a Coulomb glass are expected to lead to memory effects \shortcite{Muller-memory}, in agreement with the observations for $t_w\ll\tau_{eq}(T)$.  In that case, the final state has a large configurational similarity with the original state due to the shortness of $t_w$.  In InO$_x$ films, another well-studied electron glass, the overshooting of equilibrium has not been seen, but aging and memory effects have been observed \shortcite{films1-Zvi,films1a-Zvi,films2-Zvi,films3-Zvi,films4-Zvi}.  Those experiments were done in the regime of small perturbations, because the typical change in the carrier density due to $\Delta V_g$ was $\sim 1$\%, the system always remained deep in the insulating state, and $t_w\ll\tau_{eq}$ was satisfied.

\subsubsection{Aging effects across the MIT in 2D}
\label{aging}

Aging effects have been instrumental as a probe of complex nonequilibrium dynamics in many types of materials.  In a 2DES, where the onset of glassy dynamics takes place on the metallic side, a study of aging, especially across the MIT, is of great interest.  Aging is observed if $t_w\ll\tau_{eq}$ [Fig.~\ref{fig:T-tw}(c)].  Since the equilibration time $\tau_{eq}$ diverges exponentially as $T\rightarrow 0$ (Sec.~\ref{relax}), strictly speaking, the system can reach equilibrium at all $T>0$.  However, even at $T$ that are not too low (\textit{e.g.} $\sim 1$~K), $\tau_{eq}$ exceeds easily not only the experimental time window but also the age of the Universe \shortcite{relax-PRL}.  This makes it relatively easy to study the out-of-equilibrium relaxation of $\sigma$ at times $t\ll\tau_{eq}$, where one expects to find properties common to other types of glasses.  In strongly localized systems, such as InO$_x$ films, the aging function $\sigma(t,t_w)$ is just a function of $t/t_w$ \shortcite{films1-Zvi,films1a-Zvi,films2-Zvi,films3-Zvi,films4-Zvi}. This is known as simple, or full aging.  It is interesting that, in spin glasses, full aging has been demonstrated only relatively recently \shortcite{sg-fullaging}.  In general, however, the existence of a characteristic time scale $t_w$ does not necessarily imply simple $t/t_w$ scaling \shortcite{LesHouches-2002}.  In the mean-field models of glasses, for example, two different cases are distinguished: one, where full aging is expected, and the other, where no $t/t_w$ scaling is expected \shortcite{mean-field}. Experimentally, departures from full aging are common \shortcite{aging1,glasses}.

\begin{figure}
\centering
		\includegraphics[width=9.0cm,clip]{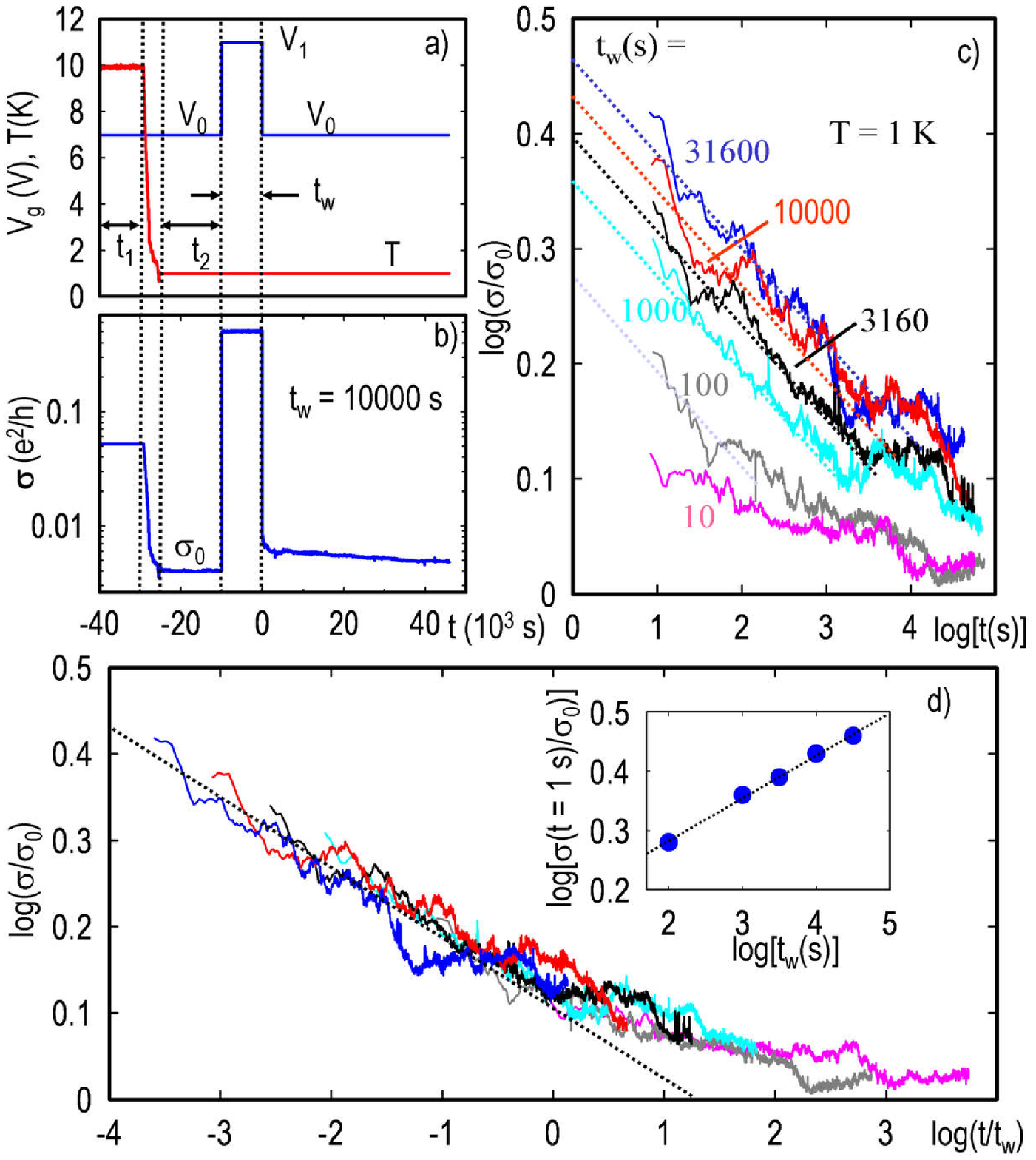}
\includegraphics[width=4.8cm,clip]{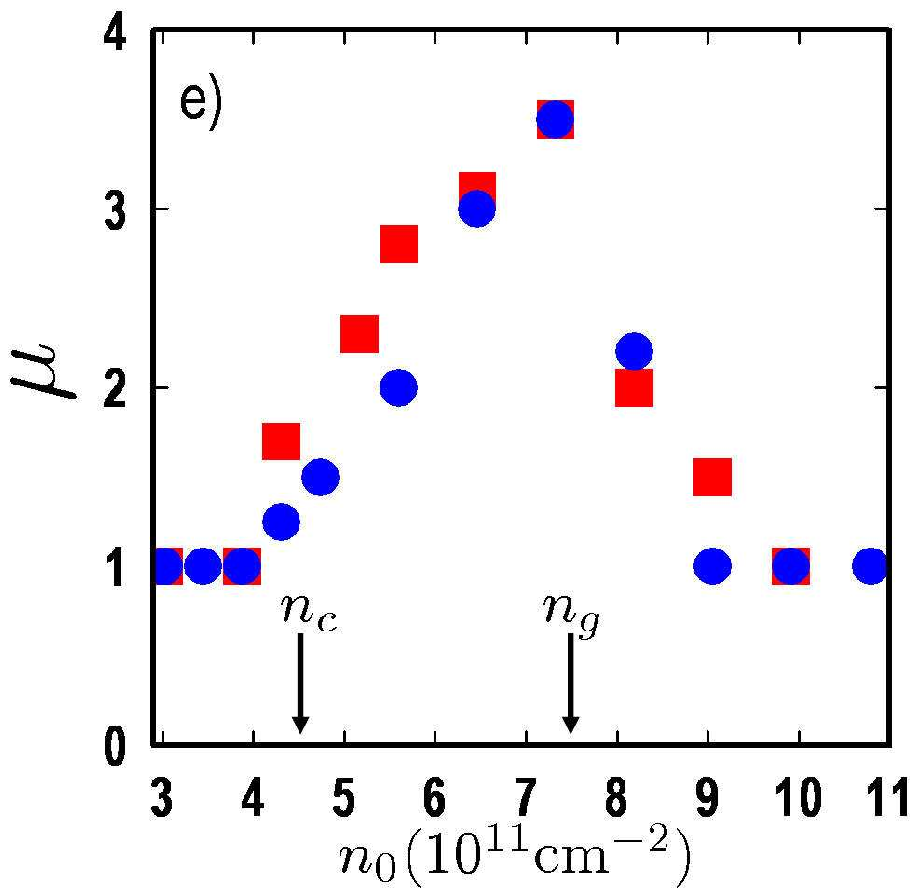}
\hspace*{-0.1cm}\includegraphics[width=8.0cm,clip]{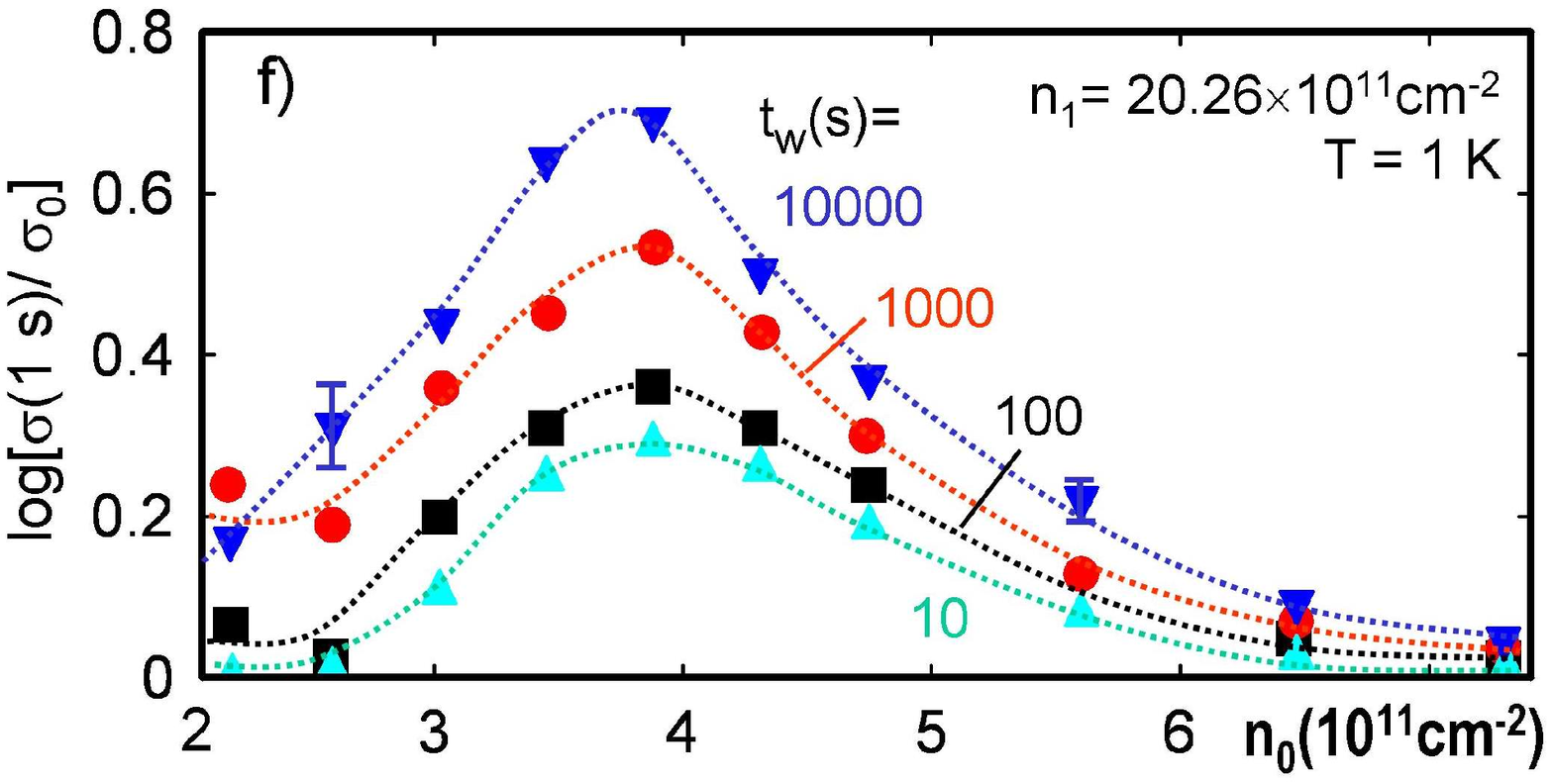}
		\caption{(a) $V_g(t)$ and $T(t)$ in a typical experimental protocol, which always starts with the 2DES in equilibrium at 10~K \protect\shortcite{tw-PRL}.
(b) The corresponding $\sigma(t)$.  The relaxation of $\sigma$ during $t_w$ is too small to be seen on this scale.  (c)  $\sigma(t>0)$ for several $t_w$, as shown; $V_{0}=7.0$~V [$n_{0}(10^{11}$cm$^{-2})=3.02<n_c$], $V_{1}=11$~V [$n_1(10^{11}$cm$^{-2})=20.26$].  The dotted lines are linear fits for $t\leq t_w$.  (d) The same data as in (c) but plotted \textit{vs.} $t/t_w$.  The dotted line is a fit for $t\leq t_w$ with the slope $-\alpha=-0.081\pm 0.005$.  Inset: $\sigma(t=1$s$)/\sigma_0$ \textit{vs.} $t_w$.  The dotted
line is a fit with the slope $\alpha=0.076\pm 0.005$.  (e) $\mu$ \textit{vs.} $n_0$ for two samples.  $\mu$ does not depend on $n_1$.  (f) Relaxation amplitudes
\textit{vs.} $n_0$ for several $t_w$.  Dotted lines guide the eye.  Adapted from \protect\shortciteN{DP-aging}, and \protect\shortciteN{aging-PhysB}.}
	\label{fig:aging}
\end{figure}

Aging was investigated in detail in a 2DES \shortcite{DP-aging} using the waiting time protocol [Sec.~\ref{tw}; Figs.~\ref{fig:aging}(a), (b)], but $T$ was kept fixed at 1~K such that $\tau_{eq}$ was astronomical and, hence, the 2DES was always deep in the $t_w\ll\tau_{eq}$ limit [Fig.~\ref{fig:T-tw}(c)].  $\sigma(t,t_w)$ were then explored systematically both as a function of final $n_s$ and of the difference in densities during and after $t_w$.  Figure~\ref{fig:aging}(c) illustrates the significant effect of $t_w$ on $\sigma(t)$.  In fact, all the $\sigma(t,t_w)$ data can be collapsed onto a single curve simply by rescaling the time axis by $t_w$ [Fig.~\ref{fig:aging}(d)].  Therefore, in this case, the system exhibits full aging at least up to $t\approx(2$--$3)t_w$.  The relaxations can be described by a power law $\sigma(t)/\sigma_0\propto (t/t_w)^{-\alpha}$ for times up to about $t_w$, followed by a slower relaxation at longer $t$.  This means that the memory of $t_w$ is imprinted on the form of each $\sigma(t)$.

For all $n_s<n_c$, $\sigma(t,t_w)$ exhibit simple or full aging.  However, as $n_s$ increases above $n_c$, there is an increasingly strong departure from full aging.  In some other glasses \shortcite{aging1,Ocio-mu1,Ocio-mu2}, it was found that the data could be scaled with a modified waiting time
$(t_{w})^{\mu}$, where $\mu$ is a fitting parameter ($\mu=1$ for full aging).  Even though $\mu$ may not have a clear physical meaning, the $\mu$-scaling approach has proved to be a useful tool for studying departures from full aging \shortcite{aging1,LesHouches-2002}.  By adopting a similar method in the study of aging in a 2DES, it was possible to achieve an approximate collapse of the data \shortcite{DP-aging}.  The plot of $\mu$ \textit{vs.} $n_0$ [Fig.~\ref{fig:aging}(e)] shows a clear distinction between the full aging regime for $n_s<n_c$, and the aging regime where significant departures from full scaling are seen.  It is striking that the largest departure occurs at $n_s\approx n_g$. For $n_s>n_g$, some small relaxations are observed (only for $k_{F}l<1$) resulting from finite-temperature effects, but they vanish in the $T\rightarrow 0$ limit (Sec.~\ref{relax}).  It was also determined that $\mu$ does not depend on $T$ \shortcite{aging-PhysB}.

These results are striking because they reveal an abrupt change in the nature of the glassy phase exactly at the 2D MIT itself, before glassiness disappears completely at a higher density $n_g$.  In other words, this is strong evidence that the insulating glassy phase and the metallic glassy phase are different.  Therefore, the difference in the aging properties below and above $n_c$ puts constraints on the theories of glassy freezing and its role in the physics of the 2D MIT.

Furthermore, for a given $t_w$, the amplitude of $\sigma(t)/\sigma_0$ has a peak at $n_s\lesssim n_c$ [Fig.~\ref{fig:aging}(f)], reflecting an interesting and surprising suppression of the relaxations on the insulating side of the 2D MIT.  While a clear understanding of this effect is lacking, it is plausible that collective charge rearrangements that are responsible for the slow dynamics will be suppressed as the 2DES becomes strongly localized.
It would be also of interest to study aging deeper in the insulator, in the variable-range hopping regime, but that is not possible because of the small relaxation amplitudes and the large intrinsic sample noise.  Of course, the effects of disorder could be explored further by extending the relaxation studies to cleaner (high-mobility) 2DES, where $n_g\gtrsim n_c$~\shortcite{JJ_PRL02,JJ_noiseB}.  Regardless of the origin of the nonmonotonic behavior in Fig.~\ref{fig:aging}(f), however, it is important to note that it reflects another change in the aging properties that occurs at the MIT.

\section{Summary}
\label{summary}

\begin{figure}
	\centering
		\includegraphics[width=9.5cm,clip]{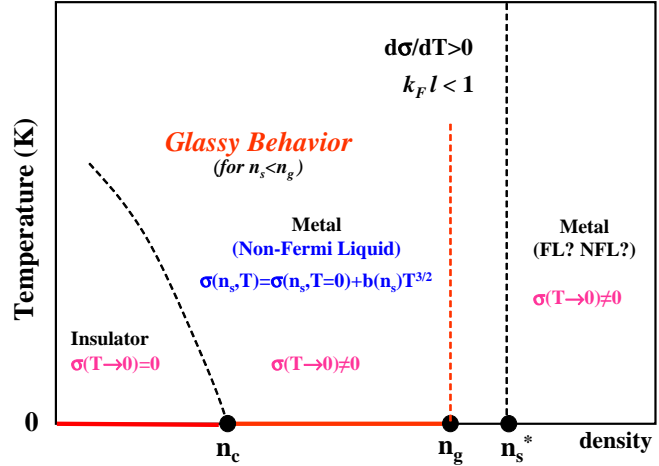}
\vspace*{-0.5cm}\caption{Experimental phase diagram of the 2DES in Si MOSFETs.  The metal-insulator transition takes place at the density $n_c$ and at $T=0$ in all samples regardless of the amount of disorder.  The nature of the metallic phase at high $n_s$, such that $k_{F}l>1$, is still under debate [see \protect\shortciteN{Sergey-review}].  The glass transition takes place at $T_g=0$ for all $n_s<n_g$.  In general, the glass transition is a precursor to the metal-insulator transition, \textit{i.e.} $n_c<n_g$, giving rise to an intermediate metallic phase with a particular form of the non-Fermi liquid temperature dependence of conductivity.  The aging properties of the glass change abruptly at $n_c$, indicating different natures of the insulating and metallic glass phases. For sufficiently low disorder, the intermediate phase vanishes: $n_c\lesssim n_{s}^{\ast}\approx n_g$.}
	\label{fig:summary}
\end{figure}

Careful studies have provided strong evidence that all 2DES in Si MOSFETs exhibit a metal-insulator transition regardless of the amount or type of disorder (Fig.~\ref{fig:summary}).  This is confirmed by extrapolating $\sigma(T)$ on both metallic and insulating sides of the MIT, and by dynamical scaling of $\sigma(n_s,T)$ over a wide range of parameters in both $B=0$ and $B\neq0$.  In $B=0$, the scaling form $\sigma (n_s,T)=\sigma_c(T)f(T/\delta_{n}^{z\nu})$ is obeyed with a temperature dependent critical conductivity $\sigma_c=\sigma(n_s=n_c,T)\propto T^x$, where the value of $x$ depends on the disorder.  Thus $\sigma_c(T)$ belongs to the insulating family of curves.  The results are consistent with general arguments near the MIT \shortcite{Belitz} and they are similar to the behavior in doped semiconductors near the 3D MIT \shortcite{Myriam_review}.  The 2D metallic phase survives in very high parallel magnetic fields, long after the 2DES becomes fully spin polarized.

Measurements of the charge dynamics in a 2DES have established that the glass transition takes place at $T_g=0$ for all $n_s<n_g$ (Fig.~\ref{fig:summary}).  In general, the glassiness sets in on the metallic side of the MIT, \textit{i.e.} at a density $n_g>n_c$, thus giving rise to an intermediate, metallic glass phase between the ``normal'' metal (\textit{i.e.} in the $k_{F}l>1$ regime) and the glassy insulator.  The glass transition and various glassy effects are observable only for $k_{F}l<1$, so that the intermediate phase is a very poor metal ($\sigma(T=0)\neq 0 \ll e^2/h$).  It is characterized by a very specific, non-Fermi liquid $T^{3/2}$ correction to $\sigma$ in both low-mobility samples in $B=0$ and high-mobility samples in a magnetic field.  In high-mobility samples in $B=0$, the intermediate phase practically vanishes and the glass transition coincides with the MIT.

The manifestations of the glass transition in a 2DES for $n_s<n_g$, as demonstrated by resistance noise measurements, include a dramatic slowing down of the electron dynamics and correlated statistics consistent with the hierarchical picture of glasses.  The results have been further confirmed and supported by the studies of relaxations, which provide evidence for the diverging equilibration time as $T\rightarrow 0$, broad distribution of relaxation times in the system, and scaling (in the time domain), consistent with a continuous phase transition at $T_g=0$.  The aging, one of the hallmarks of glassy dynamics, shows an abrupt change in its properties at the MIT, indicating that the natures of the insulating glass and metallic glass phases are different.  The 2DES in Si thus exhibits all the characteristics common to other out-of-equilibrium systems, regardless of their dimensionality.  Numerous similarities to spin glasses with long-range RKKY interaction are particularly remarkable.  Here, however, only long-range Coulomb interaction and potential scattering are present, and measurements in a parallel magnetic field confirm that charge degrees of freedom, not spin, are responsible for glassy freezing.  Therefore, the experiments show that the 2D MIT is closely related to the melting of this Coulomb glass.

\section{Discussion}
\label{disc}

\begin{figure}
	\centering
	\vspace*{-1.0cm}\includegraphics[width=12.0cm,clip]{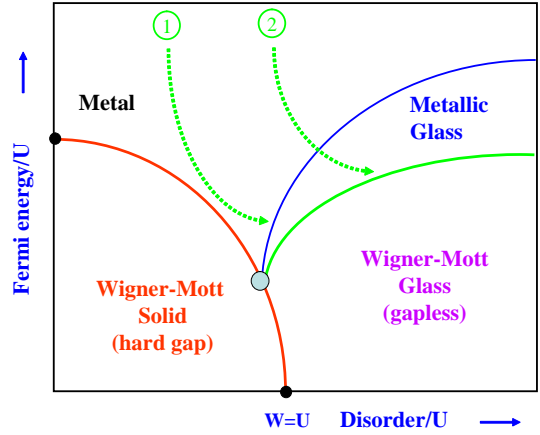}
\vspace*{-1.75cm}\caption{Theoretical phase diagram (adapted from \protect\shortciteNP{Darko-glass}).  The Fermi energy and disorder on the two axes are expressed in units of the on-site interaction $U$.  For large enough disorder, the metal-insulator transition is preceded by the glass transition, giving rise to an intermediate, metallic glass phase. Experimental trajectories, where the MIT is approached by varying $n_s$, are represented schematically by the green curved arrows for the cases of low (trajectory 1) and high (trajectory 2) disorder.}
	\label{fig:theory}
\end{figure}

Experiments on a 2DES in Si clearly provide strong support to theoretical proposals describing the 2D MIT as the melting of a Coulomb glass \shortcite{Darko-glass,MIT-glassothers1,MIT-glassothers2,MIT-glassothers3,Vlad-MITglass1,Vlad-MITglass2}.  In particular, the model of the MIT as a Mott transition with disorder \shortcite{Darko-glass} predicts the emergence of an intermediate metallic glass phase in sufficiently disordered systems (Fig.~\ref{fig:theory}).  Changing the carrier density in highly disordered samples, for example, corresponds to\footnote{Since $E_{F}\sim n_s$ and $U\sim n_{s}^{1/2}$ in 2D and the disorder $W\approx$~const, the change of $n_s$ in the experiment is described by $(E_F/U)\sim(W/U)^{-1}$ in the phase diagram.} trajectory 2 in the theoretical phase diagram in Fig.~\ref{fig:theory}, where the intermediate phase is predicted to be relatively broad.  On the other hand, for low-disordered samples (trajectory 1), the metallic glass phase is very narrow or it vanishes, in agreement with experimental observations.  The same theory predicts \shortcite{Vlad-MITglass2} the $T^{3/2}$ correction to conductivity in the metallic glass phase, which is precisely what is found in the
experiment.  The emergence of the metallic glass phase, its dependence on the disorder, and the specific form of $\sigma(T)$ in this regime were obtained using a mean-field theory approach, which is known to produce hierarchical dynamics and aging in models of other glasses.  Therefore, this approach seems promising in describing also the glassy charge dynamics in a 2DES in Si.  There is currently no other theory available that predicts any of these experimental features.

\begin{figure}
	\centering
	\vspace*{0cm}\includegraphics[width=9.0cm,clip]{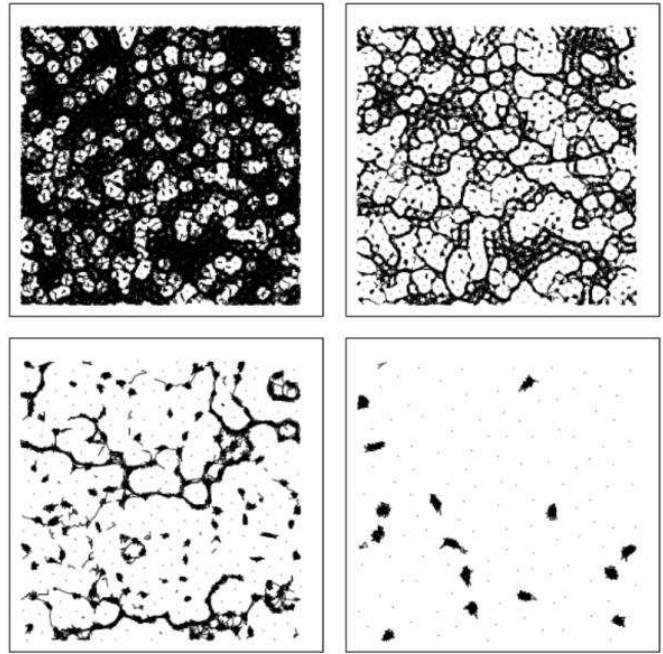}
\vspace*{0cm}\caption{Electron trajectories for a fixed period of time for a fixed $T$ and four electron densities, with the highest $n_1$ at upper left, $n_2<n_1$ upper right, $n_3<n_2$ lower left, and the lowest $n_4<n_3$ lower right.  Adapted from Ref.~\protect\shortcite{Reich-noise}.}
	\label{fig:sim}
\end{figure}

A molecular dynamics simulation of the crossover from a Wigner liquid to a Wigner glass in a 2D system of interacting electrons with disorder \shortcite{Reich-noise} has reproduced the main noise results (Sec.~\ref{fluct}).  In particular, a strong, orders of magnitude increase in the noise power and a jump in the exponent $\alpha$ was found at low $T$ and $n_s$, as well as the emergence of non-Gaussianity.  By looking at the electron trajectories for a fixed period of time (Fig.~\ref{fig:sim}), it was found that, at the highest $n_s$, electrons can flow freely throughout the sample. As $n_s$ is reduced, electron motion consists of a mixture of 2D and 1D regions (Fig.~\ref{fig:sim} upper right), but over longer times, the motion still occurs throughout the whole sample.  At even lower $n_s$ (Fig.~\ref{fig:sim} lower left), motion occurs mostly through 1D channels that percolate through the sample.  Importantly, the channel structures change very slowly with time: some channels close and others open up.  Since transport is dominated by a small number of channels, this give rise to large fluctuations in the conductivity and strong correlations.  Therefore, in this model, the large noise is due to dynamical inhomogeneities, similar to observations in other glasses \shortcite{Glotzer1,Glotzer2}.  At the lowest $n_s$, the channels disappear and transport occurs via jumps between localized regions.  This corresponds to the deep insulating regime.
In addition, a completely different technique, namely, a Monte Carlo simulation of a 3D Coulomb glass out of equilibrium has also found evidence for similar dynamical heterogeneities \shortcite{Kolton}.

\begin{figure}
	\centering
	\vspace*{-1.3cm}\includegraphics[width=12.0cm,clip]{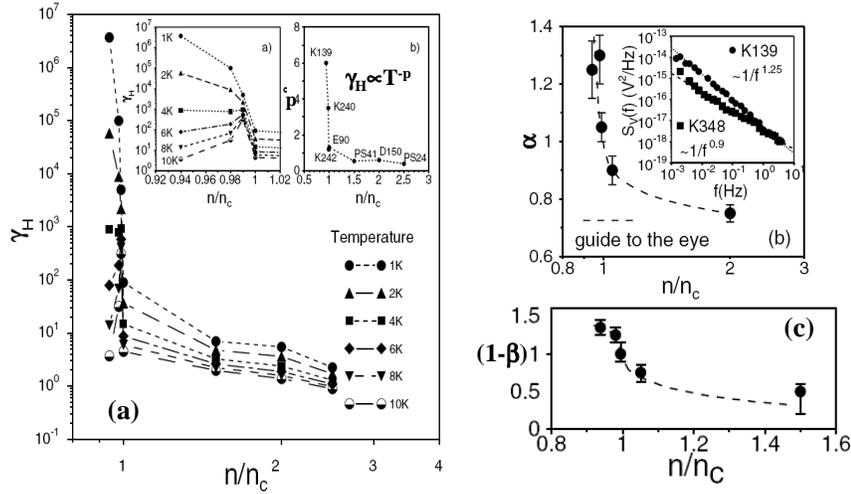}
\vspace*{-0.5cm}\caption{Resistance noise in P-doped Si.  Adapted from \protect\shortciteN{Kar}.  (a) Noise magnitude as a function of doping at different $T$.  Left inset: the same graph, magnified around $n_c$.  Right inset: the temperature dependence of the noise magnitude \textit{vs.} doping.  (b) Exponent $\alpha$ \textit{vs.} doping. Inset: the power spectrum $S(f)\propto 1/f^{\alpha}$.  (c) Exponent $(1-\beta)$, a measure of correlations, \textit{vs.} doping.  }
	\label{fig:dopedSi}
\end{figure}

There have been few experimental attempts so far to probe the charge dynamics near the MIT in other materials.  The most exciting and important has been the study of the resistance noise in bulk doped Si \shortcite{Kar}, which has been a prototypical system for investigating the critical behavior near the MIT
\shortcite{Mir-Dob-review,Myriam_review}
for several decades.  The results are strikingly similar to those on a 2DES in Si, namely: \textit{i)} the noise power increases by several orders of magnitude at $n_c$ [Fig.~\ref{fig:dopedSi}(a)], \textit{ii)} as $T$ decreases, the noise magnitude increases essentially exponentially for dopings $n/n_c\leq 1$ (Fig.~\ref{fig:dopedSi}(a) right inset), \textit{iii)} near the MIT, the exponent $\alpha$ rapidly increases to a value much larger than 1 [Fig.~\ref{fig:dopedSi}(b)], and \textit{iv)} the exponent $(1-\beta)$ of the second spectrum becomes strongly nonwhite ($\neq 0$) near $n_c$ [Fig.~\ref{fig:dopedSi}(c)], indicating the onset of correlated dynamics.  The similarity of the non-Gaussian spectra observed in both 2DES and in 3D doped Si near the MIT supports the intriguing possibility that such correlated dynamics may indeed be a universal feature of the MIT
regardless of the dimensionality.

Resistance noise was also measured in very clean 2D hole systems (2DHS) in GaAs \shortcite{LHote1,LHote2,LHote3}.  Even though an orders of magnitude enhancement of the noise was observed with decreasing $T$ and hole density $p_s$, similar to the results on 2DES in Si and bulk doped Si,
no evidence for strong correlations or glassiness was found.  It is likely that the (lack of) disorder here plays a crucial role, since glassiness is not expected theoretically in sufficiently clean systems \shortcite{Darko-glass} (Fig.~\ref{fig:theory}).  It is interesting, though, that the behavior of transport and noise in an intermediate regime of $p_s$ was attributed to the coexistence of nanoscale conducting and insulating regions, consistent with other studies that favor the percolation picture in 2DHS and 2DES in GaAs \shortcite{GaAs-percolation1,GaAs-percolation2,GaAs-percolation3,GaAs-percolationSavchenko}.  The local probes provide evidence \shortcite{GaAs-percolation1,GaAs-percolation2} that as the density approaches the critical hole density $p_c$ from the metallic side, the 2DHS fragments into localized charge configurations that are distributed in space, so that the insulating phase is spatially inhomogeneous.  The complicated structure emerges already on the metallic side of the MIT, reminiscent of the onset of glassiness in disordered 2DES in Si.  Clearly, more work is needed to examine the charge dynamics in these systems, especially in the presence of a higher amount of disorder.

The resistance noise techniques, long used in studies of various glassy systems, are starting to be recognized as powerful tools also in the investigations of the dynamics of cuprates \shortcite{IR_SPIE07,IR_PRL,IR-inplane,Weissman-YBCOnoise,Raffy-noise,YBCO-Orlyanchik}.  The currently available data, however, are scarce and difficult to compare because they probe very different materials and regions of phase space.  This warrants further systematic studies of the charge dynamics in various cuprates as a function of all the relevant parameters, such as doping and disorder, as well as different growth techniques. Comparative studies of charge dynamics in a 2DES in Si, as an excellent model system for the MIT and out-of-equilibrium behavior, and in more complex materials, such as cuprates, should provide important information on these two fundamental problems of condensed matter physics.

\section{Acknowledgments}

This work was supported by NSF Grant DMR-0905843 and the National High Magnetic Field Laboratory via NSF Grant DMR-0654118.

\bibliographystyle{OUPNamed}
\bibliography{Popovic-refs}

\thebibliography{0}

\bibitem[\protect\citeauthoryear{Alba, Hammann, Ocio, Refregier and
  Bouchiat}{Alba {\em et~al.}}{1987}]{Ocio-mu2}
Alba, M., Hammann, J., Ocio, M., Refregier, Ph., and Bouchiat, H. (1987).
\newblock Spin-glass dynamics from magnetic noise, relaxation, and
  susceptibility measurements.
\newblock {\em J. Appl. Phys.\/},~{\bf 61}, 3683.

\bibitem[\protect\citeauthoryear{Alba, Ocio and Hammann}{Alba {\em
  et~al.}}{1986}]{Ocio-mu1}
Alba, M., Ocio, M., and Hammann, J. (1986).
\newblock Ageing process and response function in spin glasses: An analysis of
  the thermoremanent magnetization decay in {A}g:{M}n (2.6\%).
\newblock {\em Europhys. Lett.\/},~{\bf 2}, 45.

\bibitem[\protect\citeauthoryear{Allison, Galaktionov, Savchenko, Safonov,
  Fogler, Simmons and Ritchie}{Allison {\em
  et~al.}}{2006}]{GaAs-percolationSavchenko}
Allison, G.~D., Galaktionov, E.~A., Savchenko, A.~K., Safonov, S.~S., Fogler,
  M.~M., Simmons, M.~Y., and Ritchie, D.~A. (2006).
\newblock Thermodynamic density of states of two-dimensional {G}a{A}s systems
  near the apparent metal-insulator transition.
\newblock {\em Phys. Rev. Lett.\/},~{\bf 96}, 216407.

\bibitem[\protect\citeauthoryear{Altshuler, Maslov and Pudalov}{Altshuler {\em
  et~al.}}{2001}]{nc4}
Altshuler, B.~L., Maslov, D.~L., and Pudalov, V.~M. (2001).
\newblock Metal-insulator transition in 2{D}: resistance in the critical
  region.
\newblock {\em Physica (Amsterdam)\/},~{\bf 9E}, 209.

\bibitem[\protect\citeauthoryear{Amir, Oreg and Imry}{Amir {\em
  et~al.}}{2011}]{ArielAmir-review}
Amir, A., Oreg, Y., and Imry, Y. (2011).
\newblock Electron glass dynamics.
\newblock {\em Annu. Rev. Condens. Matter Phys.\/},~{\bf 2}, 235.

\bibitem[\protect\citeauthoryear{Ando, Fowler and Stern}{Ando {\em
  et~al.}}{1982}]{AFS}
Ando, T., Fowler, A.~B., and Stern, F. (1982).
\newblock Electronic properties of two-dimensional systems.
\newblock {\em Rev. Mod. Phys.\/},~{\bf 54}, 437.

\bibitem[\protect\citeauthoryear{Barrat, Feigelman, Kurchan and
  Dalibard}{Barrat {\em et~al.}}{2003}]{LesHouches-2002}
Barrat, J.-L., Feigelman, M., Kurchan, J., and Dalibard, J. (ed.) (2003).
\newblock {\em Slow Relaxations and Nonequilibrium Dynamics in Condensed
  Matter}.
\newblock Springer, Berlin.

\bibitem[\protect\citeauthoryear{Belitz and Kirkpatrick}{Belitz and
  Kirkpatrick}{1994}]{Belitz}
Belitz, D. and Kirkpatrick, T.~R. (1994).
\newblock The {A}nderson-{M}ott transition.
\newblock {\em Rev. Mod. Phys.\/},~{\bf 66}, 261.

\bibitem[\protect\citeauthoryear{Belitz and Kirkpatrick}{Belitz and
  Kirkpatrick}{1995}]{noWegner3}
Belitz, D. and Kirkpatrick, T.~R. (1995).
\newblock Order parameter description of the {A}nderson-{M}ott transition.
\newblock {\em Z. Phys. B\/},~{\bf 98}, 513.

\bibitem[\protect\citeauthoryear{Ben-Chorin, Ovadyahu and Pollak}{Ben-Chorin
  {\em et~al.}}{1993}]{films1-Zvi}
Ben-Chorin, M., Ovadyahu, Z., and Pollak, M. (1993).
\newblock Nonequilibrium transport and slow relaxation in hopping conductivity.
\newblock {\em Phys. Rev. B\/},~{\bf 48}, 15025.

\bibitem[\protect\citeauthoryear{Bielejec and Wu}{Bielejec and
  Wu}{2001}]{films3-c}
Bielejec, E. and Wu, W. (2001).
\newblock Electron glass in ultrathin granular {A}l films at low temperatures.
\newblock {\em Phys. Rev. Lett.\/},~{\bf 87}, 256601.

\bibitem[\protect\citeauthoryear{Binder and Young}{Binder and
  Young}{1986}]{Binder}
Binder, K. and Young, A.~P. (1986).
\newblock Spin glasses: Experimental facts, theoretical concepts, and open
  questions.
\newblock {\em Rev. Mod. Phys.\/},~{\bf 58}, 801.

\bibitem[\protect\citeauthoryear{Bogdanovich, Dai, Sarachik, Dobrosavljevi\'c
  and Kotliar}{Bogdanovich {\em et~al.}}{1997}]{3Dbeta-2}
Bogdanovich, S., Dai, P., Sarachik, M.~P., Dobrosavljevi\'c, V., and Kotliar,
  G. (1997).
\newblock Scaling of the conductivity of {S}i:{B}: Anomalous crossover in a
  magnetic field.
\newblock {\em Phys. Rev. B\/},~{\bf 55}, 4215.

\bibitem[\protect\citeauthoryear{Bogdanovich and Popovi\'{c}}{Bogdanovich and
  Popovi\'{c}}{2002{\em a}}]{SB_PhysE}
Bogdanovich, S. and Popovi\'{c}, D. (2002{\em a}).
\newblock Glass transition in a two-dimensional electron system in silicon.
\newblock {\em Physica E\/},~{\bf 12}, 604.

\bibitem[\protect\citeauthoryear{Bogdanovich and Popovi\'{c}}{Bogdanovich and
  Popovi\'{c}}{2002{\em b}}]{SBPRL}
Bogdanovich, S. and Popovi\'{c}, D. (2002{\em b}).
\newblock Onset of glassy dynamics in a two-dimensional electron system in
  silicon.
\newblock {\em Phys. Rev. Lett.\/},~{\bf 88}, 236401.
\newblock Erratum, \textit{Phys. Rev. Lett.} \textbf{89}, 289904 (2002).

\bibitem[\protect\citeauthoryear{Bonetti, Caplan, Harlingen and
  Weissman}{Bonetti {\em et~al.}}{2004}]{Weissman-YBCOnoise}
Bonetti, J.~A., Caplan, D.~S., Harlingen, D. J.~Van, and Weissman, M.~B.
  (2004).
\newblock Electronic transport in underdoped
  {Y}{B}a$_2${C}u$_3${O}$_{7-\delta}$ nanowires: evidence for fluctuating
  domain structures.
\newblock {\em Phys. Rev. Lett.\/},~{\bf 93}, 087002.

\bibitem[\protect\citeauthoryear{Bouchaud, Cugliandolo, Kurchan and
  Mezard}{Bouchaud {\em et~al.}}{1997}]{mean-field}
Bouchaud, J.-P., Cugliandolo, L.~F., Kurchan, J., and Mezard, M. (1997).
\newblock Out of equilibrium dynamics in spin-glasses and other glassy systems.
\newblock In {\em Spin Glasses and Random Fields} (ed. A.~P. Young). World
  Scientific, Singapore.

\bibitem[\protect\citeauthoryear{Camjayi, Haule, Dobrosavljevi\'c and
  Kotliar}{Camjayi {\em et~al.}}{2008}]{Vlad-Nature}
Camjayi, A., Haule, K., Dobrosavljevi\'c, V., and Kotliar, G. (2008).
\newblock Coulomb correlations and the {W}igner–-{M}ott transition.
\newblock {\em Nature Phys.\/},~{\bf 4}, 932.

\bibitem[\protect\citeauthoryear{Caplan, Orlyanchik, Weissman, Harlingen,
  Fradkin, Hinton and Lemberger}{Caplan {\em et~al.}}{2010}]{YBCO-Orlyanchik}
Caplan, D.~S., Orlyanchik, V., Weissman, M.~B., Harlingen, D. J.~Van, Fradkin,
  E.~H., Hinton, M.~J., and Lemberger, T.~R. (2010).
\newblock Anomalous noise in the pseudogap regime of
  {Y}{B}a$_2${C}u$_3${O}$_{7-\delta}$.
\newblock {\em Phys. Rev. Lett.\/},~{\bf 104}, 177001.

\bibitem[\protect\citeauthoryear{Castellani, Kotliar and Lee}{Castellani {\em
  et~al.}}{1987}]{noWegner1}
Castellani, C., Kotliar, G., and Lee, P.~A. (1987).
\newblock Fermi-liquid theory of interacting disordered systems and the scaling
  theory of the metal-insulator transition.
\newblock {\em Phys. Rev. Lett.\/},~{\bf 59}, 323.

\bibitem[\protect\citeauthoryear{Chakravarty, Kivelson, Nayak and
  Voelker}{Chakravarty {\em et~al.}}{1999}]{MIT-glassothers3}
Chakravarty, S., Kivelson, S., Nayak, C., and Voelker, K. (1999).
\newblock Wigner glass, spin liquids and the metal-insulator transition.
\newblock {\em Philos. Mag. B\/},~{\bf 79}, 859.

\bibitem[\protect\citeauthoryear{Committee~on {CMMP}~2010}{Committee~on
  {CMMP}~2010}{2007}]{CMMP-report}
Committee~on {CMMP}~2010, {N}ational {R}esearch~{C}ouncil (2007).
\newblock {\em Condensed-Matter and Materials Physics: The Science of the World
  Around Us}.
\newblock The National Academies Press, Washington, D.C.

\bibitem[\protect\citeauthoryear{Dagotto}{Dagotto}{2002}]{phasesep3}
Dagotto, E. (2002).
\newblock {\em Nanoscale phase separation and colossal magnetoresistance}.
\newblock Springer-Verlag, Berlin.

\bibitem[\protect\citeauthoryear{Dagotto}{Dagotto}{2005}]{Elbio}
Dagotto, E. (2005).
\newblock Complexity in strongly correlated electronic systems.
\newblock {\em Science\/},~{\bf 309}, 257--262.

\bibitem[\protect\citeauthoryear{Dalidovich and Dobrosavljevi\'c}{Dalidovich
  and Dobrosavljevi\'c}{2002}]{Vlad-MITglass2}
Dalidovich, D. and Dobrosavljevi\'c, V. (2002).
\newblock Landau theory of the {F}ermi-liquid to electron-glass transition.
\newblock {\em Phys. Rev. B\/},~{\bf 66}, 081107.

\bibitem[\protect\citeauthoryear{Davies, Lee and Rice}{Davies {\em
  et~al.}}{1982}]{eglass1}
Davies, J.~H., Lee, P.~A., and Rice, T.~M. (1982).
\newblock Electron glass.
\newblock {\em Phys. Rev. Lett.\/},~{\bf 49}, 758.

\bibitem[\protect\citeauthoryear{Davies, Lee and Rice}{Davies {\em
  et~al.}}{1984}]{eglass4}
Davies, J.~H., Lee, P.~A., and Rice, T.~M. (1984).
\newblock Properties of the electron glass.
\newblock {\em Phys. Rev. B\/},~{\bf 29}, 4260.

\bibitem[\protect\citeauthoryear{Deville, Leturcq, L'H\^{o}te, Tourbot, Mellor
  and Henini}{Deville {\em et~al.}}{2005}]{LHote2}
Deville, G., Leturcq, R., L'H\^{o}te, D., Tourbot, R., Mellor, C.~J., and
  Henini, M. (2005).
\newblock $1/f$ noise in low density two-dimensional hole systems in {G}a{A}s.
\newblock {\em AIP Conf. Proc.\/},~{\bf 780}, 139.

\bibitem[\protect\citeauthoryear{Deville, Leturcq, L'H\^{o}te, Tourbot, Mellor
  and Henini}{Deville {\em et~al.}}{2006}]{LHote3}
Deville, G., Leturcq, R., L'H\^{o}te, D., Tourbot, R., Mellor, C.~J., and
  Henini, M. (2006).
\newblock $1/f$ noise in a dilute {G}a{A}s two-dimensional hole system in the
  insulating phase.
\newblock {\em Physica E\/},~{\bf 34}, 252.

\bibitem[\protect\citeauthoryear{Dobrosavljevi\'c, Tanaskovi\'c and
  Pastor}{Dobrosavljevi\'c {\em et~al.}}{2003}]{Darko-glass}
Dobrosavljevi\'c, V., Tanaskovi\'c, D., and Pastor, A.~A. (2003).
\newblock Glassy behavior of electrons near metal-insulator transitions.
\newblock {\em Phys. Rev. Lett.\/},~{\bf 90}, 016402.

\bibitem[\protect\citeauthoryear{Dolgopolov and Gold}{Dolgopolov and
  Gold}{2000}]{Bscreening1}
Dolgopolov, V.~T. and Gold, A. (2000).
\newblock Magnetoresistance of a two-dimensional electron gas in a parallel
  magnetic field.
\newblock {\em JETP Lett.\/},~{\bf 71}, 27.

\bibitem[\protect\citeauthoryear{Emery and Kivelson}{Emery and
  Kivelson}{1995}]{bad}
Emery, V.~J. and Kivelson, S.~A. (1995).
\newblock Superconductivity in bad metals.
\newblock {\em Phys. Rev. Lett.\/},~{\bf 74}, 3253.

\bibitem[\protect\citeauthoryear{Eng, Feng, Popovi\'{c} and Washburn}{Eng {\em
  et~al.}}{2001}]{Kevin_Proc}
Eng, K., Feng, X.~G., Popovi\'{c}, D., and Washburn, S. (2001).
\newblock Effects of a parallel magnetic field on the novel metallic behavior
  in two dimensions.
\newblock {\em Springer Proceedings in Physics\/},~{\bf 87}, 741.

\bibitem[\protect\citeauthoryear{Eng, Feng, Popovi\'{c} and Washburn}{Eng {\em
  et~al.}}{2002}]{Kevin_PRL}
Eng, K., Feng, X.~G., Popovi\'{c}, D., and Washburn, S. (2002).
\newblock Effects of a parallel magnetic field on the metal-insulator
  transition in a dilute two-dimensional electron system.
\newblock {\em Phys. Rev. Lett.\/},~{\bf 88}, 136402.

\bibitem[\protect\citeauthoryear{Feng, Popovi\'c and Washburn}{Feng {\em
  et~al.}}{1999}]{Feng-moments}
Feng, X.~G., Popovi\'c, D., and Washburn, S. (1999).
\newblock Effect of local magnetic moments on the metallic behavior in two
  dimensions.
\newblock {\em Phys. Rev. Lett.\/},~{\bf 83}, 368.

\bibitem[\protect\citeauthoryear{Feng, Popovi\'c, Washburn and
  Dobrosavljevi\'c}{Feng {\em et~al.}}{2001}]{Feng-novel}
Feng, X.~G., Popovi\'c, D., Washburn, S., and Dobrosavljevi\'c, V. (2001).
\newblock Novel metallic behavior in two dimensions.
\newblock {\em Phys. Rev. Lett.\/},~{\bf 86}, 2625.

\bibitem[\protect\citeauthoryear{Fletcher, Pudalov, Radcliffe and
  Possanzini}{Fletcher {\em et~al.}}{2001}]{Fletcher}
Fletcher, R., Pudalov, V.~M., Radcliffe, A. D.~B., and Possanzini, C. (2001).
\newblock Critical behaviour of thermopower and conductivity at the
  metal-insulator transition in high-mobility {S}i-{MOSFET}s.
\newblock {\em Semicond. Sci. Tech.\/},~{\bf 16}, 386.

\bibitem[\protect\citeauthoryear{Fruchter, Raffy and Li}{Fruchter {\em
  et~al.}}{2007}]{Raffy-noise}
Fruchter, L., Raffy, H., and Li, Z.~Z. (2007).
\newblock Resistance noise in {B}i$_2${S}r$_2${C}a{C}u$_2${O}$_{8+\delta}$.
\newblock {\em Phys. Rev. B\/},~{\bf 76}, 212503.

\bibitem[\protect\citeauthoryear{Gao, Mills, Ramirez, Pfeiffer and West}{Gao
  {\em et~al.}}{2002}]{GaAs-percolation3}
Gao, X. P.~A., Mills, A.~P., Ramirez, A.~P., Pfeiffer, L.~N., and West, K.~W.
  (2002).
\newblock Weak-localization-like temperature-dependent conductivity of a dilute
  two-dimensional hole gas in a parallel magnetic field.
\newblock {\em Phys. Rev. Lett.\/},~{\bf 89}, 16801.

\bibitem[\protect\citeauthoryear{Glotzer}{Glotzer}{2000}]{Glotzer1}
Glotzer, S.~C. (2000).
\newblock Spatially heterogeneous dynamics in liquids: insights from
  simulation.
\newblock {\em J. Non-Cryst. Solids\/},~{\bf 274}, 342.

\bibitem[\protect\citeauthoryear{Goldenfeld}{Goldenfeld}{1992}]{Goldenfeld}
Goldenfeld, N. (1992).
\newblock {\em Lectures on Phase Transitions and the Renormalization Group}.
\newblock Addison-Wesley.

\bibitem[\protect\citeauthoryear{Gor'kov and Sokol}{Gor'kov and
  Sokol}{1987}]{phasesep1}
Gor'kov, L.~P. and Sokol, A.~V. (1987).
\newblock Localized and delocalized states and the properties of the normal
  phase of recently discovered superconductors.
\newblock {\em JETP Lett.\/},~{\bf 46}, 420.

\bibitem[\protect\citeauthoryear{Grempel}{Grempel}{2004}]{Grempel}
Grempel, D.~R. (2004).
\newblock Off-equilibrium dynamics of the two-dimensional {C}oulomb glass.
\newblock {\em Europhys. Lett.\/},~{\bf 66}, 854.

\bibitem[\protect\citeauthoryear{Grenet}{Grenet}{2003}]{films3-e}
Grenet, T. (2003).
\newblock Symmetrical field effect and slow electron relaxation in granular
  aluminium.
\newblock {\em Eur. Phys. J. B\/},~{\bf 32}, 275.

\bibitem[\protect\citeauthoryear{Grenet, Delahaye, Sabra and Gay}{Grenet {\em
  et~al.}}{2007}]{films3-g}
Grenet, T., Delahaye, J., Sabra, M., and Gay, F. (2007).
\newblock Anomalous electric-field effect and glassy behaviour in granular
  aluminium thin films: electron glass?
\newblock {\em Eur. Phys. J. B\/},~{\bf 56}, 183.

\bibitem[\protect\citeauthoryear{Gr{\"{u}}newald, Pohlman, Schweitzer and
  W{\"{u}}rtz}{Gr{\"{u}}newald {\em et~al.}}{1982}]{eglass2}
Gr{\"{u}}newald, M., Pohlman, B., Schweitzer, L., and W{\"{u}}rtz, D. (1982).
\newblock Mean field approach to the electron glass.
\newblock {\em J. Phys. C\/},~{\bf 15}, L1153.

\bibitem[\protect\citeauthoryear{Herbut}{Herbut}{2001}]{Bscreening2}
Herbut, I.~F. (2001).
\newblock The effect of parallel magnetic field on the {B}oltzmann conductivity
  and the {H}all coefficient of a disordered two-dimensional {F}ermi liquid.
\newblock {\em Phys. Rev. B\/},~{\bf 63}, 113102.

\bibitem[\protect\citeauthoryear{Hernandez, Bhattacharya, Parendo and
  Goldman}{Hernandez {\em et~al.}}{2003}]{films3-d}
Hernandez, L.~M., Bhattacharya, A., Parendo, K.~A., and Goldman, A.~M. (2003).
\newblock Electrical transport of spin-polarized carriers in disordered
  ultrathin films.
\newblock {\em Phys. Rev. Lett.\/},~{\bf 91}, 126801.

\bibitem[\protect\citeauthoryear{Hewson}{Hewson}{1993}]{Hewson}
Hewson, A.~C. (1993).
\newblock {\em The {K}ondo Problem to Heavy Fermions}.
\newblock Cambridge Univ. Press, Cambridge, England.

\bibitem[\protect\citeauthoryear{Hodge}{Hodge}{1995}]{aging2}
Hodge, I.~M. (1995).
\newblock Physical aging in polymer glasses.
\newblock {\em Science\/},~{\bf 267}, 1945.

\bibitem[\protect\citeauthoryear{Hohenberg and Halperin}{Hohenberg and
  Halperin}{1977}]{scaling}
Hohenberg, P.~C. and Halperin, B.~I. (1977).
\newblock Theory of dynamic critical phenomena.
\newblock {\em Rev. Mod. Phys.\/},~{\bf 49}, 435.

\bibitem[\protect\citeauthoryear{Hooge}{Hooge}{1976}]{Hooge}
Hooge, F.~N. (1976).
\newblock $1/f$ noise.
\newblock {\em Physica (Amsterdam)\/},~{\bf 83B}, 14.

\bibitem[\protect\citeauthoryear{Ilani, Yacoby, Mahalu and Shtrikman}{Ilani
  {\em et~al.}}{2000}]{GaAs-percolation1}
Ilani, S., Yacoby, A., Mahalu, D., and Shtrikman, H. (2000).
\newblock Unexpected behavior of the local compressibility near the ${B}=0$
  metal-insulator transition.
\newblock {\em Phys. Rev. Lett.\/},~{\bf 84}, 3133.

\bibitem[\protect\citeauthoryear{Ilani, Yacoby, Mahalu and Shtrikman}{Ilani
  {\em et~al.}}{2001}]{GaAs-percolation2}
Ilani, S., Yacoby, A., Mahalu, D., and Shtrikman, H. (2001).
\newblock Microscopic structure of the metal-insulator transition in two
  dimensions.
\newblock {\em Science\/},~{\bf 292}, 1354.

\bibitem[\protect\citeauthoryear{Jaroszy\'nski and Popovi\'c}{Jaroszy\'nski and
  Popovi\'c}{2006}]{relax-PRL}
Jaroszy\'nski, J. and Popovi\'c, D. (2006).
\newblock Nonexponential relaxations in a two-dimensional electron system in
  silicon.
\newblock {\em Phys. Rev. Lett.\/},~{\bf 96}, 037403.

\bibitem[\protect\citeauthoryear{Jaroszy\'nski and Popovi\'c}{Jaroszy\'nski and
  Popovi\'c}{2007{\em a}}]{DP-aging}
Jaroszy\'nski, J. and Popovi\'c, D. (2007{\em a}).
\newblock Aging effects across the metal-insulator transition in two
  dimensions.
\newblock {\em Phys. Rev. Lett.\/},~{\bf 99}, 216401.

\bibitem[\protect\citeauthoryear{Jaroszy\'nski and Popovi\'c}{Jaroszy\'nski and
  Popovi\'c}{2007{\em b}}]{tw-PRL}
Jaroszy\'nski, J. and Popovi\'c, D. (2007{\em b}).
\newblock Nonequilibrium relaxations and aging effects in a two-dimensional
  {C}oulomb glass.
\newblock {\em Phys. Rev. Lett.\/},~{\bf 99}, 046405.

\bibitem[\protect\citeauthoryear{Jaroszy\'nski and Popovi\'c}{Jaroszy\'nski and
  Popovi\'c}{2009}]{aging-PhysB}
Jaroszy\'nski, J. and Popovi\'c, D. (2009).
\newblock Aging and memory in a two-dimensional electron system in {S}i.
\newblock {\em Physica B\/},~{\bf 404}, 466.

\bibitem[\protect\citeauthoryear{Jaroszy\'nski, Popovi\'{c} and
  Klapwijk}{Jaroszy\'nski {\em et~al.}}{2002{\em a}}]{JJ_PhysE02}
Jaroszy\'nski, J., Popovi\'{c}, D., and Klapwijk, T.~M. (2002{\em a}).
\newblock Low-frequency resistance noise studies across the metal-insulator
  transition in silicon {MOSFET}s.
\newblock {\em Physica E\/},~{\bf 12}, 612.

\bibitem[\protect\citeauthoryear{Jaroszy\'nski, Popovi\'c and
  Klapwijk}{Jaroszy\'nski {\em et~al.}}{2002{\em b}}]{JJ_PRL02}
Jaroszy\'nski, J., Popovi\'c, D., and Klapwijk, T.~M. (2002{\em b}).
\newblock Universal behavior of the resistance noise across the metal-insulator
  transition in silicon inversion layers.
\newblock {\em Phys. Rev. Lett.\/},~{\bf 89}, 276401.

\bibitem[\protect\citeauthoryear{Jaroszy\'nski, Popovi\'c and
  Klapwijk}{Jaroszy\'nski {\em et~al.}}{2004{\em a}}]{JJ_SPIE04}
Jaroszy\'nski, J., Popovi\'c, D., and Klapwijk, T.~M. (2004{\em a}).
\newblock Glassy behavior of a two-dimensional electron system in {S}i in
  parallel magnetic fields.
\newblock {\em Proceedings of SPIE\/},~{\bf 5469}, 95.

\bibitem[\protect\citeauthoryear{Jaroszy\'nski, Popovi\'c and
  Klapwijk}{Jaroszy\'nski {\em et~al.}}{2004{\em b}}]{JJ_noiseB}
Jaroszy\'nski, J., Popovi\'c, D., and Klapwijk, T.~M. (2004{\em b}).
\newblock Magnetic field dependence of the anomalous noise behavior in a
  two-dimensional electron system in silicon.
\newblock {\em Phys. Rev. Lett.\/},~{\bf 92}, 226403.

\bibitem[\protect\citeauthoryear{Jaroszy\'nski, Wr\'obel, G.Karczewski,
  Wojtowicz and Dietl}{Jaroszy\'nski {\em et~al.}}{1998}]{jjprl98}
Jaroszy\'nski, J., Wr\'obel, J., G.Karczewski, Wojtowicz, T., and Dietl, T.
  (1998).
\newblock Magnetoconductance noise and irreversibilities in submicron wires of
  spin-glass \textit{n}$^{+}$-{C}d$_{1-x}${M}n$_{x}${T}e.
\newblock {\em Phys. Rev. Lett.\/},~{\bf 80}, 5635.

\bibitem[\protect\citeauthoryear{Jelbert, Sasagawa, Fletcher, Park, Thompson
  and Panagopoulos}{Jelbert {\em et~al.}}{2008}]{Glenton}
Jelbert, G.~R., Sasagawa, T., Fletcher, J.~D., Park, T., Thompson, J.~D., and
  Panagopoulos, C. (2008).
\newblock Measurement of low energy charge correlations in underdoped
  spin-glass {L}a-based cuprates using impedance spectroscopy.
\newblock {\em Phys. Rev. B\/},~{\bf 78}, 132513.

\bibitem[\protect\citeauthoryear{J{\"{o}}nsson and Takayama}{J{\"{o}}nsson and
  Takayama}{2005}]{sg-nonmon2}
J{\"{o}}nsson, P.~E. and Takayama, H. (2005).
\newblock ``{G}lassy dynamics'' in {I}sing spin glasses -- experiment and
  simulation.
\newblock {\em J. Phys. Soc. Jpn\/},~{\bf 74}, 1131.

\bibitem[\protect\citeauthoryear{Jonsson, Jonason and Nordblad}{Jonsson {\em
  et~al.}}{1999}]{sg-nonmon1}
Jonsson, T., Jonason, K., and Nordblad, P. (1999).
\newblock Relaxation of the field-cooled magnetization of an {I}sing spin
  glass.
\newblock {\em Phys. Rev. B\/},~{\bf 59}, 9402.

\bibitem[\protect\citeauthoryear{Kar, Raychaudhuri, Ghosh, v.~L{\"{o}}hneysen
  and Weiss}{Kar {\em et~al.}}{2003}]{Kar}
Kar, S., Raychaudhuri, A.~K., Ghosh, A., v.~L{\"{o}}hneysen, H., and Weiss, G.
  (2003).
\newblock Observation of non-{G}aussian conductance fluctuations at low
  temperatures in {S}i:{P}({B}) at the metal-insulator transition.
\newblock {\em Phys. Rev. Lett.\/},~{\bf 91}, 216603.

\bibitem[\protect\citeauthoryear{Kim, Popovi\'c and Washburn}{Kim {\em
  et~al.}}{1998}]{njk1}
Kim, N.-J., Popovi\'c, D., and Washburn, S. (1998).
\newblock Phase diagram and validity of one-parameter scaling near the
  two-dimensional metal-insulator transition.
\newblock cond-mat/9809357.

\bibitem[\protect\citeauthoryear{Kirkpatrick and Belitz}{Kirkpatrick and
  Belitz}{1994}]{noWegner2}
Kirkpatrick, T.~R. and Belitz, D. (1994).
\newblock Anderson-{M}ott transition as a random-field problem.
\newblock {\em Phys. Rev. Lett.\/},~{\bf 74}, 1178.

\bibitem[\protect\citeauthoryear{Kivelson, Bindloss, Fradkin, Oganesyan,
  Tranquada, Kapitulnik and Howald}{Kivelson {\em et~al.}}{2003}]{phasesep2}
Kivelson, S.~A., Bindloss, I.~P., Fradkin, E., Oganesyan, V., Tranquada, J.~M.,
  Kapitulnik, A., and Howald, C. (2003).
\newblock How to detect fluctuating stripes in the high-temperature
  superconductors.
\newblock {\em Rev. Mod. Phys.\/},~{\bf 75}, 1201.

\bibitem[\protect\citeauthoryear{Kohsaka, Taylor, Fujita, Schmidt, Lupien,
  Hanaguri, Azuma, Takano, Eisaki, Takagi, Uchida and Davis}{Kohsaka {\em
  et~al.}}{2007}]{STM-Seamus}
Kohsaka, Y., Taylor, C., Fujita, K., Schmidt, A., Lupien, C., Hanaguri, T.,
  Azuma, M., Takano, M., Eisaki, H., Takagi, H., Uchida, S., and Davis, J.~C.
  (2007).
\newblock An intrinsic bond-centered electronic glass with unidirectional
  domains in underdoped cuprates.
\newblock {\em Science\/},~{\bf 315}, 1380.

\bibitem[\protect\citeauthoryear{Kolton, Grempel and Dom\'{\i}nguez}{Kolton
  {\em et~al.}}{2005}]{Kolton}
Kolton, A.~B., Grempel, D.~R., and Dom\'{\i}nguez, D. (2005).
\newblock Heterogeneous dynamics of the three-dimensional {C}oulomb glass out
  of equilibrium.
\newblock {\em Phys. Rev. B\/},~{\bf 71}, 024206.

\bibitem[\protect\citeauthoryear{Kravchenko, Kravchenko, Furneaux, Pudalov and
  D'Iorio}{Kravchenko {\em et~al.}}{1994}]{Krav-scal1}
Kravchenko, S.~V., Kravchenko, G.~V., Furneaux, J.~E., Pudalov, V.~M., and
  D'Iorio, M. (1994).
\newblock Possible metal-insulator transition at ${B}=0$ in two dimensions.
\newblock {\em Phys. Rev. B\/},~{\bf 50}, 8039.

\bibitem[\protect\citeauthoryear{Kravchenko, Mason, Bowker, Furneaux, Pudalov
  and D'Iorio}{Kravchenko {\em et~al.}}{1995}]{Krav-scal2}
Kravchenko, S.~V., Mason, W.~E., Bowker, G.~E., Furneaux, J.~E., Pudalov,
  V.~M., and D'Iorio, M. (1995).
\newblock Scaling of an anomalous metal-insulator transition in a
  two-dimensional system in silicon at ${B}=0$.
\newblock {\em Phys. Rev. B\/},~{\bf 51}, 7038.

\bibitem[\protect\citeauthoryear{Kravchenko and Sarachik}{Kravchenko and
  Sarachik}{2004}]{Sergey-review}
Kravchenko, S.~V. and Sarachik, M.~P. (2004).
\newblock Metal-insulator transition in two-dimensional electron systems.
\newblock {\em Rep. Prog. Phys.\/},~{\bf 67}, 1.

\bibitem[\protect\citeauthoryear{Kurzweil and Frydman}{Kurzweil and
  Frydman}{2007}]{Aviad-nonmon}
Kurzweil, N. and Frydman, A. (2007).
\newblock Inverse slow relaxation in granular hopping systems.
\newblock {\em Phys. Rev. B\/},~{\bf 75}, 020202(R).

\bibitem[\protect\citeauthoryear{Lebanon and M{\"u}ller}{Lebanon and
  M{\"u}ller}{2005}]{Muller-memory}
Lebanon, E. and M{\"u}ller, M. (2005).
\newblock Memory effect in electron glasses: theoretical analysis via a
  percolation approach.
\newblock {\em Phys. Rev. B\/},~{\bf 72}, 174202.

\bibitem[\protect\citeauthoryear{Lee, Oikonomou, Segalova, Rosenbaum, Hoekstra
  and Littlewood}{Lee {\em et~al.}}{2005}]{films3-f}
Lee, M., Oikonomou, P., Segalova, P., Rosenbaum, T.~F., Hoekstra, A. F.~Th.,
  and Littlewood, P.~B. (2005).
\newblock The electron glass in a switchable mirror: relaxation, ageing and
  universality.
\newblock {\em J. Phys. Condens. Matter\/},~{\bf 17}, L439.

\bibitem[\protect\citeauthoryear{Lee and Ramakrishnan}{Lee and
  Ramakrishnan}{1985}]{LR}
Lee, P.~A. and Ramakrishnan, T.~V. (1985).
\newblock Disordered electronic systems.
\newblock {\em Rev. Mod. Phys.\/},~{\bf 57}, 287.

\bibitem[\protect\citeauthoryear{Leturcq, L'H\^{o}te, Tourbot, Mellor and
  Henini}{Leturcq {\em et~al.}}{2003}]{LHote1}
Leturcq, R., L'H\^{o}te, D., Tourbot, R., Mellor, C.~J., and Henini, M. (2003).
\newblock Resistance noise scaling in a dilute two-dimensional hole system in
  {G}a{A}s.
\newblock {\em Phys. Rev. Lett.\/},~{\bf 90}, 076402.

\bibitem[\protect\citeauthoryear{Leuzzi and Nieuwenhuizen}{Leuzzi and
  Nieuwenhuizen}{2008}]{nieuw-book}
Leuzzi, L. and Nieuwenhuizen, T.~M. (2008).
\newblock {\em Thermodynamics of the Glassy State}.
\newblock Taylor \& Francis, New York.

\bibitem[\protect\citeauthoryear{Levy, Parisi, Granja, Indelicato and
  Polla}{Levy {\em et~al.}}{2002}]{mang-nonmon}
Levy, P., Parisi, F., Granja, L., Indelicato, E., and Polla, G. (2002).
\newblock Novel dynamical effects and persistent memory in phase separated
  manganites.
\newblock {\em Phys. Rev. Lett.\/},~{\bf 89}, 137001.

\bibitem[\protect\citeauthoryear{Martinez-Arizala, Christiansen, Grupp,
  Markovic, Mack and Goldman}{Martinez-Arizala {\em et~al.}}{1998}]{films3-b}
Martinez-Arizala, G., Christiansen, C., Grupp, D.~E., Markovic, N., Mack,
  A.~M., and Goldman, A.~M. (1998).
\newblock Coulomb-glass-like behavior of ultrathin films of metals.
\newblock {\em Phys. Rev. B\/},~{\bf 57}, R670.

\bibitem[\protect\citeauthoryear{Martinez-Arizala, Grupp, Christiansen, Mack,
  Markovic, Seguchi and Goldman}{Martinez-Arizala {\em
  et~al.}}{1997}]{films3-a}
Martinez-Arizala, G., Grupp, D.~E., Christiansen, C., Mack, A.~M., Markovic,
  N., Seguchi, Y., and Goldman, A.~M. (1997).
\newblock Anomalous field effect in ultrathin films of metals near the
  superconductor-insulator transition.
\newblock {\em Phys. Rev. Lett.\/},~{\bf 78}, 1130.

\bibitem[\protect\citeauthoryear{Miranda and Dobrosavljevi\'c}{Miranda and
  Dobrosavljevi\'c}{2005}]{Mir-Dob-review}
Miranda, E. and Dobrosavljevi\'c, V. (2005).
\newblock Disorder-driven non-{F}ermi liquid behaviour of correlated electrons.
\newblock {\em Rep. Prog. Phys.\/},~{\bf 68}, 2337.

\bibitem[\protect\citeauthoryear{Miyashita, Tanaka and Hirano}{Miyashita {\em
  et~al.}}{2007}]{th-nonmon2}
Miyashita, S., Tanaka, S., and Hirano, M. (2007).
\newblock Nonmonotonic relaxation in systems with reentrant-type interaction.
\newblock {\em J. Phys. Soc. Jpn\/},~{\bf 76}, 083001.

\bibitem[\protect\citeauthoryear{Monroe}{Monroe}{1990}]{Monroe2}
Monroe, D. (1990).
\newblock Capacitance measurements of the dynamics of screening in the electron
  glass.
\newblock In {\em Hopping and Related Phenomena 5} (ed. H.~Fritzsche and
  M.~Pollak). World Scientific, Singapore.

\bibitem[\protect\citeauthoryear{Monroe, Gossard, English, Golding, Haemmerle
  and Kastner}{Monroe {\em et~al.}}{1987}]{Monroe1}
Monroe, D., Gossard, A.~C., English, J.~H., Golding, B., Haemmerle, W.~H., and
  Kastner, M.~A. (1987).
\newblock Long-lived {C}oulomb gap in a compensated semiconductor -- the
  electron glass.
\newblock {\em Phys. Rev. Lett.\/},~{\bf 59}, 1148.

\bibitem[\protect\citeauthoryear{Morita and Kaneko}{Morita and
  Kaneko}{2005}]{th-nonmon1}
Morita, H. and Kaneko, K. (2005).
\newblock Roundabout relaxation: collective excitation requires a detour to
  equilibrium.
\newblock {\em Phys. Rev. Lett.\/},~{\bf 94}, 087203.

\bibitem[\protect\citeauthoryear{M{\"u}ller and Ioffe}{M{\"u}ller and
  Ioffe}{2004}]{Muller-scr}
M{\"u}ller, M. and Ioffe, L.~B. (2004).
\newblock Glass transition and the {C}oulomb gap in electron glasses.
\newblock {\em Phys. Rev. Lett.\/},~{\bf 93}, 256403.

\bibitem[\protect\citeauthoryear{M{\"u}ller and Lebanon}{M{\"u}ller and
  Lebanon}{2005}]{Muller-French}
M{\"u}ller, M. and Lebanon, E. (2005).
\newblock History dependence, memory and metastability in electron glasses.
\newblock {\em J. Phys. IV France\/},~{\bf 131}, 167.

\bibitem[\protect\citeauthoryear{Nelson}{Nelson}{2003}]{bio-nonmon}
Nelson, P. (2003).
\newblock {\em Biological physics: energy, information, life}, Chapter~12.
\newblock W. H. Freeman \& Co., New York.

\bibitem[\protect\citeauthoryear{Neuttiens, Strunk, Haesendonck and
  Bruynseraede}{Neuttiens {\em et~al.}}{2000}]{Neutt}
Neuttiens, G., Strunk, C., Haesendonck, C.~V., and Bruynseraede, Y. (2000).
\newblock Universal conductance fluctuations and low-temperature \textit{1/f}
  noise in mesoscopic {A}u{F}e spin glasses.
\newblock {\em Phys. Rev. B\/},~{\bf 62}, 3905.

\bibitem[\protect\citeauthoryear{Ogielski}{Ogielski}{1985}]{Ogielski}
Ogielski, A.~T. (1985).
\newblock Dynamics of three-dimensional {I}sing spin glasses in thermal
  equilibrium.
\newblock {\em Phys. Rev. B\/},~{\bf 32}, 7384.

\bibitem[\protect\citeauthoryear{Okamoto, Hosoya, Kawaji and Yagi}{Okamoto {\em
  et~al.}}{1999}]{polarization1}
Okamoto, T., Hosoya, K., Kawaji, S., and Yagi, A. (1999).
\newblock Spin degree of freedom in a two-dimensional electron liquid.
\newblock {\em Phys. Rev. Lett.\/},~{\bf 82}, 3875.

\bibitem[\protect\citeauthoryear{Orenstein and Millis}{Orenstein and
  Millis}{2000}]{Millis}
Orenstein, J. and Millis, A.~J. (2000).
\newblock Advances in the physics of high-temperature superconductivity.
\newblock {\em Science\/},~{\bf 288}, 468.

\bibitem[\protect\citeauthoryear{Orlyanchik and Ovadyahu}{Orlyanchik and
  Ovadyahu}{2004}]{films5-Zvi}
Orlyanchik, V. and Ovadyahu, Z. (2004).
\newblock Stress aging in the electron glass.
\newblock {\em Phys. Rev. Lett.\/},~{\bf 92}, 066801.

\bibitem[\protect\citeauthoryear{Ovadyahu}{Ovadyahu}{2006{\em a}}]{films6-Zvi}
Ovadyahu, Z. (2006{\em a}).
\newblock Quench-cooling procedure compared with the gate protocol for aging
  experiments in electron glasses.
\newblock {\em Phys. Rev. B\/},~{\bf 73}, 214204.

\bibitem[\protect\citeauthoryear{Ovadyahu}{Ovadyahu}{2006{\em b}}]{films7-Zvi}
Ovadyahu, Z. (2006{\em b}).
\newblock Temperature- and field-dependence of dynamics in electron glasses.
\newblock {\em Phys. Rev. B\/},~{\bf 73}, 214208.

\bibitem[\protect\citeauthoryear{Ovadyahu and Pollak}{Ovadyahu and
  Pollak}{1997}]{films1a-Zvi}
Ovadyahu, Z. and Pollak, M. (1997).
\newblock Disorder and magnetic field dependence of slow electronic relaxation.
\newblock {\em Phys. Rev. Lett.\/},~{\bf 79}, 459.

\bibitem[\protect\citeauthoryear{Paalanen, Rosenbaum, Thomas and
  Bhatt}{Paalanen {\em et~al.}}{1983}]{Rosenbaum}
Paalanen, M.~A., Rosenbaum, T.~F., Thomas, G.~A., and Bhatt, R.~N. (1983).
\newblock Critical scaling of the conductance in a disordered insulator.
\newblock {\em Phys. Rev. Lett.\/},~{\bf 51}, 1896.

\bibitem[\protect\citeauthoryear{Pankov and Dobrosavljevi\'c}{Pankov and
  Dobrosavljevi\'c}{2005}]{Pankov-scr}
Pankov, S. and Dobrosavljevi\'c, V. (2005).
\newblock Nonlinear screening theory of the {C}oulomb glass.
\newblock {\em Phys. Rev. Lett.\/},~{\bf 94}, 046402.

\bibitem[\protect\citeauthoryear{Pappas, Mezei, Ehlers, Manuel and
  Campbell}{Pappas {\em et~al.}}{2003}]{Pappas}
Pappas, C., Mezei, F., Ehlers, G., Manuel, P., and Campbell, I.~A. (2003).
\newblock Dynamic scaling in spin glasses.
\newblock {\em Phys. Rev. B\/},~{\bf 68}, 054431.

\bibitem[\protect\citeauthoryear{Pastor and Dobrosavljevi\'c}{Pastor and
  Dobrosavljevi\'c}{1999}]{Vlad-MITglass1}
Pastor, A.~A. and Dobrosavljevi\'c, V. (1999).
\newblock Melting of the electron glass.
\newblock {\em Phys. Rev. Lett.\/},~{\bf 83}, 4642.

\bibitem[\protect\citeauthoryear{Pollak}{Pollak}{1984}]{eglass5}
Pollak, M. (1984).
\newblock Non-ergodic behaviour of {A}nderson insulators with and without
  {C}oulomb interactions.
\newblock {\em Philos. Mag. B\/},~{\bf 50}, 265.

\bibitem[\protect\citeauthoryear{Pollak and Ortu{\~{n}}o}{Pollak and
  Ortu{\~{n}}o}{1982}]{eglass3}
Pollak, M. and Ortu{\~{n}}o, M. (1982).
\newblock Coulomb interactions in {A}nderson localized disordered systems.
\newblock {\em Sol. Energy Mater.\/},~{\bf 8}, 81.

\bibitem[\protect\citeauthoryear{Popovi\'c, Bogdanovich, Jaroszy\'nski and
  Klapwijk}{Popovi\'c {\em et~al.}}{2003}]{SB_SPIE03}
Popovi\'c, D., Bogdanovich, S., Jaroszy\'nski, J., and Klapwijk, T.~M. (2003).
\newblock Metal-insulator transition and glassy behavior in two-dimensional
  electron systems.
\newblock {\em Proceedings of SPIE\/},~{\bf 5112}, 99.

\bibitem[\protect\citeauthoryear{Popovi\'{c}, Fowler and Washburn}{Popovi\'{c}
  {\em et~al.}}{1997}]{DP_PRL}
Popovi\'{c}, D., Fowler, A.~B., and Washburn, S. (1997).
\newblock Metal-insulator transition in two dimensions: effects of disorder and
  magnetic field.
\newblock {\em Phys. Rev. Lett.\/},~{\bf 79}, 1543.

\bibitem[\protect\citeauthoryear{Pudalov, Brunthaler, Prinz and Bauer}{Pudalov
  {\em et~al.}}{1998{\em a}}]{nc3}
Pudalov, V.~M., Brunthaler, G., Prinz, A., and Bauer, G. (1998{\em a}).
\newblock Lack of universal one-parameter scaling in the two-dimensional
  metallic regime.
\newblock {\em JETP Lett.\/},~{\bf 68}, 442.

\bibitem[\protect\citeauthoryear{Pudalov, Brunthaler, Prinz and Bauer}{Pudalov
  {\em et~al.}}{1998{\em b}}]{nc5}
Pudalov, V.~M., Brunthaler, G., Prinz, A., and Bauer, G. (1998{\em b}).
\newblock Logarithmic temperature dependence of the conductivity of the
  two-dimensional metal.
\newblock {\em JETP Lett.\/},~{\bf 68}, 534.

\bibitem[\protect\citeauthoryear{Pudalov, D'{I}orio, Kravchenko and
  Campbell}{Pudalov {\em et~al.}}{1993}]{nc1}
Pudalov, V.~M., D'{I}orio, M., Kravchenko, S.~V., and Campbell, J.~W. (1993).
\newblock Zero-magnetic-field collective insulator phase in a dilute 2{D}
  electron system.
\newblock {\em Phys. Rev. Lett.\/},~{\bf 70}, 1866.

\bibitem[\protect\citeauthoryear{Rai\v{c}evi\'c, Jaroszy\'nski, Popovi\'c,
  Jelbert, Panagopoulos and Sasagawa}{Rai\v{c}evi\'c {\em
  et~al.}}{2007}]{IR_SPIE07}
Rai\v{c}evi\'c, I., Jaroszy\'nski, J., Popovi\'c, D., Jelbert, G.,
  Panagopoulos, C., and Sasagawa, T. (2007).
\newblock Low-temperature resistance noise in lightly doped
  {L}a$_{2-x}${S}r$_x${C}u{O}$_4$.
\newblock {\em Proceedings of SPIE\/},~{\bf 6600}, 660020.

\bibitem[\protect\citeauthoryear{Rai\v{c}evi\'c, Jaroszy\'nski, Popovi\'c,
  Panagopoulos and Sasagawa}{Rai\v{c}evi\'c {\em et~al.}}{2008}]{IR_PRL}
Rai\v{c}evi\'c, I., Jaroszy\'nski, J., Popovi\'c, D., Panagopoulos, C., and
  Sasagawa, T. (2008).
\newblock Evidence for charge glasslike behavior in lightly doped
  {L}a$_{2-x}${S}r$_x${C}u{O}$_4$ at low temperatures.
\newblock {\em Phys. Rev. Lett.\/},~{\bf 101}, 177004.

\bibitem[\protect\citeauthoryear{Rai\v{c}evi\'c, Popovi\'c, Panagopoulos and
  Sasagawa}{Rai\v{c}evi\'c {\em et~al.}}{2011}]{IR-inplane}
Rai\v{c}evi\'c, I., Popovi\'c, D., Panagopoulos, C., and Sasagawa, T. (2011).
\newblock Non-{G}aussian noise in the in-plane transport of lightly doped
  {L}a$_{2-x}${S}r$_x${C}u{O}$_4$: Evidence for a collective state of charge
  clusters.
\newblock {\em Phys. Rev. B\/},~{\bf 83}, 195133.

\bibitem[\protect\citeauthoryear{Reichhardt and Reichhardt}{Reichhardt and
  Reichhardt}{2004}]{Reich-noise}
Reichhardt, C. and Reichhardt, C. J.~Olson (2004).
\newblock Noise at the crossover from {W}igner liquid to {W}igner glass.
\newblock {\em Phys. Rev. Lett.\/},~{\bf 93}, 176405.

\bibitem[\protect\citeauthoryear{Richert}{Richert}{2002}]{Glotzer2}
Richert, R. (2002).
\newblock Heterogeneous dynamics in liquids: fluctuations in space and time.
\newblock {\em J. Phys.: Condens. Matter\/},~{\bf 14}, R703.

\bibitem[\protect\citeauthoryear{Rodriguez, Kenning and Orbach}{Rodriguez {\em
  et~al.}}{2003}]{sg-fullaging}
Rodriguez, G.~F., Kenning, G.~G., and Orbach, R. (2003).
\newblock Full aging in spin glasses.
\newblock {\em Phys. Rev. Lett.\/},~{\bf 91}, 037203.

\bibitem[\protect\citeauthoryear{Rosenbaum, Andres, Thomas and Bhatt}{Rosenbaum
  {\em et~al.}}{1980}]{Rosenbaum2}
Rosenbaum, T.~F., Andres, K., Thomas, G.~A., and Bhatt, R.~N. (1980).
\newblock Sharp metal-insulator transition in a random solid.
\newblock {\em Phys. Rev. Lett.\/},~{\bf 45}, 1723.

\bibitem[\protect\citeauthoryear{Rosenbaum, Field and Bhatt}{Rosenbaum {\em
  et~al.}}{1989}]{3Dbeta-1}
Rosenbaum, T.~F., Field, S.~B., and Bhatt, R.~N. (1989).
\newblock Variation of the metallic onset with magnetic field in doped
  germanium.
\newblock {\em Europhys. Lett.\/},~{\bf 10}, 269.

\bibitem[\protect\citeauthoryear{Rubi and Perez-Vicente}{Rubi and
  Perez-Vicente}{1997}]{glasses}
Rubi, M. and Perez-Vicente, C. (ed.) (1997).
\newblock {\em Complex behavior of glassy systems}.
\newblock Volume 492, Lecture Notes in Physics.
\newblock Springer, Berlin.

\bibitem[\protect\citeauthoryear{Sachdev}{Sachdev}{1999}]{Sachdev-book}
Sachdev, S. (1999).
\newblock {\em Quantum Phase Transitions}.
\newblock Cambridge University Press, UK.

\bibitem[\protect\citeauthoryear{Sarachik}{Sarachik}{1995}]{Myriam_review}
Sarachik, M.~P. (1995).
\newblock Transport studies in doped semiconductors near the metal-insulator
  transition.
\newblock In {\em The Metal-Nonmetal Transition Revisited: A Tribute to Sir
  Nevill Mott} (ed. P.~P. Edwards and C.~N. Rao). Francis and Taylor Ltd.

\bibitem[\protect\citeauthoryear{Sarachik, Simonian, Kravchenko, Bogdanovich,
  Dobrosavljevi\'c and Kotliar}{Sarachik {\em et~al.}}{1998}]{3Dbeta-3}
Sarachik, M.~P., Simonian, D., Kravchenko, S.~V., Bogdanovich, S.,
  Dobrosavljevi\'c, V., and Kotliar, G. (1998).
\newblock Metal-insulator transition in {S}i:{X} ({X}={P},{B}): Anomalous
  response to a magnetic field.
\newblock {\em Phys. Rev. B\/},~{\bf 58}, 6692.

\bibitem[\protect\citeauthoryear{Schmalian and Wolynes}{Schmalian and
  Wolynes}{2000}]{Schmalian}
Schmalian, J. and Wolynes, P.~G. (2000).
\newblock Stripe glasses: Self-generated randomness in a uniformly frustrated
  system.
\newblock {\em Phys. Rev. Lett.\/},~{\bf 85}, 836.

\bibitem[\protect\citeauthoryear{Scofield}{Scofield}{1987}]{Sco87}
Scofield, J.~H. (1987).
\newblock ac method for measuring low-frequency resistance fluctuation spectra.
\newblock {\em Rev. Sci. Instrum.\/},~{\bf 58}, 985--993.

\bibitem[\protect\citeauthoryear{Shashkin, Kravchenko and Klapwijk}{Shashkin
  {\em et~al.}}{2001}]{nc2}
Shashkin, A.~A., Kravchenko, S.~V., and Klapwijk, T.~M. (2001).
\newblock Metal-insulator transition in a 2{D} electron gas: Equivalence of two
  approaches for determining the critical point.
\newblock {\em Phys. Rev. Lett.\/},~{\bf 87}, 266402.

\bibitem[\protect\citeauthoryear{Sj{\"{o}}strand, Cole and
  Stiles}{Sj{\"{o}}strand {\em et~al.}}{1976}]{Stiles2-T2}
Sj{\"{o}}strand, M.~E., Cole, T., and Stiles, P.~J. (1976).
\newblock Low temperature saturation of the channel conductivity in silicon
  inversion layers.
\newblock {\em Surf. Sci.\/},~{\bf 58}, 72.

\bibitem[\protect\citeauthoryear{Sj{\"{o}}strand and Stiles}{Sj{\"{o}}strand
  and Stiles}{1975}]{Stiles1-T2}
Sj{\"{o}}strand, M.~E. and Stiles, P.~J. (1975).
\newblock The channel conductivity in n-type {S}i inversion layers at very low
  electron densities.
\newblock {\em Solid State Commun.\/},~{\bf 16}, 903.

\bibitem[\protect\citeauthoryear{Struik}{Struik}{1978}]{aging1}
Struik, L. C.~E. (1978).
\newblock {\em Physical aging in amorphous polymers and other materials}.
\newblock Elsevier, Amsterdam.

\bibitem[\protect\citeauthoryear{Thakur and Neilson}{Thakur and
  Neilson}{1996}]{MIT-glassothers1}
Thakur, J.~S. and Neilson, D. (1996).
\newblock Frozen electron solid in the presence of small concentrations of
  defects.
\newblock {\em Phys. Rev. B\/},~{\bf 54}, 7674.

\bibitem[\protect\citeauthoryear{Thakur and Neilson}{Thakur and
  Neilson}{1999}]{MIT-glassothers2}
Thakur, J.~S. and Neilson, D. (1999).
\newblock Phase diagram of the metal-insulator transition in two-dimensional
  electronic systems.
\newblock {\em Phys. Rev. B\/},~{\bf 59}, R5280.

\bibitem[\protect\citeauthoryear{Thorsm{\o}lle and Armitage}{Thorsm{\o}lle and
  Armitage}{2010}]{Armitage}
Thorsm{\o}lle, V.~K. and Armitage, N.~P. (2010).
\newblock Ultrafast (but many-body) relaxation in a low-density electron glass.
\newblock {\em Phys. Rev. Lett.\/},~{\bf 105}, 086601.

\bibitem[\protect\citeauthoryear{Tutuc, {D}e Poortere, Papadakis and
  Shayegan}{Tutuc {\em et~al.}}{2001}]{polarization3}
Tutuc, E., {D}e Poortere, E.~P., Papadakis, S.~J., and Shayegan, M. (2001).
\newblock In-plane magnetic field-induced spin polarization and transition to
  insulating behavior in two-dimensional hole systems.
\newblock {\em Phys. Rev. Lett.\/},~{\bf 86}, 2858.

\bibitem[\protect\citeauthoryear{Vaknin, Ovadyahu and Pollak}{Vaknin {\em
  et~al.}}{1998}]{films2-Zvi}
Vaknin, A., Ovadyahu, Z., and Pollak, M. (1998).
\newblock Evidence for interactions in nonergodic electronic transport.
\newblock {\em Phys. Rev. Lett.\/},~{\bf 81}, 669.

\bibitem[\protect\citeauthoryear{Vaknin, Ovadyahu and Pollak}{Vaknin {\em
  et~al.}}{2000}]{films3-Zvi}
Vaknin, A., Ovadyahu, Z., and Pollak, M. (2000).
\newblock Aging effects in an {A}nderson insulator.
\newblock {\em Phys. Rev. Lett.\/},~{\bf 84}, 3402.

\bibitem[\protect\citeauthoryear{Vaknin, Ovadyahu and Pollak}{Vaknin {\em
  et~al.}}{2002}]{films4-Zvi}
Vaknin, A., Ovadyahu, Z., and Pollak, M. (2002).
\newblock Nonequilibrium field effect and memory in the electron glass,.
\newblock {\em Phys. Rev. B\/},~{\bf 65}, 134208.

\bibitem[\protect\citeauthoryear{Verbruggen, Stoll, Heeck and Koch}{Verbruggen
  {\em et~al.}}{1989}]{Verbruggen89}
Verbruggen, A.~H., Stoll, H., Heeck, K., and Koch, R.~H. (1989).
\newblock A novel technique for measuring resistance fluctuations independently
  of background noise.
\newblock {\em Appl. Phys. A\/},~{\bf 48}, 233--236.

\bibitem[\protect\citeauthoryear{Vincent}{Vincent}{2007}]{Vincent-school}
Vincent, E. (2007).
\newblock Aging, rejuvenation and memory: the example of spin glasses.
\newblock In {\em Ageing and the Glass Transition} (ed. M.~Henkel,
  M.~Pleimling, and R.~Sanctuary), Volume 716, Lecture Notes in Physics, pp.\
  7--60. Springer.

\bibitem[\protect\citeauthoryear{Vitkalov, Sarachik and Klapwijk}{Vitkalov {\em
  et~al.}}{2001}]{polarization4}
Vitkalov, S.~A., Sarachik, M.~P., and Klapwijk, T.~M. (2001).
\newblock Spin polarization of two-dimensional electrons determined from
  {S}hubnikov–-de {H}aas oscillations as a function of angle.
\newblock {\em Phys. Rev. B\/},~{\bf 64}, 073101.

\bibitem[\protect\citeauthoryear{Vitkalov, Zheng, Mertes and Sarachik}{Vitkalov
  {\em et~al.}}{2000}]{polarization2}
Vitkalov, S.~A., Zheng, H., Mertes, K.~M., and Sarachik, M.~P. (2000).
\newblock Small angle {S}hubnikov-de {H}aas measurements in a 2{D} electron
  system: the effect of a strong in-plane magnetic field.
\newblock {\em Phys. Rev. Lett.\/},~{\bf 85}, 2164.

\bibitem[\protect\citeauthoryear{Washburn, Kim, Feng and Popovi\'c}{Washburn
  {\em et~al.}}{1999{\em a}}]{njk2}
Washburn, S., Kim, N.~J., Feng, X.~G., and Popovi\'c, D. (1999{\em a}).
\newblock Scaling laws, phase diagram, localized magnetic moments and {K}ondo
  effect in two-dimensional metals.
\newblock {\em Ann. Phys. (Leipzig)\/},~{\bf 8}, 569.

\bibitem[\protect\citeauthoryear{Washburn, Kim, Li and Popovi\'c}{Washburn {\em
  et~al.}}{1999{\em b}}]{njk3}
Washburn, S., Kim, N.-J., Li, K.~P., and Popovi\'c, D. (1999{\em b}).
\newblock Scaling and universal behavior near the two-dimensional
  metal-insulator transition.
\newblock {\em Mol. Phys. Rept.\/},~{\bf 24}, 150.

\bibitem[\protect\citeauthoryear{Watanabe, Itoh, Ootuka and Haller}{Watanabe
  {\em et~al.}}{1999}]{3Dbeta-4}
Watanabe, M., Itoh, K.~M., Ootuka, Y., and Haller, E.~E. (1999).
\newblock Metal-insulator transition of isotopically enriched
  neutron-transmutation-doped 70{G}e:{G}a in magnetic fields.
\newblock {\em Phys. Rev. B\/},~{\bf 60}, 15817.

\bibitem[\protect\citeauthoryear{Weissman}{Weissman}{1988}]{Weiss88}
Weissman, M.~B. (1988).
\newblock \textit{1/f} noise and other slow, nonexponential kinetics in
  condensed matter.
\newblock {\em Rev. Mod. Phys.\/},~{\bf 60}, 537.

\bibitem[\protect\citeauthoryear{Weissman}{Weissman}{1993}]{Weiss93}
Weissman, M.~B. (1993).
\newblock What is a spin glass? {A} glimpse via mesoscopic noise.
\newblock {\em Rev. Mod. Phys.\/},~{\bf 65}, 829.

\bibitem[\protect\citeauthoryear{Weissman, Israeloff and Alers}{Weissman {\em
  et~al.}}{1992}]{WeissMMM}
Weissman, M.~B., Israeloff, N.~E., and Alers, G.~B. (1992).
\newblock Spin-glass fluctuation statistics: mesoscopic experiments in
  {C}u{M}n.
\newblock {\em J. Magn. Magn. Mater.\/},~{\bf 114}, 87.

\bibitem[\protect\citeauthoryear{Wr\'obel, Jaroszy\'nski, Dietl, Regi\'nski and
  Bugajski}{Wr\'obel {\em et~al.}}{1998}]{Wrobel}
Wr\'obel, J., Jaroszy\'nski, J., Dietl, T., Regi\'nski, K., and Bugajski, M.
  (1998).
\newblock Conductance noise of submicron wires in the regime of quantum {H}all
  effect.
\newblock {\em Physica (Amsterdam)\/},~{\bf 256B-258B}, 69.

\endthebibliography

\end{document}